\documentclass[11pt,a4paper]{article}

\usepackage[T1]{fontenc}
\usepackage[utf8]{inputenc}
\DeclareUnicodeCharacter{2013}{\textendash}
\usepackage[a4paper,margin=1in]{geometry}
\usepackage{amsmath,amssymb,mathtools,mathrsfs}
\usepackage{graphicx}
\usepackage{booktabs}
\usepackage{siunitx}
\usepackage{makecell}
\usepackage{tikz}
\usepackage{xcolor}
\definecolor{AJ}{rgb}{0,0,0.6}
\usepackage{float}
\usepackage{url}
\usepackage{caption}
\usepackage[mathlines]{lineno}
\usepackage[numbers,sort&compress]{natbib}
\usepackage[hidelinks]{hyperref}

\makeatletter
\newcommand{\sectionpartlabel}[2]{%
  \phantomsection
  \begingroup
  \protected@edef\@currentlabel{\thesection(#1)}%
  \label{#2}%
  \endgroup
}
\newcommand{\bibentrylabel}[1]{%
  \begingroup
  \edef\@currentlabel{\NAT@num}%
  \edef\@currentHref{cite.#1}%
  \label{bib:#1}%
  \endgroup
}
\makeatother

\definecolor{AJ}{RGB}{0,40,180} 

\definecolor{BKorange}{HTML}{D55E00}

\newcommand{\be}{\begin{equation}}
\newcommand{\ee}{\end{equation}}

\begin{document}
\title{Spin--Momentum Impedance and Filtering at a Spin-Coupled Absorbing Boundary}
\author{Alireza Jozani\thanks{jozani.alireza@gmail.com; alireza.jozani@uni-tuebingen.de}\\
\small Departments of Mathematics and Physics, Eberhard-Karls-Universit\"at T\"ubingen\\
\small Auf der Morgenstelle 10, 72076 Tübingen, Germany}
\date{\today}
\maketitle

\begin{abstract}

In a separable harmonic waveguide, transverse confinement leaves the longitudinal bulk Hamiltonian unchanged. Earlier simulations nevertheless found that stronger confinement suppresses prompt detection of a spin-1/2 particle at a spin-coupled absorbing boundary. Here we identify the local boundary response underlying this effect. We treat the boundary as an effective detector, identifying the squared-norm loss rate with the detection-time probability density. The incident transverse ground state has a distribution of momenta parallel to the detecting plane, with characteristic scale proportional to \(\sqrt{\omega}\), where \(\omega\) is the confinement frequency. For each parallel momentum, we resolve the boundary relation into two spin channels. At fixed normal wave number and detector parameter, local plane-wave matching gives these channels equal absorption probabilities but generally different reflection phases. Their common absorption probability decreases with the magnitude of the parallel momentum. A stationary multichannel scattering calculation including the harmonic confinement confirms that absorption decreases with increasing confinement over the sampled range, even without a reflecting lower wall. Time-dependent simulations show that stronger confinement reduces detection probability and, at longer observation horizons, increases the conditional mean of recorded click times. The mean detection time capped at horizon \(T\) is summarized by the empirical fit \(A_T+B_T\sqrt{\omega}\) over the sampled range, with setup- and horizon-dependent coefficients. The local response makes detection sensitive to transverse confinement; propagation, interference, and repeated encounters of the undetected wave shape the detection-time distribution.

\end{abstract}
\vspace{1em}
\section{Introduction}
Predicting when a quantum detector clicks requires more than the usual Born rule for position at a prescribed time. Proposed descriptions of arrival and detection times include flux-based rules, detector-free trajectory arrival times, models using complex absorbing potentials or absorbing boundary conditions~\cite{ML2000,All69b,VHD13,DD19,Tum22,Tum23,RSM23}, and, more recently, quantum-stroboscopic measurement protocols~\cite{Lloyd26}. An important question is how these descriptions relate to the click-time statistics recorded by a physical detector.

A concrete example is the harmonically confined waveguide studied by Das and D\"urr ~\cite{DD19}. They calculated spin-dependent first-arrival distributions from Bohmian trajectories without including detector-induced changes in the Pauli evolution. They proposed these intrinsic distributions as experimental benchmarks, expecting a suitable detector to cause only a small disturbance. However, a detector-free arrival time and a detector-present click time need not coincide: the detector can alter the particle’s evolution before a click, so its response must be included when predicting the measured distribution~\cite{VHD13,GTZ24, GTZ24Spin}.

The relation between these intrinsic arrival times and recorded clicks is constrained by standard quantum measurement theory. For a fixed detector arrangement, click-time probabilities are described by a positive-operator-valued measure (POVM)~\cite{VHD13,Tum22,GTZ24Spin}. In particular, no single fixed detector can reproduce the intrinsic arrival-time distributions of Ref.~\cite{DD19} exactly for all initial spin orientations~\cite{GTZ24, GTZ24Spin}. These constraints do not, however, determine the response of a particular detector, which must be specified separately~\cite{Tum22,JT26}.

Two effective approaches to specifying such a detector response are absorbing boundary conditions (ABCs) and complex absorbing potentials (CAPs)~\cite{Tum22, Tum23, HY09, JT26}.  A CAP represents absorption within a spatial region through a negative imaginary potential, whereas an ABC represents it by a boundary law at the detecting surface~\cite{Werner1987}. Both modify the particle’s pre-click evolution and can produce reflection as well as absorption~\cite{JT26}. For scalar ABCs, Cavendish and Das found scattering examples in which boundary reflection leads to detection probabilities different from standard scattering-theory predictions even in the far-field limit~\cite{CD25}. 

We adopt the spin-coupled ABC as an effective detector
model, identifying the squared norm of the evolving
subnormalized state with the no-click probability.
With no other loss channels, its loss rate equals the
outward Pauli flux integrated over the detecting surface
and defines the click-time density. This interpretation
does not resolve microscopic apparatus dynamics and is
distinct from the numerical use of absorbers for
unbound-particle analysis~\cite{Sel21} or radiation conditions~\cite{SST23}.

In our recent study~\cite{JT26}, we compared these detector
models in the same waveguide geometry. For the separable
preparation, the purely longitudinal CAP and componentwise
scalar ABC yielded detector-position-integrated click-time
distributions independent of transverse confinement.
The spin-coupled ABC instead produced confinement-dependent
redistribution and delayed structure, with no noticeable
dependence on the initial spin orientation in the cases
tested. Since the longitudinal bulk Hamiltonian remains
unchanged, the question is how the spin-coupled boundary
introduces this confinement dependence and how reflected,
undetected amplitude contributes to later detection.

The present setup keeps the same waveguide geometry, with harmonic
transverse confinement of frequency $\omega$, a reflecting end at
$z=0$, and a detecting plane at $z=L$. The spin-$1/2$ particle is
initially prepared with the transverse ground-state profile and a
longitudinal Gaussian wavepacket centered at $z=L/2$, with central
momentum along $+z$. At the detecting plane, we adopt the spin-coupled
ABC 
\begin{equation*}
\left.(\boldsymbol{\sigma}\cdot\boldsymbol \nabla)\Psi\right|_{z=L}
=
i\kappa(\mathbf n\cdot\boldsymbol{\sigma})
\left.\Psi\right|_{z=L},
\qquad \mathbf n=\hat{\mathbf{z}},
\end{equation*}
where $\mathbf n$ is the outward unit normal,
$\boldsymbol{\sigma}=(\sigma_x,\sigma_y,\sigma_z)$ is the vector
of Pauli matrices, and $\kappa>0$ is an inverse-length parameter
governing absorption and reflection.  For each tangential wave vector $\boldsymbol{\xi}$,
the spin-coupled boundary relates the normal derivative
of the corresponding Fourier component to its boundary
value through a matrix-valued impedance with eigenvalues
$i\kappa\pm|\boldsymbol{\xi}|$. We examine the resulting
momentum-dependent absorption and spin-dependent
reflection phases through local plane-wave matching and
stationary multichannel scattering in a harmonic guide.

We also derive a covariance representation of the first
inward variation of an auxiliary time moment, separating
the characteristic transverse momentum scale from the
correlations encoded in the evolved boundary trace.
Time-dependent simulations and boundary controls then
investigate how this response, propagation, interference,
and subsequent detector encounters shape the click-time
distribution.

Together, these analyses extend the numerical findings
of Ref.~\cite{JT26} and help distinguish detector-induced changes
from bulk propagation effects. The boundary impedance also
provides a theoretical target for momentum-selective
absorption and spin-dependent phase control of reflected
matter waves. Supporting derivations, numerical details,
control calculations, and physical scales are given in
the Supplementary Material.

\section{Model and Detector-Present Observable}
\label{sec:setup}

We work in units \(\hbar=m=1\). The spinor obeys
\(i\partial_t\Psi=H\Psi\), with spin-independent bulk
differential expression
\begin{equation}
H=-\frac12\Delta
+\frac12\omega^2
\left[(x-x_c)^2+(y-y_c)^2\right],
\end{equation}
where $(x_c,y_c)$ is the transverse trap center.
The numerical transverse domain has Dirichlet walls at
fixed distances in oscillator units; its scaling and
resolution are specified in Supplementary Material,
Sec.~S2.  

The initial preparation has the factorized profile
\begin{equation}
\Psi_0(x,y,z)=\chi_\omega(x,y)\,\phi_0(z)\,\eta,
\label{eqn:seperable}
\end{equation}
where \(\eta\in\mathbb C^2\) is a constant normalized spinor
and \(\chi_\omega\) is the normalized transverse harmonic
ground-state profile, with oscillator length
\(\ell_\perp=\omega^{-1/2}\).
For a guide of length \(L>0\), the longitudinal profile is
\begin{equation}
\phi_0(z)=\mathcal{N}_{\parallel}
\exp\left[
-\frac{(z-z_c)^2}{2\sigma_{\parallel}^2}
+ik_z(z-z_c)
\right],
\qquad 0<z<L,
\label{eqn:refereeSlove}
\end{equation}
where \(\mathcal N_\parallel\) is the normalization constant,
\(z_c=L/2\) is the packet center, \(k_z>0\) is the central
wave number along \(+z\), and \(\sigma_\parallel\) is the
Gaussian amplitude-width parameter.

At the reflecting end, we impose the Dirichlet condition \(\Psi(x,y,0,t)=0\). Equation (3) specifies the interior preparation profile. For the reference parameters, the unmodified Gaussian has a nonzero lower-wall amplitude equal to \(e^{-200}\) times its central amplitude. Numerically, the sampled spinor is set to zero at all Dirichlet boundary nodes before normalization in the discrete norm (Supplementary Material, Sec. S2). This numerical prescription does not imply that the unmodified Gaussian satisfies the continuum boundary conditions pointwise. A separate test uses a smooth preparation that vanishes near every boundary. On the reference grid, at \(\omega=300\) and \(T=40\), the resulting changes in the detected fraction and capped mean are much smaller than the changes observed under numerical refinement (Supplementary Material, Sec. S6.4).

The upper end of the guide is the detecting plane
\begin{equation}
\Sigma_L=\{z=L\}.
\end{equation}
At this plane, we adopt the spin-coupled absorbing boundary
condition as a defining assumption of the effective
detector model:

\begin{equation}
(\boldsymbol{\sigma}\!\cdot\!\boldsymbol{\nabla})\Psi
=
i\kappa\sigma_z\Psi,
\,\quad \kappa>0 .
\label{eqn:spinorABC}
\end{equation}
Equation~\eqref{eqn:spinorABC} also arises from the semi-ideal Dirac absorbing-boundary model in the nonrelativistic limit with the boundary-parameter scaling specified in Ref.~\cite{Tum16}. This is a limiting relation between idealized models. Derivations of scalar absorbing-boundary detection laws from detector models have also been investigated~\cite{Tum24Derivation,Frolov26Arrival}. Mathematical results for related scalar absorbing-boundary problems are given in Refs.~\cite{FTT25, Frolov25}. For sufficiently regular spinor solutions, the Pauli current is
\begin{equation}
\boldsymbol j^P=\operatorname{Im}(\Psi^\dagger\boldsymbol \nabla\Psi)
+\frac12\boldsymbol \nabla\times(\Psi^\dagger\boldsymbol \sigma\Psi).
\end{equation}
The spinor ABC gives the outward normal probability-current density
\begin{equation}
j_z^P(x,y,L,t)=\kappa\rho(x,y,L,t),\qquad \rho=\Psi^\dagger\Psi.
\label{eqn:flux}
\end{equation}
Let \(\tau\in[0,\infty]\) denote the random time of the detector
click in a single realization, with \(\tau=\infty\) if no detection
ever occurs. By contrast, \(t\geq 0\) is the deterministic evolution
time and serves as the integration variable below; \(T>0\) is a
fixed observation horizon. For normalized initial data, the model
identifies
\(\|\Psi_t\|^2=\mathbb P(\tau>t)\).
With no flux through the nondetecting boundaries, the continuity
equation then gives the density of finite click times,
\begin{equation}
g(t;\omega)
=
\kappa\int_{\Sigma_L}\rho(x,y,L,t)\,dxdy
=
-\frac{d}{dt}\|\Psi_t\|^2 .
\end{equation}
For a finite observation horizon \(T\), define the detected fraction
\begin{equation}
D_{T}(\omega)=\int_0^{T}g(t;\omega)\,dt,
\end{equation}
and the finite-window restricted mean detection time, with click times capped at $T$,
\begin{equation}
\mu^*( T;\omega)
=
\int_0^{ T}S(t;\omega)\,dt,
\qquad
S(t;\omega)=\|\Psi_t\|^2.
\end{equation}
Equivalently,
\begin{equation}
\mu^*(T;\omega)=\mathbb{E}[\min(\tau,T)].
\end{equation}
For $D_T(\omega)>0$, define $M_T(\omega)=\int_0^T t\,g(t;\omega)\,dt$.
The mean click time conditioned on detection by $T$ is
$\overline{\tau}^{\,\mathrm{cond}}_T(\omega)=M_T(\omega)/D_T(\omega)$.
Then
\begin{equation}
\mu^*(T;\omega)
 =D_T(\omega)\,
  \overline{\tau}^{\,\mathrm{cond}}_T(\omega)
  +T[1-D_T(\omega)].
\label{eq:restricted-mean-decomposition}
\end{equation}
The detection-time density is not renormalized over the observation window; its integral is \(D_T(\omega)=1-S(T;\omega)\), which generally depends on confinement. The restricted mean summarizes this finite window without extrapolating the unobserved tail and includes the contribution \(TS(T;\omega)\) from realizations undetected by \(T\). Since \(\partial_T\mu^*(T;\omega)=S(T;\omega)\), it increases with the observation horizon while survival probability remains positive. Numerical convergence is assessed separately at fixed \(T\). Reporting \(D_T\) and the conditional mean alongside \(\mu^*\) distinguishes reduced detection probability from changes in recorded-click timing (Supplementary Material, Sec. S3).

The principal numerical results in the following section
are obtained using Cartesian finite differences and
Crank--Nicolson time stepping. The spinor ABC is implemented
using a centered ghost-point approximation and a compatible
discrete norm, replacing the earlier one-sided normal-boundary
closure. The continuum boundary condition and detector-loss
law remain unchanged (Supplementary Material, Sec.~S2).

\section{Confinement-dependent detection statistics}

\begin{figure}[H]
  \centering
  \includegraphics[width=0.82\linewidth]{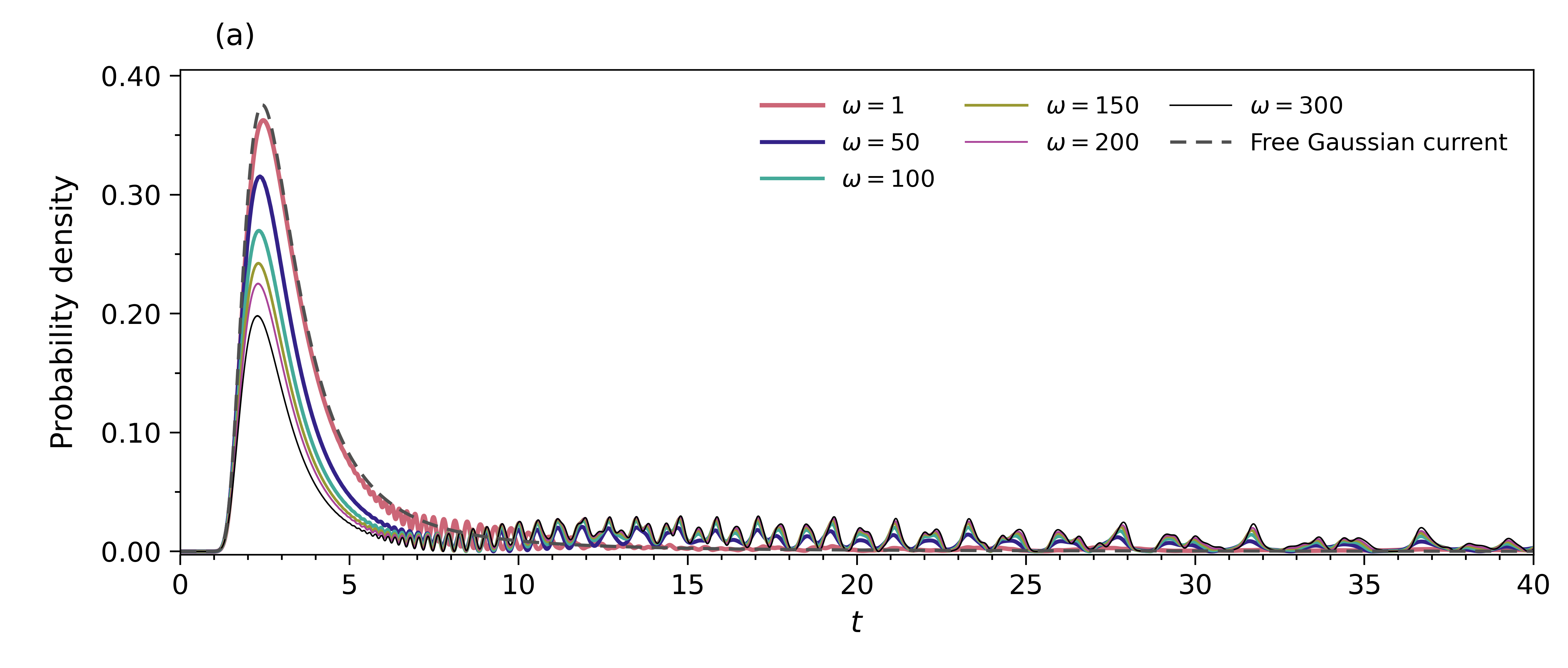}
  \includegraphics[width=0.82\linewidth]{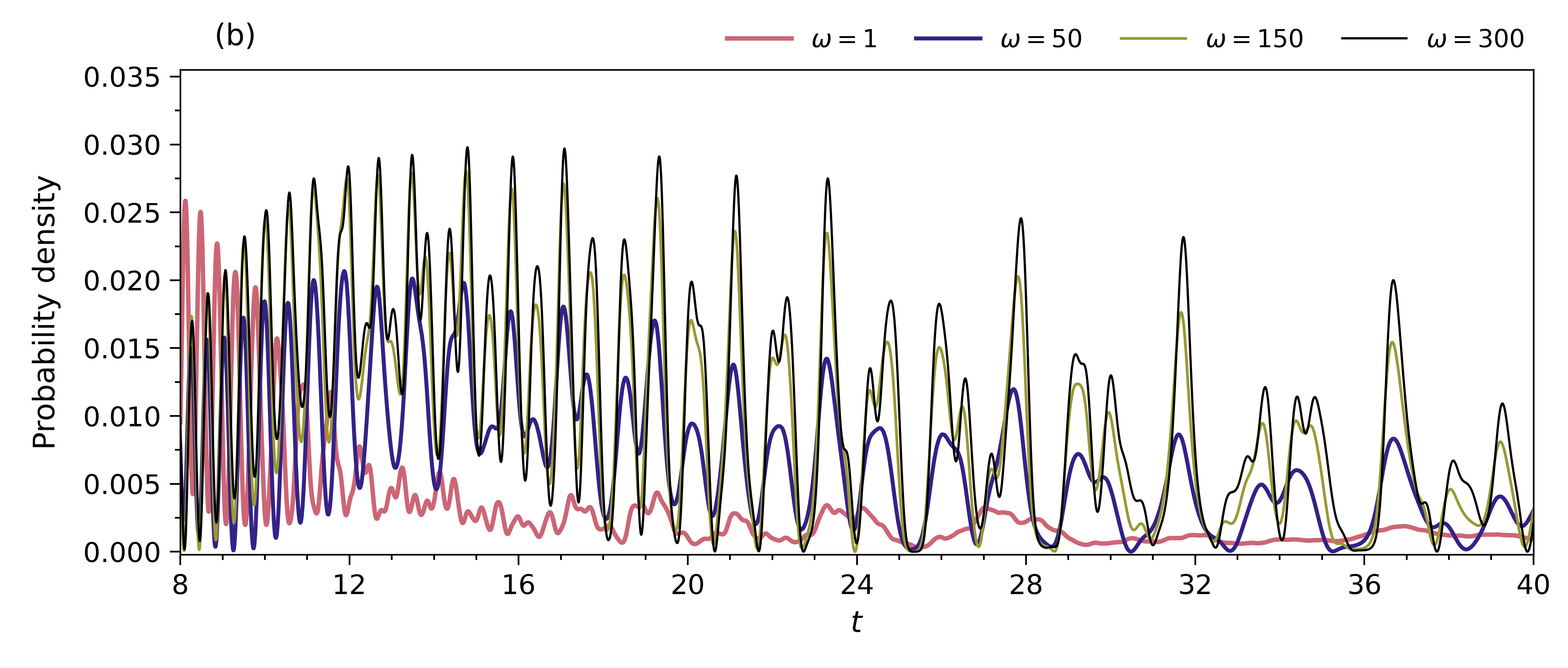}
  \caption{Detector-present click-time density $g(t;\omega)$ for
representative confinements, with $L=20$, $\kappa=k_z=\pi$,
$\sigma_\parallel=0.5$, and $\eta=(1,0)^{\mathsf T}$.
(a) Profiles for $\omega=1,50,100,150,200,300$ over
$0\le t\le40$; the black dashed curve is the detector-free
Gaussian comparison current.
(b) Late-time detail over $8\le t\le40$ for
$\omega=1,50,150,300$, with the same colors as in (a).
The detector-present curves are not conditioned on detection
by $t=40$; their areas in (a) equal the detected fractions
$D_{40}(\omega)$.
The choice $\kappa=k_z$ matches the scalar boundary to the
central longitudinal plane-wave component. 
Numerical settings and detected fractions are given in
Supplementary Material, Sec.~S2 and Table~S2, respectively.}
\label{fig:click-time-profiles}
\end{figure}

Figure~\ref{fig:click-time-profiles} shows the click-time density for representative
confinements, with the longitudinal preparation, detector
parameter, initial spinor, and guide length fixed.
Increasing $\omega$ suppresses the prompt peak and
reshapes the subsequent oscillatory signal.

To quantify this redistribution, we use a common cutoff
$t_{\mathrm{cut}}=3.6$, defining early and late observation
windows $[0,t_{\mathrm{cut}}]$ and $(t_{\mathrm{cut}},T]$.
This is a diagnostic division, not an encounter-resolved
decomposition.

Between $\omega=1$ and $300$, the early detected probability
decreases from approximately $0.58$ to $0.30$, while its
conditional mean changes only from $2.60$ to $2.50$.
At $T=40$, the fraction of recorded clicks in the late
window increases from $0.39$ to $0.59$, and their conditional
mean increases from $7.84$ to $17.09$.
Stronger confinement therefore suppresses prompt detection
rather than uniformly delaying the initial wavepacket
(Supplementary Material, Secs.~S3.2--S3.3 and Table~S2).

The lower panel in Fig.~\ref{fig:click-time-profiles} resolves the late-time structure; the dashed detector-free Gaussian current in the upper panel provides the corresponding free-crossing comparison. Numerical settings are given in Supplementary Material, Sec.~S2.

\begin{figure}[H]
\centering
\setlength{\tabcolsep}{3pt}
\begin{tabular}{@{}cc@{}}
\multicolumn{2}{c}{
\includegraphics[width=0.7\textwidth]{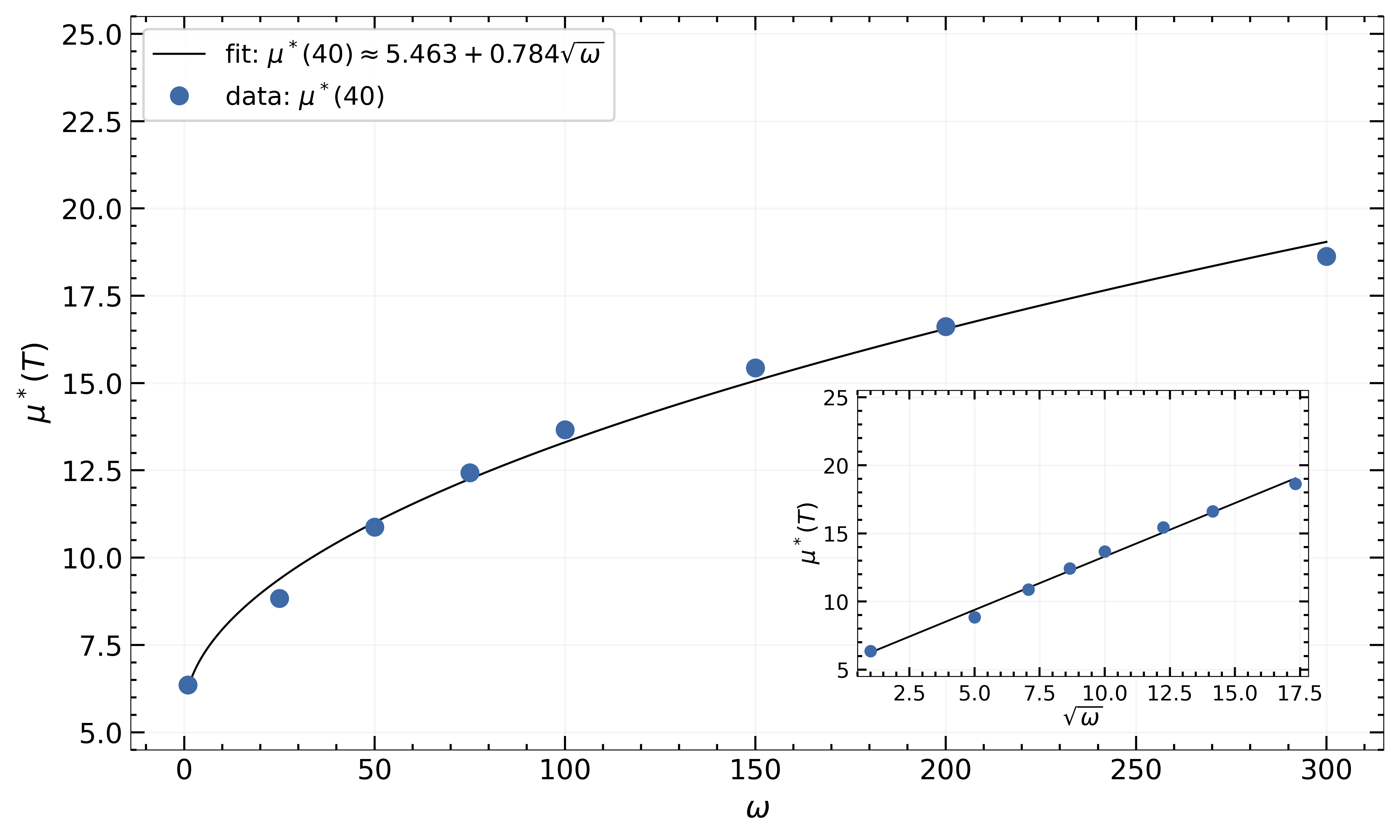}
}
\\[-1mm]
\multicolumn{2}{c}{\small restricted mean detection time, \(T=40\)}
\end{tabular}
\caption{Restricted mean detection time at \(T=40\) for \(\omega\in\{1,25,50,75,100,150,200,300\}\), with the preparation of Fig.~\ref{fig:click-time-profiles}. Markers are calculated values; the solid line is the unweighted fit in Eq.~\eqref{eqn:mu}. The inset shows the same data against \(\sqrt{\omega}\). Fit residuals and numerical integration conventions are given in Supplementary Material, Sec.~S3.}
\label{fig:confinement-sweep}
\end{figure}

Figure~\ref{fig:confinement-sweep} summarizes the restricted mean at \(T=40\) for the eight sampled confinements. An unweighted least-squares fit gives
\begin{equation}
\mu^*(40;\omega)\simeq5.463+0.784\sqrt{\omega}.
\label{eqn:mu}
\end{equation}
Since $0\leq\mu^*(T;\omega)\leq T$, a fit of this form with nonzero slope can only describe a finite confinement range at fixed $T$. Equation~\eqref{eqn:mu} describes the sampled confinements at $T=40$; extrapolation beyond this range or horizon is not established.

\begin{figure}[H]
 \centering
 \includegraphics[width=\linewidth]{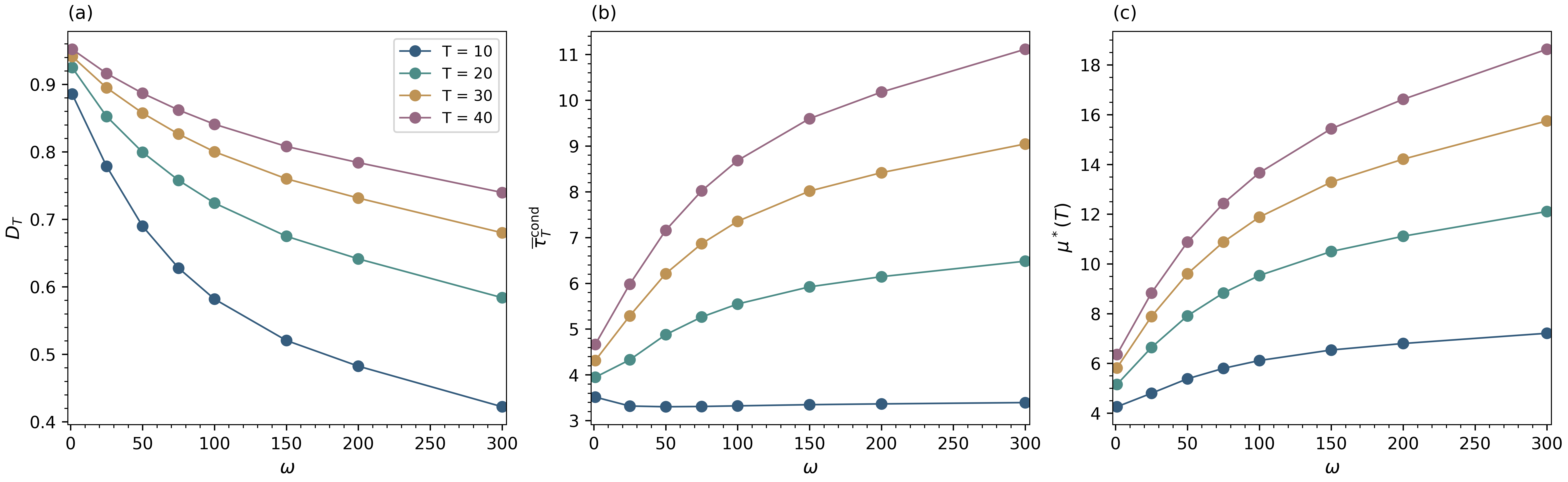}
 \caption{Observation-window dependence for the eight reference confinements: (a) detected fraction \(D_T(\omega)\); (b) conditional mean detection time \(\bar\tau_T^{\mathrm{cond}}(\omega)=M_T(\omega)/D_T(\omega)\), where \(M_T(\omega)=\int_0^T t\,g(t;\omega)\,dt\); and (c) restricted mean \(\mu^*(T;\omega)\). Each panel shows \(T=10,20,30,40\), evaluated from prefixes of the same recorded evolutions. Markers are calculated values; connecting lines guide the eye.}
 \label{fig:comparision_horizon}
\end{figure}

To distinguish changes in detection probability from changes in the timing of recorded clicks, Fig.~\ref{fig:comparision_horizon} compares \(D_T(\omega)\), \(\bar\tau_T^{\mathrm{cond}}(\omega)\), and \(\mu^*(T;\omega)\) at \(T=10,20,30,40\), using prefixes of the same recorded evolutions. Across the sampled confinements, \(D_T\) decreases and \(\mu^*\) increases at every displayed horizon. The conditional mean increases with confinement at \(T=20,30,40\), but varies weakly and nonmonotonically at \(T=10\). This illustrates Eq.~\eqref{eq:restricted-mean-decomposition}: a larger undetected fraction can raise \(\mu^*\) even when the conditional mean changes little.

At $\omega=300$, about 26\% of the probability remains undetected at $T=40$; the survival contribution $40[1-D_{40}(300)]$ accounts for about 56\% of $\mu^*(40;300)$.

Spatial and temporal refinement checks at $\omega=300$ and $T=40$ support the stability of the principal detection statistics. Increasing $N_z$ from 1500 to 4000 changes the restricted mean from 18.63 to 19.15. Comparing the two finer longitudinal grids ($N_z=2999$ and $4000$) gives discrepancies of 0.89\% in the restricted mean and 0.50\% in the normalized integrated absolute detector-rate difference over $8\leq t\leq40$. The detailed time profiles are more sensitive to resolution on the reference grid (Supplementary Material, Sec.~S6.3 and Table~S5). At the reference spatial grid, time-step halving changes the same late-window rate measure by 5.33\%, while changing the restricted mean by 0.031\%; these tested-grid differences are not continuum error bounds (Supplementary Material, Sec.~S6.3 and Table~S5).

At the baseline resolution for \(\omega=300\), replacing the reference Gaussian with a smooth, boundary-compatible preparation changes \(D_{40}\) and \(\mu^*(40)\) by only \(1.03\times10^{-6}\) and \(1.59\times10^{-5}\), respectively. The normalized integrated absolute detector-rate discrepancy over \(8\leq t\leq40\) is \(3.51\times10^{-4}\%\) (Supplementary Material, Sec.~S6.4). These differences are much smaller than those observed under spatial and temporal refinement.

An additional long-time calculation at $\omega=200$ gives $D_{40}=0.784$ and $D_{200}=0.943$, with the window-averaged detector rate decreasing by more than an order of magnitude between the intervals $20\leq t\leq40$ and $160\leq t\leq200$ (Supplementary Material, Sec. S9.6 and Fig. S13). The unrestricted mean remains undetermined by these finite-window data.

\section{Spin--Momentum Boundary Impedance and Scattering}
\label{sec:SpinMomentum_Boundary}

The numerical results above raise two related questions: why stronger
transverse confinement suppresses prompt detection, and how the
undetected amplitude contributes to the later detector signal.
We first isolate the separable bulk contribution and examine
reflection and absorption at the spin-coupled boundary, both at
fixed tangential momentum and with the transverse harmonic
confinement included explicitly.

\textbf{(i) A separable first-pass bulk reference.}
For comparison, consider an auxiliary bulk evolution whose
Hamiltonian acts identically on the two spin components,
\(H_{\mathrm{bulk}}=[H_{xy}(\omega)+H_z]I_2\).
Here \(H_{xy}(\omega)\) contains the transverse kinetic and harmonic
terms, while \(H_z\) is the longitudinal kinetic operator.
For the factorized initial state in Eq.~\eqref{eqn:seperable}, separability preserves
the form
\begin{equation}
\Psi_{\rm fp}(x,y,z,t)=\chi_\omega(x,y,t)\phi_{\rm fp}(z,t)\eta,
\end{equation}
where \(\phi_{\mathrm{fp}}(z,0)=\phi_0(z)\) and the prepared
spinor \(\eta\) is unchanged. Here ``spin independent'' refers to
the Hamiltonian, not to the state, which still depends on \(\eta\).
The corresponding Pauli current is
\begin{equation}
\boldsymbol j_{\rm fp}^{P}
=
\boldsymbol j_{\rm fp}^{\rm conv}
+
\frac12\boldsymbol \nabla\times(\rho_{\rm fp}\boldsymbol s_\eta),
\quad
\boldsymbol s_\eta:=\eta^\dagger\boldsymbol\sigma\eta,
\quad
\rho_{\rm fp}:=|\Psi_{\rm fp}|^2 .
\end{equation}
For this auxiliary separable field, $\mathbf s_\eta$ is
spatially constant, so the normal spin-curl term is a
transverse divergence. By Stokes' theorem on the transverse
cross-section, together with the lateral Dirichlet conditions,
its cross-section integral vanishes. Hence, using transverse normalization, the flux through an interior plane \(\Sigma_b=\{z=z_b\}\), with \(z_b<L\), satisfies
\begin{equation}
\int_{\Sigma_b}j^{P}_{{\rm fp},z}\,dA
=
\int_{\Sigma_b}
\operatorname{Im}(\Psi_{\rm fp}^\dagger\partial_z\Psi_{\rm fp})\,dA
=
\operatorname{Im}
(\phi_{\rm fp}^*\partial_z\phi_{\rm fp})(z_b,t).
\end{equation}
This plane-integrated flux is independent of \(\omega\) and of the
prepared spin orientation, although the local current may depend
on spin. The full spinor-ABC evolution need not preserve the
factorized form. Details of the auxiliary bulk calculation are
given in Supplementary Material, Sec.~S4.

\sectionpartlabel{ii}{sec:boundary-scattering}
\noindent\textbf{(ii) Boundary spin channels, reflection, and absorption.}
The separable reference isolates the longitudinal bulk flux.
In the detector-present problem, the boundary condition
couples the normal derivative to transverse motion and spin.
At $z=L$, the spinor ABC in Eq.~\eqref{eqn:spinorABC} reads
\begin{equation}
\begin{aligned}
\partial_z \psi_\uparrow
&=
i\kappa\,\psi_\uparrow
-
(\partial_x-i\partial_y)\psi_\downarrow,
\\
\partial_z \psi_\downarrow
&=
i\kappa\,\psi_\downarrow
+
(\partial_x+i\partial_y)\psi_\uparrow.
\end{aligned}
\label{eq:abc-comp}
\end{equation}
Equivalently,
\begin{equation}
\left.\partial_z\Psi\right|_{z=L}
=
\left. C\,\Psi\right|_{z=L},
\quad
 C=
\begin{pmatrix}
i\kappa & -D_-\\
D_+ & i\kappa
\end{pmatrix},
\quad
D_\pm=\partial_x\pm i\partial_y .
\label{eqn:boundary_symbol}
\end{equation}
Let \(\widehat\Psi(\boldsymbol\xi,z,t)\) be the Fourier transform
of the spinor in the tangential coordinates \((x,y)\), where
\(\boldsymbol\xi=(\xi_x,\xi_y)\) is the corresponding tangential
wave vector. At the detecting plane,
\begin{equation}
\partial_z\widehat\Psi(\boldsymbol\xi,L,t)
=
\mathcal C(\boldsymbol\xi)\widehat\Psi(\boldsymbol\xi,L,t).
\end{equation}
We call \(\mathcal C(\boldsymbol{\xi})\) the matrix-valued boundary impedance: it relates each tangential Fourier component of the boundary spinor to its outward normal derivative. From Eq.~\eqref{eqn:boundary_symbol},
\begin{equation}
\mathcal C(\boldsymbol \xi)=i\kappa I+(\hat{\mathbf{z}}\times \boldsymbol \xi)\cdot\boldsymbol \sigma .
\label{eqn:C}
\end{equation}
For \(R:=|\boldsymbol\xi|>0\), write
\begin{equation}
\Gamma(\boldsymbol \xi)=\frac{(\hat{\mathbf{z}}\times \boldsymbol \xi)\cdot\boldsymbol\sigma}{R},
\qquad
\Gamma(\boldsymbol\xi)^\dagger=\Gamma(\boldsymbol\xi),
\qquad
\Gamma(\boldsymbol\xi)^2=I.
\end{equation}
Therefore Eq.~\eqref{eqn:C} can be rewritten as
\begin{equation}
\mathcal C(\boldsymbol \xi)=i\kappa I+R\Gamma(\boldsymbol \xi).
\end{equation}
The corresponding orthogonal spectral projectors are
\begin{equation}
\Pi_\pm(\boldsymbol \xi)=\frac12\bigl(I\pm\Gamma(\boldsymbol \xi)\bigr).
\label{eqn:proj}
\end{equation}
The projectors \(\Pi_\pm(\boldsymbol{\xi})\) select particle-spin orientations along and opposite to \(\boldsymbol n_\xi=(\hat{\mathbf{z}}\times\boldsymbol{\xi})/R\), respectively. They diagonalize the local boundary symbol, not the full harmonically confined Hamiltonian, whose transverse potential couples different Fourier components. At \(R=0\), the branches are degenerate. The boundary-symbol eigenvalues are
\begin{equation}
\lambda_\pm(\boldsymbol \xi)=i\kappa\pm R.
\end{equation}
The boundary therefore introduces two distinct scales:
the detector parameter $\kappa$, which determines the
local probability-loss law, and the tangential wave-number
magnitude $R$, which enters through the spin-dependent
Hermitian part of the impedance. The same boundary
eigenvalues govern both reflection and absorption and
the local normal response of the evolved boundary trace.
We first examine their scattering consequences.

\noindent\textbf{(iii) Local reflection and absorption.} To interpret the scattering significance of these boundary eigenvalues, consider an auxiliary locally planar matching problem with fixed tangential wave vector \(\boldsymbol{\xi}\) and positive normal wave number \(k\), neglecting the coupling between Fourier components induced by the transverse harmonic potential. For either boundary-spin branch, we write the normal amplitude as a superposition of incident and reflected waves,
\begin{equation}
\psi_\pm(z)
 = e^{ik(z-L)}+r_\pm e^{-ik(z-L)},
\qquad z\le L.
\end{equation}
Imposing the corresponding boundary relation gives
\begin{equation}
ik(1-r_\pm)=(i\kappa\pm R)(1+r_\pm),
\end{equation}
and hence
\begin{equation}    
r_\pm(k,R)
 =\frac{i(k-\kappa)\mp R}{i(k+\kappa)\pm R},
\qquad
A_\pm(k,R)
 =1-|r_\pm(k,R)|^2
 =\frac{4k\kappa}{(k+\kappa)^2+R^2}.
 \label{eqn:scattering}
\end{equation}
Here \(A_\pm\) is the fraction of incident normal flux absorbed at the boundary. For unit incident amplitude, the detector law \(j_z^P=\kappa\rho\) gives \(A_\pm=(\kappa/k)|1+r_\pm|^2\). Thus, the imaginary impedance term \(i\kappa I\) fixes the positive local detector-flux law, while the Hermitian spin–momentum term \(R\Gamma\) modifies the boundary density through matching. At fixed \(k\) and \(\kappa\), absorption decreases with \(R\) equally in both branches.

More precisely, the local reflection operator can be written as

\begin{equation}
\mathcal R_{\mathrm{loc}}=r_+\Pi_++r_-\Pi_-,
\qquad
\mathcal R_{\mathrm{loc}}^\dagger\mathcal R_{\mathrm{loc}}=(1- A)I,
\quad A=A_\pm.
\end{equation}
Its generally different branch phases can rotate a coherent incident spinor about \(\boldsymbol n_\xi\) (Supplementary Material, Sec. S5.3). Here ``spin–momentum filtering'' denotes this combination of tangential-momentum-dependent absorption and spin-dependent reflection phases, not preferential local absorption of one spin branch at fixed incident wave vector.

By contrast, the componentwise scalar ABC gives \(r_{\mathrm{sc}}(k)=(k-\kappa)/(k+\kappa)\), independent of tangential momentum at fixed \(k\).

For the incident harmonic ground state, the characteristic tangential wave-number magnitude grows as $\ell_\perp^{-1}=\sqrt{\omega}$. Equation~\eqref{eqn:scattering} thus explains why stronger confinement favors components with lower absorption probabilities in the local matching problem.

\sectionpartlabel{iv}{sec:harmonic-scattering}
\noindent\textbf{(iv) Stationary scattering in the harmonic waveguide.}
The local plane-wave matching treats the tangential momentum
as fixed, whereas the transverse harmonic potential couples
different tangential Fourier components. To examine the boundary
response with the confinement included explicitly, we consider
stationary scattering in a semi-infinite harmonic guide,
$z\leq L$, terminated by the same spinor ABC.
The transverse domain is $\mathbb{R}^2$, and the lower
reflecting wall is absent.

Let $\chi_{n_x,n_y}^{(\omega)}$ denote the normalized Cartesian
eigenfunctions of the transverse harmonic Hamiltonian
$H_{xy}(\omega)$, with energies
\begin{equation}
E_{n_x,n_y}^{\perp}=\omega(n_x+n_y+1).
\end{equation}
We consider a particle incident on the detector in the transverse
ground state with spin up and longitudinal wave number $k>0$,
\begin{equation}
\Psi_E^{\mathrm{in}}(x,y,z)
=
\chi_{00}^{(\omega)}(x,y)
|\uparrow\rangle e^{ik(z-L)},
\label{eqn:30}
\end{equation}
where the total energy is $E=\omega+k^2/2$.
The transverse ground state is the untruncated counterpart of
the profile used in Eq.~\eqref{eqn:seperable}.

Although the bulk Hamiltonian is spin independent, the tangential
part of the boundary operator in Eq.~\eqref{eqn:boundary_symbol},
$\mathcal J=C-i\kappa I=\sigma_y p_x-\sigma_x p_y$,
with $p_j=-i\partial_j$, couples transverse oscillator states
with different spin components. In particular,
\begin{equation}
\mathcal J\left(\chi_{00}^{(\omega)}|\uparrow\rangle\right)
=
-\sqrt{\omega}\,\chi_+^{(\omega)}|\downarrow\rangle,
\qquad
\chi_+^{(\omega)}
=
\frac{\chi_{10}^{(\omega)}
+i\chi_{01}^{(\omega)}}{\sqrt{2}}.
\end{equation}
Thus, the boundary couples the incident spin-up ground state
to a spin-down first excited oscillator state.
More generally, it couples neighboring oscillator shells with
accompanying spin flips. For fixed oscillator indices, the
coupling matrix elements are proportional to $\sqrt{\omega}$,
reflecting the characteristic tangential momentum scale of
the incident ground state.

For a transverse channel $\boldsymbol{n}=(n_x,n_y)$,
the longitudinal wave number at the incident energy is
\begin{equation}
q_{\boldsymbol{n}}
=
\sqrt{k^2-2\omega(n_x+n_y)},
\end{equation}
with the square-root branch chosen such that
$\operatorname{Re}q_{\boldsymbol{n}}\geq0$ and
$\operatorname{Im}q_{\boldsymbol{n}}\geq0$.
Channels with positive real $q_{\boldsymbol{n}}$ describe
propagating waves, whereas those with imaginary wave numbers
$q_{\boldsymbol{n}}=i\gamma_{\boldsymbol{n}}$,
$\gamma_{\boldsymbol{n}}>0$, are energetically closed.
Their scattered contributions decay away from the detector
as $e^{-\gamma_{\boldsymbol{n}}(L-z)}$.

To calculate the scattered state, we retain oscillator shells
with $n_x+n_y\leq M$ and expand the stationary spinor in the
corresponding transverse spin-orbital basis. The longitudinal
amplitude in each channel has the form
\begin{equation}
u_{\boldsymbol{n},\alpha}(z)
=
a_{\boldsymbol{n},\alpha}e^{iq_{\boldsymbol{n}}(z-L)}
+
b_{\boldsymbol{n},\alpha}e^{-iq_{\boldsymbol{n}}(z-L)},
\end{equation}
where $\alpha\in\{\uparrow,\downarrow\}$ and the incident
coefficients are nonzero only in the spin-up ground-state
channel, with unit amplitude.
Let $K_M$ be the diagonal matrix of longitudinal wave numbers
and $\mathcal C_M=i\kappa I+\mathcal J_M$ the boundary operator projected onto
the retained oscillator states.
We write $\mathbf{a}$ and $\mathbf{b}$ for the channel
coefficient vectors; the projected spinor ABC gives
\begin{equation}
(iK_M+\mathcal C_M)\mathbf{W}_M=2iK_M\mathbf{a},
\qquad
\mathbf{W}_M=\mathbf{a}+\mathbf{b},
\end{equation}
where $\mathbf{W}_M$ represents the stationary boundary trace in the retained oscillator basis.

Using the detector flux law in Eq.~\eqref{eqn:flux}, the absorbed and reflected fractions of the incident longitudinal flux are
\begin{equation}
A_{\mathrm{osc}}
=
\frac{\kappa}{k}\|\mathbf{W}_M\|^2,
\qquad
R_{\mathrm{refl}}
=
\frac{1}{k}
\sum_{\boldsymbol{n},\alpha:\,\mathrm{open}}
q_{\boldsymbol{n}}|b_{\boldsymbol{n},\alpha}|^2,
\end{equation}
with $ A_{\mathrm{osc}}+ R_{\mathrm{refl}}=1$.

For \(\omega>k^2/2\), all excited transverse channels are energetically closed. Boundary coupling nevertheless excites these channels, which contribute to the boundary density without carrying asymptotic reflected longitudinal flux. In the ideal rotationally symmetric guide, conservation of \(L_z+\sigma_z/2\), where \(L_z\) is the axial orbital angular momentum, leaves the spin-up transverse ground state as the only open reflected channel for the incident state in Eq.~\eqref{eqn:30}. Spin-down evanescent amplitudes can therefore contribute to detector loss while the outgoing reflected flux remains spin up. This shows how the particle-spin content associated with detector loss can differ from that of the reflected wave.

For $k=\kappa=\pi$, the calculated absorption probability
decreases across the eight sampled confinements from $0.976$ at
$\omega=1$ to $0.406$ at $\omega=300$. By contrast, the
componentwise scalar ABC gives unit absorption at $k=\kappa$.
Increasing the oscillator cutoff from $M=127$ to $255$ changes
the calculated absorption probabilities by less than $1.6\times10^{-13}$.
Independent Cartesian-basis calculations agree with the symmetry-reduced
representation at matched smaller cutoffs (Supplementary Material, Sec.~S5.5).

The confinement-dependent suppression of absorption therefore persists
when the transverse harmonic potential is retained explicitly and the
lower reflecting wall is absent. A complementary time-dependent control
removes the tangential spin coupling at the detecting plane while including
the bulk Hamiltonian, preparation and reference geometry. At $T=40$, the
componentwise scalar ABC gives $D_{40}\simeq0.956$ at both $\omega=1$
and $300$, whereas the spinor ABC gives $0.952$ and $0.740$, respectively
(Supplementary Material, Table~S2).  The stationary absorption probability
and the finite-window detected fraction describe different observables,
but both comparisons support the boundary mechanism underlying the
suppression of prompt detection in Fig.~\ref{fig:click-time-profiles}.

\section{Local Boundary Response and Finite-Guide Dynamics}

We now examine the local normal
response and spin composition of the evolved boundary
trace, followed by the role of subsequent propagation,
interference, and detector encounters in the finite guide.

\subsection{Auxiliary normal response and moment variation}
The boundary operator also governs the local normal
variation of the evolved boundary trace
$W(\boldsymbol{\xi},t)=
\widehat{\Psi}(\boldsymbol{\xi},L,t)$. For an inward distance $\epsilon=L-z>0$, variation of constants
gives the exact decomposition
\begin{equation}
\widehat{\Psi}(\boldsymbol{\xi},L-\epsilon,t)
=
e^{-i\kappa\epsilon}
B_{\mathrm{br}}(\epsilon,\boldsymbol{\xi})
W(\boldsymbol{\xi},t)
+\widehat{\mathcal R}_{\mathrm{Duh}},
\end{equation}
where
\begin{equation}
B_{\mathrm{br}}(\epsilon,\boldsymbol{\xi})
=
e^{-\epsilon R\Gamma(\boldsymbol{\xi})}
=
e^{-R\epsilon}\Pi_+(\boldsymbol{\xi})
+
e^{R\epsilon}\Pi_-(\boldsymbol{\xi}).
\end{equation}
The two exponential factors characterize the homogeneous
inward continuation, with characteristic normal length
$R^{-1}$ for $R>0$. They are distinct from the stationary
evanescent channels discussed in Sec.~4(iv), whose
decay lengths are determined by the transverse excitation
energies. The growing factor does not
represent physical amplification. Under the regularity
assumptions stated in Supplementary Material, Sec.~S6.1,
the Duhamel remainder is $O(\epsilon^2)$ modewise. Consequently,
\begin{equation}
B_{\mathrm{br}}(0,\boldsymbol{\xi})=I,
\qquad
\left.
\partial_\epsilon B_{\mathrm{br}}
\right|_{\epsilon=0}
=-R\Gamma(\boldsymbol{\xi}).
\end{equation}

For the incident transverse ground state, the characteristic
tangential momentum scale is
$R_{\mathrm{typ}}\sim\ell_\perp^{-1}=\sqrt{\omega}$.
The evolved boundary trace, however, need not retain the
incident momentum distribution.

To characterize the local response of this trace, we introduce
the auxiliary measure
\begin{equation}
d\mu_{\omega,\epsilon}^{\mathrm{bl}}
=
\left|
B_{\mathrm{br}}(\epsilon,\boldsymbol{\xi})
W(\boldsymbol{\xi},t)
\right|^2
d^2\xi\,dt,
\qquad 0\leq t\leq T,
\end{equation}
and its normalized counterpart
\begin{equation}
d\nu_{\omega,\epsilon}^{\mathrm{bl}}
=
\frac{d\mu_{\omega,\epsilon}^{\mathrm{bl}}}
{\int d\mu_{\omega,\epsilon}^{\mathrm{bl}}}.
\end{equation}
Here $\mathrm{bl}$ labels the auxiliary inward continuation;
the physical detector remains at $z=L$.

At $\epsilon=0$, write
$\nu_\omega^{(0)}=\nu_{\omega,0}^{\mathrm{bl}}$.
For $D_T>0$, Parseval's identity and the detector-flux law
give the temporal marginal $g(t;\omega)/D_T(\omega)$, so that
\begin{equation}
\mathbb{E}_{\nu_\omega^{(0)}}[t]
=
\bar{\tau}_T^{\mathrm{cond}}(\omega).
\end{equation}

Define the normalized branch-imbalance factor
\begin{equation}
a_\omega(\boldsymbol{\xi},t)
=
-2\frac{W^\dagger\Gamma(\boldsymbol{\xi})W}
{W^\dagger W},
\qquad |a_\omega|\leq 2,
\end{equation}
with $a_\omega=0$ where $W=0$ or $R=0$. The first inward
variation of the auxiliary density is therefore
\begin{equation}
\left.
\partial_\epsilon |B_{\mathrm{br}}W|^2
\right|_{\epsilon=0}
=
R a_\omega |W|^2.
\end{equation}

Assuming normalizability of the auxiliary measures near
$\epsilon=0^+$ and that differentiation can be interchanged
with integration, the first variation of the normalized
time moment is
\begin{equation}
\Lambda_\omega
:=
\left.
\partial_\epsilon
\mathbb{E}_{\nu_{\omega,\epsilon}^{\mathrm{bl}}}[t]
\right|_{\epsilon=0}
=
\operatorname{Cov}_{\nu_\omega^{(0)}}(t,Ra_\omega)
=
\sqrt{\omega}\,\beta_\omega^{\mathrm{bl}},
\label{eqn:cov}
\end{equation}
where
\begin{equation}
\beta_\omega^{\mathrm{bl}}
=
\operatorname{Cov}_{\nu_\omega^{(0)}}(t,sa_\omega),
\qquad s=\frac{R}{\sqrt{\omega}}.
\end{equation}

This identity separates the characteristic transverse
momentum scale from the correlations encoded in the evolved
boundary trace. The general covariance derivation and
associated bounds are given in Supplementary Material,
Sec.~S5.4.

The finite-grid evaluation of the covariance, the
boundary-weighted momentum moments, and the comparison
of the homogeneous continuation with the recorded field
on the neighboring longitudinal plane are reported in
Supplementary Material, Sec.~S6 and Table~S3.
The covariance magnitude remains sensitive to longitudinal
resolution (Sec.~S6.3 and Table~S5). Equation~\eqref{eqn:cov}
characterizes an auxiliary local response rather than
independently predicting the confinement dependence of
the restricted mean.

\paragraph{Boundary-branch composition and spin polarization} The local reflection coefficients describe the response to
incident spin channels at fixed wave vector. To examine
the spin structure of the full confined evolution, we
resolve the evolved boundary trace into the tangential
branches defined by the projectors $\Pi_\pm$.

For the spin-up preparation at $T=40$, increasing $\omega$
from $1$ to $300$ increases the positive-branch fraction
of the time-integrated boundary density from approximately
$44\%$ to $95\%$. The integrated detector-associated
$z$-polarization, defined as the spin-up minus spin-down
contribution to detector loss divided by the total
detector loss, decreases from approximately $0.91$ to
$0.10$, whereas the final survivor polarization remains
close to unity at $\omega=300$.

The definitions and time-dependent diagnostics are given
in Supplementary Material, Sec.~S6.2, Fig.~S1, and
Table~S4. Both the positive-branch predominance and
the detector--survivor polarization contrast persist
under the tested spatial and temporal refinements
(Sec.~S6.3 and Table~S5).

These quantities characterize the particle-spin content
within the adopted detector model. Unequal branch
weights in the evolved boundary trace are compatible
with equal local absorption probabilities for equally
normalized incident branches at fixed wave vector,
as established in Sec.~\ref{sec:SpinMomentum_Boundary}.

\subsection{Lower-boundary controls and late-time oscillations}
The reflected, undetected amplitude carries the momentum and spin
dependence of the boundary response into its subsequent propagation
and interference. After reflection at the lower wall, it can reach
the detector again. We call the influence of earlier boundary
encounters on the later detector signal \emph{finite-guide memory}.

To test the role of the lower wall, we displace it at $\omega=200$
while keeping the preparation, source and detector positions, and
spatial resolution fixed. Moving the wall from $z_B=0$ to $-15$
reduces $D_{40}$ from $0.784$ to $0.687$ and increases the late-window
conditional mean from $16.234$ to $18.125$, while the early detected
probabilities agree at the reported precision.
At the same confinement, replacing the lower Dirichlet wall by a
partially absorbing scalar ABC, with the upper detector unchanged,
reduces the upper-detection probability in $(3.6,40]$ from $0.438$
to $0.130$, with lower escape recorded separately.
The window-integrated $+$-branch fraction and detector-associated
$z$-polarization differ little between the two lower terminations,
while the final in-guide $z$-polarization remains close to unity.
These controls establish the lower-boundary sensitivity of the
late detector signal (Supplementary Material,
Secs.~S9.4--S9.5, Tables~S11--S12, and Figs.~S11--S12).

A separate analysis at \(\omega=300\) shows that the median spacing between prominent detector-rate maxima increases over time. On the $N_z=4000$ grid, the median spacings are $0.524$, $0.753$, and $1.214$ in the windows $8\leq t<16$, $16\leq t<27$, and $27\leq t<40$, respectively. The increase by a factor of approximately 2.3 from the first to the last window is accompanied by a shift toward lower dominant spectral frequencies. This increase persists under the tested spatial and time-step refinements and variations of the peak-selection criteria (Supplementary Material, Sec. S6.3). The spectral evolution is consistent with changing contributions from interfering, differently damped modes, although individual maxima are not identified with particular round trips and a unique recurrence period is not established.

Supplementary Material, Sec.~S7, separates the exact near-detector
probability balance from an assumed scalar return-kernel closure.
Under its stated assumptions, the latter relates the normalized
first moments of the untruncated fluxes through an effective return
weight and mean lag. The covariance in Eq.~\eqref{eqn:cov}
does not determine the kernel moments. The closure is not calibrated to the simulations;
the reported finite-window statistics are obtained directly from the full
spinor-ABC evolution.

\section{Conclusion}

We have identified how the spin-coupled absorbing boundary introduces the confinement dependence of the detection statistics previously reported in Ref.~\cite{JT26}, even though the longitudinal bulk Hamiltonian remains unchanged. Its matrix-valued impedance gives rise to momentum-dependent absorption and spin-dependent reflection phases, with equal local absorption probabilities in the two spin channels at fixed incident wave vector. Since the characteristic tangential momentum of the incident transverse ground state grows as \(\sqrt{\omega}\), stronger confinement favors momentum components that are less efficiently absorbed in the local matching problem. Stationary multichannel scattering in a semi-infinite harmonic waveguide confirms the decrease in absorption over the sampled confinement range, without a reflecting lower wall.

The stationary analysis also helps explain the spin structure of the detector response. For the spin-up incident ground state, when all excited transverse channels are energetically closed, spin-down evanescent amplitudes can contribute to detector loss while the asymptotic reflected flux remains spin up. A related contrast appears in the time-dependent simulations: with increasing confinement, the positive boundary branch becomes dominant, and at strong confinement the surviving state remains nearly spin up while the detector-associated longitudinal spin polarization is considerably smaller. The covariance representation of the first inward variation of an auxiliary time moment provides a complementary characterization of the evolved boundary trace, separating the characteristic transverse momentum scale from its time-dependent spin–momentum correlations.

Reduced absorption leaves more undetected amplitude available for subsequent propagation and interference in the guide. In the time-dependent simulations, stronger confinement primarily reduces the probability of early detection rather than shifting the early conditional mean to later times. Moving the lower reflecting wall or replacing it with a partially absorbing boundary substantially changes the late detector signal while leaving the early detected probability nearly unchanged in the tested cases. The principal branch and spin contrasts also persist under partial absorption at the lower boundary. At strong confinement, the increasing spacing between prominent late-time maxima, together with the shift toward lower dominant spectral frequencies, is consistent with interference between differently damped modes, although a unique recurrence period is not established.

For the longer observation windows considered in the confinement sweep, the conditional mean of recorded clicks also increases with confinement. The increase in the restricted mean thus reflects both later recorded clicks on average and a larger fraction of realizations remaining undetected at the observation horizon. Its confinement dependence is summarized by the empirical fit $A_T+B_T\sqrt{\omega}$, with setup- and horizon-dependent coefficients. A separate extended evolution at fixed confinement shows that appreciable detection continues beyond the main observation window, despite the decrease in the window-averaged detector rate. Neither the local covariance identity nor the available finite-time data establishes an unrestricted mean-time law.

The confinement trend persists across the tested detector parameters, longitudinal preparations, and guide lengths. Comparisons at two confinements also show that the contrast in detection statistics persists under the selected bulk-spin couplings and for the excited transverse preparation. Spatial and temporal refinements, together with the smooth boundary-compatible preparation test, support the stability of the principal detection statistics, although detailed time profiles and auxiliary boundary diagnostics remain sensitive to numerical resolution (Supplementary Material, Secs.~S6, S8, and S9).

Beyond its role in detection-time statistics, the complex boundary impedance provides a calibration target for absorbing layers and a possible basis for momentum-selective absorption and spin-dependent phase control of reflected matter waves. The layer comparisons show that matching the entrance impedance over a range of longitudinal momenta improves agreement in the restricted mean, without establishing equivalence of the full detection-time laws. Optical analogues could be explored through polarization- and wave-vector-dependent interfaces~\cite{Rubin21Jones,Wetter25Selective}. 

Further work is needed to develop a microscopic detector model and investigate its physical realization. Another direction is to examine how an unmonitored transverse environment modifies the populations and coherences reaching the detector, while detection itself remains governed by the absorbing boundary.

\section*{Declarations and acknowledgments}

\paragraph{Data availability.}
Selected processed numerical data and representative density animations
are publicly available at
\url{https://github.com/jloOop/Boundary-Spin-Momentum-Filtering-Data}.
Additional numerical data supporting the results reported in this study
are available from the corresponding author on reasonable request.

\paragraph{Code availability.}
Selected Python scripts for the simulations, diagnostics, and
post-processing are available in the same repository. The code and
parameters required to reproduce the reported calculations are available
from the corresponding author on reasonable request.  The solver follows the numerical framework described in Appendix A of Ref.~\cite{JT26}, with the boundary implementation specified in Supplementary Material, Sec. S2.

\paragraph{Author contributions and use of AI-assisted tools.}   The author conceived the study, developed the theoretical framework and analysis, including the matrix-valued boundary symbol, local scattering, and the variation-of-constants formulation, carried out the numerical investigations, and interpreted the results.

ChatGPT (OpenAI) assisted with conceptual and mathematical discussions, the development and verification of analyses, particularly in the Supplementary Material, Python programming, numerical analysis, figure preparation, reference checks, and manuscript drafting and revision. The author directed and critically reviewed all AI-assisted work and takes full responsibility for the methods, results, and conclusions.

\paragraph{Acknowledgments.}
The author is grateful to Roderich Tumulka for insightful discussions, particularly concerning the scattering interpretation of the spin-coupled absorbing boundary condition. The author also thanks Navid Javid for helpful discussions. The author acknowledges computing resources on Otus at the Paderborn Center for Parallel Computing (PC2) and on Lise at NHR@ZIB. The NHR centers are jointly funded by the participating German states and the Federal Ministry of Research, Technology and Space.

\urlstyle{same}
\providecommand{\doilink}[1]{\href{https://doi.org/#1}{doi:#1}}
\providecommand{\arxivlink}[1]{\href{https://arxiv.org/abs/#1}{arXiv:#1}}

\end{document}


\title{Supplementary Material for\texorpdfstring{\\}{ }
``Spin--Momentum Impedance and Filtering at a Spin-Coupled Absorbing Boundary''}

\author{Alireza Jozani}
\affiliation{Departments of Mathematics and Physics, Eberhard-Karls-Universit\"at T\"ubingen,
Auf der Morgenstelle 10, 72076 Tübingen, Germany}

\date{\today}

\maketitle
\pagestyle{fancy}
\thispagestyle{fancy}
\fancyhf{}
\fancyhead[R]{\thepage}
\renewcommand{\headrulewidth}{0pt}
\renewcommand{\footrulewidth}{0pt}

\section{Scope, notation, and relation to the main text}
\label{sec:SM_scope}

This Supplement specifies the numerical model and develops the boundary analysis of the confinement-dependent detection statistics reported in Ref.~[11] of the main text. We adopt the spinor absorbing boundary condition (ABC) as an effective detector model. In the reference geometry, with no other loss channels, the outward detector flux equals the loss rate of \(\|\Psi_t\|^2\) and defines the detection-time probability density. Spin–momentum filtering denotes momentum-dependent absorption and spin-dependent reflection phases; the two local spin channels have equal absorption probabilities at fixed incident wave vector.

Section~\ref{sec:SM_numerics} gives the numerical specification;~\ref{sec:SM_finite_window} separates finite-window detection probability from detection timing. Sections~\ref{sec:SM_bulk}–\ref{sec:SM_boundary_symbol} develop the separable bulk reference,
the boundary response, and the complementary stationary
multichannel scattering calculation, and~\ref{sec:S7_boundary_diagnostics} reports boundary and spin diagnostics and refinement checks. Section~\ref{sec:S8_return_closure} gives exact probability balance and a reduced model of repeated encounters under stated closure assumptions. Sections~\ref{sec:SM_controls}–\ref{sec:SM_physical_scales} collect controls, test robustness via parameter scans, and give physical scales.

\section{Numerical model and grid scaling}
\label{sec:SM_numerics}

In units \(\hbar=m=1\), we evolve the two-component spinor \(\Psi=(\psi_\uparrow,\psi_\downarrow)^T\) in the domain
\begin{equation}
        \Omega_\omega=[0,L_x(\omega)]\times[0,L_y(\omega)]\times[0,L],
\end{equation}
according to
\begin{equation}
 i\partial_t\Psi
 =
 \left[
 -\frac12\Delta
 +\frac12\omega^2\bigl((x-x_c)^2+(y-y_c)^2\bigr)
 \right]\Psi,
 \qquad
 \Psi=
 \begin{pmatrix}
 \psi_\uparrow\\
 \psi_\downarrow
 \end{pmatrix}.
\end{equation}
The initial state is a normalized factorized packet,
\begin{equation}
\Psi_0(x,y,z)=\chi_\omega(x,y)\,\phi_0(z)\,\eta ,
\end{equation}
where the transverse factor is the harmonic ground-state profile, restricted and normalized in the
finite transverse box,
\begin{equation}
\chi_\omega(x,y)=\mathcal N_\perp(\omega)
\exp\!\left[-\frac{\omega}{2}\bigl((x-x_c)^2+(y-y_c)^2\bigr)\right],
\end{equation}
and the longitudinal factor is the right-moving Gaussian
\begin{equation}
\phi_0(z)=\mathcal{N}_{\parallel}
\exp\left[
-\frac{(z-z_c)^2}{2\sigma_{\parallel}^2}
+ik_z(z-z_c)
\right],
\qquad 0<z<L.
\label{eq:SM_longitudinal_packet}
\end{equation}
The constants $\mathcal N_\perp(\omega)$ and $\mathcal{N}_{\parallel}$ normalize the factors on their finite domains. For the numerical evolution, the full sampled state is normalized using Eq.~\eqref{eqn:CenteredP} after imposing the Dirichlet values specified below. The longitudinal Gaussian has positive central wave number \(k_z\) and amplitude-width parameter \(\sigma_\parallel\); its untruncated standard deviations are \(\Delta z=\sigma_\parallel/\sqrt2\) and \(\Delta k_z=1/(\sqrt2\,\sigma_\parallel)\). The constant spinor is
\begin{equation}
\eta=
\bigl(\cos(\theta/2),e^{i\phi}\sin(\theta/2)\bigr)^T,
\end{equation}
with \(\theta=\phi=0\) for the reference sweep. Across that sweep, the longitudinal preparation, detector parameter, initial spinor, and spatial grid prescription are fixed as listed in Table~\ref{tab:SM_reference_numerics}. The discrete initialization is specified below.

\noindent\textit{Boundary conditions and observables.} The lower face \(z=0\) is a reflecting Dirichlet wall; the artificial transverse walls are also Dirichlet. At the detecting plane \(\Sigma_L=\{z=L\}\), the spinor ABC and its outward Pauli current are
\begin{equation}
(\boldsymbol{\sigma}\cdot\boldsymbol\nabla)\Psi=i\kappa\sigma_z\Psi,
\qquad
j_z^{\mathrm P}(x,y,L,t)=\kappa\Psi^\dagger\Psi\ge0,
\end{equation}
where \(\boldsymbol{\sigma}\) denotes the Pauli matrices and \(\kappa>0\). For normalized initial data, the survival probability and detection-time density are
\begin{equation}
S(t;\omega)=\|\Psi_t\|^2,
\qquad
g(t;\omega)
=\kappa\int_{\Sigma_L}\Psi^\dagger\Psi\,dx\,dy
=-\partial_tS(t;\omega).
\end{equation}
For an observation horizon \(T_{\mathrm{obs}}\), we use
\begin{equation}
\begin{aligned}
D_{T_{\mathrm{obs}}}(\omega)
&=\mathbb{P}(\tau\leq T_{\mathrm{obs}})= \int_0^{T_{\mathrm{obs}}}g(t;\omega)\,dt,\\
\mu^*(T_{\mathrm{obs}})
&=\mathbb E[\min(\tau,T_{\mathrm{obs}})]
=\int_0^{T_{\mathrm{obs}}} S(t)\,dt.
\end{aligned}
\end{equation}
where \(\tau\) is the random detector click time, $D_{T_{\mathrm{obs}}}$ is the probability of detection by \(T_{\mathrm{obs}}\), assumed positive. The restricted mean detection time \(\mu^*\) includes realizations undetected at the horizon without extrapolating the unobserved tail.

\noindent\textit{Discretization and recorded statistics}.  The Cartesian evolution uses the Crank–Nicolson and GMRES framework of Ref. [11], with a centered ghost-point approximation replacing the earlier one-sided normal-boundary closure. The continuum spinor ABC and surface detector-loss law are unchanged. The centered derivative is second-order accurate for smooth fields, and the compatible discrete norm assigns half weight to the terminal longitudinal plane. Sensitivity of the computed observables is assessed separately in Sec.~\ref{sec:refinement-T40}.  The compatible discrete norm is given in Eq.~\eqref{eqn:CenteredP}.

 Table~\ref{tab:SM_reference_numerics} gives the preparation and numerical settings.
The transverse box scales with the oscillator length \(\ell_\perp=\omega^{-1/2}\). With \(h_\perp=h_x=h_y\),
\begin{equation}
L_x=L_y=\ell_{\mathrm{box}}\ell_\perp,
\qquad
\frac{h_\perp}{\ell_\perp}
=\frac{\ell_{\mathrm{box}}}{N_x},
\qquad
\ell_{\mathrm{box}}=10,\quad N_x=N_y.
\end{equation}
Thus the nominal wall distances and transverse resolution in oscillator units are fixed across the sweep.

\begin{table}[H]
\centering
\begingroup

\makeatletter
\expandafter\def\expandafter\@mkpream@relax\expandafter{%
  \@mkpream@relax\let\@endpbox\relax
}
\makeatother

\small
\setlength{\tabcolsep}{5pt}
\renewcommand{\arraystretch}{0.85}

\setlength{\heavyrulewidth}{0.10em}
\setlength{\lightrulewidth}{0.045em}

\caption{Physical preparation and numerical settings for the eight
reference evolutions used in main-text Fig.~2 and
Tables~\ref{tab:SM_R4_reference_scalar_T40}--\ref{tab:SM_branch_spin_rechecked}. For the time-step-refined $\omega=300$ history
used in main-text Figs.~1 and~3, $\Delta t=10^{-4}$; other
settings are unchanged. Units are $\hbar=m=1$.}
\label{tab:SM_reference_numerics}

\begin{tabular}{
  @{}p{0.33\linewidth}
  @{\hspace{0.025\linewidth}}
  p{0.645\linewidth}@{}
}
\toprule
Quantity & Value or prescription \\
\midrule

Confinement frequencies &
$\omega\in\{1,25,50,75,100,150,200,300\}$ \\

Time-step  &
$\Delta t\in\{10,8,8,4,5,2.5,2.5,2\}\times10^{-4}$,
respectively, \\

Nominal box sizes &
$L_x=L_y=10/\sqrt{\omega}$; $L=20$ \\

Trap and source centers &
$x_c=L_x/2$, $y_c=L_y/2$; $z_c=10$ \\

Longitudinal preparation &
$k_z=\pi$, $\sigma_\parallel=0.5$ \\

Detector and initial spin &
$\kappa=\pi$; $\eta=(1,0)^{\mathsf T}$
($\theta=\phi=0$) \\

Stored spatial grid &
$N_x=N_y=100$, $N_z=1500$;
nominal upper endpoints excluded \\

Mesh spacings &
$h_x=h_y\simeq 0.1/\sqrt{\omega}$;
$h_z\simeq 0.013$ \\

Numerical detecting plane &
$z_D=z_{N_z-1}\simeq 19.987$ \\

Evolution scheme &
Cartesian finite differences;
Crank--Nicolson (Cayley) steps \\

Linear solver &
Jacobi-preconditioned GMRES;
$\mathrm{rtol}=10^{-8}$,
$\mathrm{atol}=0$, restart $30$,
\texttt{maxiter}$=1000$ \\

Numerical precision &
Coordinates: \texttt{float32};
stored state and matrix assembly:
\texttt{complex64};
linear solves and additional spin/branch diagnostics:
\texttt{complex128}/\texttt{float64} \\

\bottomrule
\end{tabular}

\endgroup
\end{table}

The coordinate arrays exclude the nominal upper endpoints, and the numerical detecting plane is the last stored longitudinal plane, $z_D=z_{N_z-1}$. Unless otherwise stated, the spinor ABC is implemented through the
centered ghost-point relation
\begin{equation}
\frac{\Psi_{N+1}-\Psi_{N-1}}{2h_z}
=\mathsf C_h\Psi_N,
\qquad N=N_z-1,
\end{equation}

where $\Psi_N$ is the grid spinor on the numerical detecting plane
and $\mathsf C_h$ is the centered tangential finite-difference form
of the boundary operator in main-text Eq.~(18). This relation
eliminates the ghost value $\Psi_{N+1}$ from the longitudinal
Laplacian and gives a second-order approximation to the normal
derivative for smooth fields.

Before evolution, the sampled Gaussian is set to zero on the stored
lateral boundary planes and at $z=0$, then normalized.
These Dirichlet values are enforced after every step, without
renormalizing the evolved state. Equation~\eqref{eq:SM_longitudinal_packet} specifies the interior profile. The simulated
initial state has the imposed Dirichlet grid values and
normalization described here; the unmodified smooth Gaussian
is not asserted to satisfy all continuum boundary conditions
at $t=0$ or to belong to the continuum generator domain.

The numerical squared norm is
\begin{equation}
P(t)=h_xh_yh_z
\sum_{i,j}
\left[
\sum_{k=0}^{N_z-2}\Psi_{ijk}^{\dagger}\Psi_{ijk}
+\frac12\Psi_{ij,N_z-1}^{\dagger}\Psi_{ij,N_z-1}
\right].
\label{eqn:CenteredP}
\end{equation}
The terminal half weight makes this discrete norm compatible
with the centered boundary closure. For the reference geometry,
the unrounded stencil and exact Crank--Nicolson solves give a
decrease of $P$ equal to the detector flux integrated over the
detecting plane, evaluated on the midpoint spinor
$(\Psi^{n+1}+\Psi^n)/2$, multiplied by the step duration.
The same volume weights are used for normalization, survival
probabilities, and survivor-spin matrices.

Mass sums are accumulated in double precision, and probability
statistics are normalized by the recorded launch norm
$P_0=P(0)$. Detector-event probabilities are recorded separately
from casting and boundary-clamp corrections, which are excluded
from detector events. Each reference evolution reaches
$T_{\max}=40$, with $T=10,20,30,40$ evaluated from the same history.
Event probabilities and stored-state masses are recorded at
every accepted step, plane currents every 20 steps, and spin/branch
diagnostics at nominal intervals of $0.02$. All integrations use
the actual recorded times.

Refinement checks for the detection statistics, detector rates, and
boundary-branch diagnostics at $\omega=300$ and $T=40$ are reported
in Sec.~\ref{sec:refinement-T40}.

\section{Finite-window detection statistics}
\label{sec:SM_finite_window}

This section separates the probability of detection by a finite horizon from the timing of the recorded clicks. We use the eight reference evolutions of Table~\ref{tab:SM_reference_numerics} and report the $T=40$ statistics in Table~\ref{tab:SM_R4_reference_scalar_T40}.

\subsection{Conditional detection-time statistics and restricted-mean identity}
\label{subsec:SM_RMST_identity}

Using $g$, $S$, $D_T$, and $\mu^*$ defined in Sec.~\ref{sec:SM_numerics}, the density conditioned on detection by $T$, for $D_T(\omega)>0$, and its mean are
\begin{equation}
\begin{gathered}
f_T(t;\omega)=\frac{g(t;\omega)}{D_T(\omega)},
\qquad 0\le t\le T,\\[4pt]
\bar\tau_T^{\mathrm{cond}}(\omega)
=\mathbb E[\tau\mid\tau\leq T]=\int_0^T t\,f_T(t;\omega)\,dt
=\frac{M_T(\omega)}{D_T(\omega)},\\[4pt]
M_T(\omega)=\int_0^T t\,g(t;\omega)\,dt.
\end{gathered}
\label{eq:s34draft:conditional}
\end{equation}
Recorded detections contribute their click times, while realizations undetected at \(T\) contribute the horizon itself,
\begin{equation}
\begin{aligned}
\mu^*(T;\omega)
&= \mathbb{P}(\tau\le T)\,
   \mathbb{E}[\tau\mid\tau\le T] \\
&\quad + \mathbb{P}(\tau>T)\,
   \mathbb{E}[T\mid\tau>T] \\
&= \mathbb{P}(\tau\le T)\,
   \mathbb{E}[\tau\mid\tau\le T]
   + T\,\mathbb{P}(\tau>T) \\
&= D_T(\omega)\,\bar\tau_T^{\mathrm{cond}}(\omega)
   + [1-D_T(\omega)]\,T.
\end{aligned}
\label{eqn:balance}
\end{equation}

Equivalently,
\begin{equation}
\bar\tau_T^{\mathrm{cond}}(\omega)
=\frac{\mu^*(T;\omega)-[1-D_T(\omega)]T}{D_T(\omega)}.
\label{eq:s34draft:conditional_from_survival}
\end{equation}
Since $0\le\mu^*(T;\omega)\le T$, an affine fit in $\sqrt{\omega}$ with nonzero slope can describe only a finite confinement range at a specified horizon.

For the spinor reference evolutions, $D_T$ and $M_T$ are evaluated from recorded Crank--Nicolson boundary probabilities, using a constant rate within each accepted step and interval overlap at window endpoints. The restricted mean is independently integrated from the piecewise-linear stored survival history. All probability quantities are divided by the recorded launch norm. The largest absolute detector--survival budget discrepancy over
the eight histories is below $3.68\times10^{-7}$.
The largest absolute residual in Eq.~\eqref{eqn:balance}, evaluated at
$T=10,20,30,40$, is below $1.18\times10^{-5}$.
These measure consistency of the recorded statistics, not
discretization accuracy.

\subsection{Common diagnostic cutoff}
\label{subsec:SM_cutoff}

The early/late split uses one cutoff time \(t_{\rm cut}\), common to all values of
\(\omega\). The cutoff is not fitted from the detector-flux data and is not chosen
separately for each run. It is used only to define a fixed diagnostic window for comparing the early and late contributions to the detector signal.  Let
\begin{equation}
        \Sigma_i=\{z=z_i\}
\end{equation}
be an interior monitoring plane below the detector. From the main upward-flux peak
times on two such planes, we estimate an effective axial speed \(v_{\rm est}(\omega)\).
If \(t_{\rm pk,up}^{(i)}(\omega)\) is the main upward peak time on \(\Sigma_i\), then the estimated arrival time at \(\Sigma_i\) of the signal reflected from the detecting plane is
\begin{equation}
        t_{\rm return}^{(i)}(\omega)
        =
        t_{\rm pk,up}^{(i)}(\omega)
        +
        \frac{2d_R^{(i)}}{v_{\rm est}(\omega)},
        \qquad
        d_R^{(i)}=z_{\rm detector}-z_i .
\end{equation}
A subsequent signal reflected from the lower wall would
require an additional round trip from the monitoring plane
to the bottom and back. We estimate the corresponding
cycle-completion time by
\begin{equation}
        t_{\rm cyc}^{(i)}(\omega)
        =
        t_{\rm return}^{(i)}(\omega)
        +
        \gamma
        \frac{2d_B^{(i)}}{v_{\rm est}(\omega)},
        \qquad
        d_B^{(i)}=z_i-z_{\rm bottom},
        \qquad
        \gamma=0.95,
\end{equation}
The factor \(\gamma=0.95\) is a fixed margin in this timing prescription, not a reflection coefficient. For the closer monitor \(\Sigma_1\), the estimates are
\begin{equation}
\begin{array}{c|cccccc}
\omega & 1 & 50 & 100 & 150 & 200 & 300\\
\hline
t_{\rm cyc}^{(1)}(\omega)
& 13.58 & 7.30 & 6.02 & 6.34 & 5.06 & 3.21 

\end{array}
\label{tab:earlyLate}
\end{equation}
The smallest peak-based cycle estimate is approximately 3.21. The timing estimates provide guidance for choosing a common early/late diagnostic cutoff, rather than a bound on the onset of returned amplitude. For the reference decomposition, we use the common cutoff
\begin{equation}
        t_{\rm cut}=3.6 .
        \label{eq:SM_tcut}
\end{equation}

This diagnostic division is not a strict return-free interval or an encounter-resolved decomposition of the wavefunction. Earlier cutoffs are tested in Sec.~\ref{sec:s4v40-earlylate}.

\subsection{Early/late identity and reference values}
\label{sec:s4v40-earlylate}
For \(T>t_{\mathrm{cut}}\), define
\begin{equation}
 I_E=[0,t_{\mathrm{cut}}],\qquad I_L=(t_{\mathrm{cut}},T] .
 \label{eq:s4v40-intervals}
\end{equation}
The conditional weights are
\begin{equation}
 w_L(T;\omega)=\int_{t_{\mathrm{cut}}}^T f_T(t;\omega)\,dt,
 \qquad w_E(T;\omega)=1-w_L(T;\omega).
 \label{eq:s4v40-weights}
\end{equation}
For positive early and late weights, as in all cases reported here, the corresponding conditional means are
\begin{equation}
 t_E(\omega)=\frac{1}{w_E(T;\omega)}
             \int_0^{t_{\mathrm{cut}}}t f_T(t;\omega)\,dt ,
 \label{eq:s4v40-tE}
\end{equation}
and
\begin{equation}
 t_L(T;\omega)=\frac{1}{w_L(T;\omega)}
             \int_{t_{\mathrm{cut}}}^T t f_T(t;\omega)\,dt .
 \label{eq:s4v40-tL}
\end{equation}
They satisfy
\begin{equation}
\begin{aligned}
\bar\tau_T^{\mathrm{cond}}(\omega)
&=w_E(T;\omega)t_E(\omega)+w_L(T;\omega)t_L(T;\omega)\\
&=t_E(\omega)+w_L(T;\omega)
\bigl[t_L(T;\omega)-t_E(\omega)\bigr].
\end{aligned}
\label{eq:s4v40-earlylate-identity}
\end{equation}
We define the late-window contribution\footnote{For the six confinements in Eq.~\eqref{tab:earlyLate}, reducing the common cutoff to \(3.21\) preserves the decrease in early detected probability, the weak variation of its conditional mean, and the increase in both the late fraction among detections and the late conditional mean.} to the conditional mean,
\begin{equation}
 \Delta t_L^{\mathrm{cond}}(T;\omega)
 :=w_L(T;\omega)[t_L(T;\omega)-t_E(\omega)] .
 \label{eq:s4v40-delta}
\end{equation}
The absolute early detected probability is
\begin{equation}
 D_E(\omega):=\int_0^{t_{\mathrm{cut}}}g(t;\omega)\,dt
             =D_T(\omega)w_E(T;\omega).
 \label{eq:s4v40-DE}
\end{equation}
Both \(D_E\) and \(t_E\) depend only on the prefix up to
\(t_{\mathrm{cut}}\), and hence are unchanged by choosing
\(T=10,20,30,\) or \(40\). The late quantities and the conditional weights
do depend on \(T\).

\begin{table}[H]
\centering
\caption{Finite-window statistics at $T=40$, with $t_{\mathrm{cut}}=3.6$. The upper block contains spinor-ABC reference evolutions; the lower block contains componentwise scalar-ABC controls. Weights and window means refer to the corresponding conditional density $f_{40}$. Scalar statistics are reconstructed from normalized survival histories. The scalar control values were obtained with the earlier one-sided boundary closure and full terminal norm weight; the centered spinor results use Eq.~\eqref{eqn:CenteredP}. Entries are rounded independently; displayed digits do not specify discretization accuracy.}
\label{tab:SM_R4_reference_scalar_T40}

\begingroup


\setlength{\tabcolsep}{5pt}

\renewcommand{\arraystretch}{0.80}

\setlength{\heavyrulewidth}{0.12em}
\setlength{\lightrulewidth}{0.05em}


\begin{tabular*}{0.78\linewidth}
{@{\extracolsep{\fill}}rrrrrrrr@{}}

\toprule

$\omega$
& $D_{40}$
& $\bar\tau_{40}^{\rm cond}$
& $\mu^*(40)$
& $t_E$
& $w_L(40)$
& $t_L(40)$
& $\Delta t_L^{\rm cond}(40)$ \\

\midrule

\addlinespace[1pt]
\multicolumn{2}{c}{\textit{Spinor ABC:}}\\[0pt]
\addlinespace[1pt]

1
& 0.952
& 4.663
& 6.358
& 2.595
& 0.394
& 7.839
& 2.067 \\

25
& 0.916
& 5.980
& 8.832
& 2.561
& 0.405
& 10.999
& 3.419 \\

50
& 0.887
& 7.158
& 10.878
& 2.543
& 0.447
& 12.864
& 4.615 \\

75
& 0.862
& 8.020
& 12.433
& 2.533
& 0.479
& 13.982
& 5.488 \\

100
& 0.841
& 8.683
& 13.666
& 2.526
& 0.504
& 14.735
& 6.157 \\

150
& 0.808
& 9.598
& 15.436
& 2.517
& 0.538
& 15.677
& 7.081 \\

200
& 0.784
& 10.178
& 16.618
& 2.511
& 0.559
& 16.234
& 7.666 \\

300
& 0.740
& 11.113
& 18.631
& 2.501
& 0.590
& 17.088
& 8.613 \\

\midrule

\addlinespace[1pt]
\multicolumn{2}{c}{\textit{Scalar ABC:}}\\[0pt]
\addlinespace[1pt]

1
& 0.956
& 4.372
& 5.923
& 2.598
& 0.386
& 7.198
& 1.774 \\

300
& 0.956
& 4.375
& 5.926
& 2.600
& 0.386
& 7.195
& 1.775 \\

\bottomrule

\end{tabular*}

\endgroup
\end{table}

Table~\ref{tab:SM_R4_reference_scalar_T40} reports the reference values at $T = 40$. Across the sweep, $D_E$ decreases from approximately $0.58$ at $\omega = 1$ to $0.30$ at $\omega = 300$, whereas $t_E$ changes only from $2.59$ to $2.50$. The main prompt-sector effect is therefore a reduction in detected probability rather than a shift of its conditional mean to later times. At $T = 40$, the late fraction among detected events increases from $0.39$ to $0.59$, and the conditional mean within that sector increases from $7.84$ to $17.09$. This conditional fraction is distinct from the absolute late detection probability $D_L(T;\omega)=D_T(\omega)w_L(T;\omega)$.

Combining the conditional split with
Eq.~\eqref{eqn:balance} gives
\begin{equation}
 \mu^*(T;\omega)
 =D_T(\omega)\bigl[t_E(\omega)+
       \Delta t_L^{\mathrm{cond}}(T;\omega)\bigr]
  +[1-D_T(\omega)]T .
 \label{eq:s4v40-combined}
\end{equation}
Consequently the confinement dependence of the restricted mean cannot be
identified with the delayed conditional contribution alone. These
identities describe the signal; possible dynamical contributions from
returns are discussed separately in Sec.~\ref{sec:S8_return_closure}.

\noindent\textit{Scalar-boundary control.} The lower block of Table~\ref{tab:SM_R4_reference_scalar_T40} removes the tangential coupling by imposing
$\partial_z\Psi=i\kappa\Psi$ componentwise, retaining the bulk Hamiltonian,
physical preparation, and reference geometry and spatial grid. The scalar rows are excluded from the spinor confinement fits.

\noindent\textit{Cutoff sensitivity.} Repeating the decomposition at $T=40$ with $t_{\mathrm{cut}}=3.0$ for the same eight spinor-ABC reference evolutions preserves the suppression of early detected probability and the weak variation of its conditional mean. Between $\omega=1$ and $300$, $D_E$ decreases from $0.426$ to $0.239$, $t_E$ changes from $2.353$ to $2.295$, and $t_L$ increases from $6.532$ to $15.328$. The late fraction among detections increases overall, with a small initial decrease between $\omega=1$ and $25$; the absolute late probability is nonmonotonic for both cutoffs. The tabulated early/late statistics use $t_{\mathrm{cut}}=3.6$.

\subsection{Observation-window comparison}
\label{sec:s4v40-horizons}

Main-text Fig.~3 compares $T=10,20,30,40$ evaluated from a common history at each confinement. At $\omega=300$, extending $T$ from $20$ to $40$ adds detected probability $D_{40}-D_{20}\simeq 0.156$, corresponding to $37.4\%$ of the probability surviving at $t=20$. The survival contribution $TS(T)$ accounts for approximately $68.7\%$ of $\mu^*(20)$ and $55.9\%$ of $\mu^*(40)$. With $S(40)\simeq 0.260$, the unrestricted mean remains undetermined by the recorded histories.

For the eight spinor reference values in Table~\ref{tab:SM_R4_reference_scalar_T40}, the unweighted fit $\mu^*(40;\omega)\simeq 5.463+0.784\sqrt{\omega}$, reported in main-text Eq.~(13), has a root-mean-square residual of $0.32$ and a maximum absolute residual of $0.55$, at $\omega=25$. These residuals quantify agreement with the empirical fit.

\section{First-pass bulk independence of \texorpdfstring{\(\omega\)}{omega}}
\label{sec:SM_bulk}

We first examine an auxiliary separable bulk evolution. For a fixed longitudinal preparation and evolution, changing $\omega$ changes the normalized transverse factor, but not the cross-section-integrated normal Pauli flux.

We now show this explicitly. Let \(z_b=L-\delta\) be an interior counting plane below the detector, and let
\begin{equation}
        \Sigma_b=\{(x,y,z):z=z_b\}.
\end{equation}
In the region below the detecting plane, assume the first-pass separable bulk Hamiltonian
\begin{equation}
        H_{\rm bulk}=H_{xy}(\omega)+H_z ,
\end{equation}
with
\begin{equation}
        H_{xy}(\omega)
        =
        -\frac12(\partial_x^2+\partial_y^2)+V_{xy}(x,y;\omega),
        \qquad
        H_z=-\frac12\partial_z^2 .
\end{equation}
We also assume that the absorbing boundary has not yet significantly modified the state at the interior plane.  For a prepared state with spatially constant spinor \(\eta\), the auxiliary first-pass bulk field is
\begin{equation}
\Psi_{\rm fp}(x,y,z,t)
=
\chi_\omega(x,y,t)\,\phi_{\rm fp}(z,t)\,\eta,
\qquad
\rho_{\rm fp}=|\Psi_{\rm fp}|^2
=
|\chi_\omega(x,y,t)|^2|\phi_{\rm fp}(z,t)|^2 .
\label{eq:SM_separable_bulk}
\end{equation}
Here ``first-pass'' means the separable, spin-independent bulk evolution before appreciable feedback
from the absorbing boundary. The longitudinal factor is initialized by $
\phi_{\rm fp}(z,0)=\phi_0(z),$ and \(\Psi_{\rm fp}\) is not the full spinor-ABC solution. It is used only to test whether ordinary bulk propagation can generate the observed \(\omega\)-dependence. The spin density of this auxiliary field is
\begin{equation}
\boldsymbol S_{\rm fp}
=
\Psi_{\rm fp}^\dagger\boldsymbol \sigma\Psi_{\rm fp}
=
\rho_{\rm fp}\boldsymbol s_\eta,
\qquad
\boldsymbol s_\eta=\eta^\dagger\boldsymbol \sigma\eta .
\end{equation}
The corresponding Pauli current is
\begin{equation}
\boldsymbol j^P_{\rm fp}
=
\boldsymbol j^{\rm conv}_{\rm fp}
+
\frac12\boldsymbol\nabla\times \boldsymbol S_{\rm fp},
\qquad
\boldsymbol j^{\rm conv}_{\rm fp}
=
\operatorname{Im}(\Psi_{\rm fp}^\dagger\boldsymbol \nabla\Psi_{\rm fp}) .
\end{equation}
Since \(\boldsymbol s_{\eta}\) is constant in the first-pass separable approximation,
\begin{equation}
        (\boldsymbol \nabla\times\boldsymbol S_{\rm fp})_z
        =
        \partial_x(\rho_{\rm fp} s_{\rm \eta, y})-\partial_y(\rho_{\rm fp} s_{\rm \eta, x}).
\end{equation}
Therefore
\begin{equation}
\int_{\Sigma_b}
\boldsymbol (\boldsymbol \nabla\times\boldsymbol S_{\rm fp})_z\,dx\,dy
=\int_{\partial\Sigma_b}
\bigl(S_{{\rm fp},y}\nu_x-S_{{\rm fp},x}\nu_y\bigr)\,d\ell
=0,
\end{equation}
where \(\boldsymbol\nu=(\nu_x,\nu_y)\) is the outward unit normal to \(\partial\Sigma_b\) in the transverse plane. The last equality follows from the lateral boundary conditions in the finite box,
or from transverse decay in the ideal waveguide. Hence the plane-integrated Pauli
flux equals the plane-integrated convective flux:
\begin{equation}
        \int_{\Sigma_b}j_{\rm fp,z}^P\,dx\,dy
        =
        \int_{\Sigma_b}j_{\rm fp,z}^{\rm conv}\,dx\,dy .
        \label{eq:SM_pauli_convective_plane}
\end{equation}
Define the first-pass bulk flux through \(\Sigma_b\) by
\begin{equation}
        \Phi_{\rm bulk}(t)
        =
        \int_{\Sigma_b}j_{\rm fp,z}^P(\boldsymbol  r,t)\,dA .
\end{equation}
Using Eq.~\eqref{eq:SM_separable_bulk},
\begin{equation}
\begin{aligned}
        \Phi_{\rm bulk}(t)
        &=
        \int_{\Sigma_b}
        \Im(\Psi_{\rm fp}^\dagger\partial_z\Psi_{\rm fp})\,dA
        \\
        &=
        \left(
        \int |\chi_\omega(x,y,t)|^2\,dx\,dy
        \right)
        \Im\bigl(\phi_{\rm fp}^*\partial_z\phi_{\rm fp}\bigr)(z_b,t).
\end{aligned}
\end{equation}
Since \(H_{xy}(\omega)\) is self-adjoint with the chosen transverse boundary
conditions, the transverse factor remains normalized. Therefore
\begin{equation}
        \Phi_{\rm bulk}(t)
        =
        \Im\bigl(\phi_{\rm fp}^*\partial_z\phi_{\rm fp}\bigr)(z_b,t),
        \label{eq:SM_bulk_flux_independent}
\end{equation}
which is independent of \(\omega\) and independent of the detector parameter \(\kappa\). When the integrals converge absolutely and the denominator is
nonzero, the corresponding flux-weighted time is
\begin{equation}
        \bar\tau_{\rm bulk}
        =
        \frac{\int_0^\infty t\,\Phi_{\rm bulk}(t)\,dt}
        {\int_0^\infty \Phi_{\rm bulk}(t)\,dt}
\end{equation}
which is \(\omega\)-independent under the stated assumptions. This does not claim that
the full finite-box detector signal is \(\omega\)-independent. It only shows that
the leading first-pass bulk transport does not generate the confinement trend;
the effect must enter through the spinor absorbing boundary.

\section{Boundary-symbol mechanism}
\label{sec:SM_boundary_symbol}
This section develops the spin--momentum boundary symbol,
its local reflection and normal-response properties,
and its stationary scattering consequences in a harmonically
confined guide. The auxiliary inward continuation introduced
below is a probe of the boundary relation, not a second
absorber or a physical detector layer. The physical detector
observable remains the flux through the detecting plane
at $z=L$. At the detecting plane the spinor ABC is
\begin{equation}
        (\boldsymbol\sigma\cdot\boldsymbol \nabla)\Psi=i\kappa\boldsymbol \sigma_z\Psi,
        \qquad
        \Psi=
        \begin{pmatrix}
        \psi_\uparrow\\
        \psi_\downarrow
        \end{pmatrix}.
        \label{eq:SM_spinor_ABC}
\end{equation}
In components,
\begin{equation}
\begin{aligned}
        \partial_z\psi_\uparrow
        &=
        i\kappa\psi_\uparrow
        -
        (\partial_x-i\partial_y)\psi_\downarrow,\\
        \partial_z\psi_\downarrow
        &=
        i\kappa\psi_\downarrow
        +
        (\partial_x+i\partial_y)\psi_\uparrow .
\end{aligned}
        \label{eq:SM_spinor_ABC_components}
\end{equation}
Equivalently, the boundary condition is the first-order normal boundary relation
\begin{equation}
        \partial_z\Psi\big|_{z=L}
        =
        C\Psi\big|_{z=L},
        \qquad
        C=
        \begin{pmatrix}
        i\kappa & -D_-\\
        D_+ & i\kappa
        \end{pmatrix},
        \qquad
        D_\pm=\partial_x\pm i\partial_y .
        \label{eq:SM_C_operator}
\end{equation}
Therefore the detector does not impose only a scalar normal impedance. It couples the
normal boundary relation to tangential derivatives of the boundary trace.

\subsection{Exact near-boundary identity}
\label{subsec:SM_exact_near_roof_identity}

We now isolate the local normal response generated by the boundary relation itself.
Fix a time \(t\) and take the boundary trace \(\Psi(\cdot,L,t)\) as given. We use the
homogeneous relation \(\partial_z\Psi=C\Psi\) only to continue this trace a small
auxiliary distance inward, to \(z=L-\epsilon\) with \(\epsilon>0\). This continuation
is a local probe of the boundary symbol, not a physical detector layer; the true
time dependent Schr\"odinger equation (TDSE) solution need not satisfy \(\partial_z\Psi=C\Psi\) away from the detecting plane. Define the boundary-condition defect/error in a small interval \(L-\delta<z\le L\) by
\begin{equation}
        \mathcal E_{\rm bc}(z,t)
        :=
        \partial_z\Psi(\cdot,z,t)-C\Psi(\cdot,z,t).
        \label{eq:SM_Ebc_def}
\end{equation}
The spinor ABC implies
\begin{equation}
        \mathcal E_{\rm bc}(L,t)=0 .
        \label{eq:SM_Ebc_zero}
\end{equation}
After tangential Fourier transform, each mode satisfies
\begin{equation}
        \partial_z\widehat\Psi(\boldsymbol\xi,z,t)
        =
        \mathcal C(\boldsymbol\xi)\widehat\Psi(\boldsymbol\xi,z,t)
        +
        \widehat{\mathcal E}_{\rm bc}(\boldsymbol\xi,z,t).
        \label{eq:SM_mode_normal_ode}
\end{equation}
Variation of constants, integrated backward from \(L\) to \(z\), gives the exact
near-detector identity
\begin{equation}
        \widehat\Psi(\boldsymbol\xi,z,t)
        =
        e^{(z-L)\mathcal C(\boldsymbol\xi)}
        \widehat\Psi(\boldsymbol\xi,L,t)\\
        -
        \int_z^L
        e^{(z-s)\mathcal C(\boldsymbol\xi)}
        \widehat{\mathcal E}_{\rm bc}(\boldsymbol\xi,s,t)\,ds.
        \label{eq:SM_Duhamel_identity}
\end{equation}
The first term is the homogeneous normal continuation generated by the spinor ABC.
The second term is the Duhamel remainder. We denote the boundary trace by
\begin{equation}
        W_\omega(\boldsymbol\xi,t)
        :=
        \widehat\Psi_\omega(\boldsymbol\xi,L,t).
        \label{eq:SM_roof_trace}
\end{equation}
For \(z=L-\epsilon\), the Duhamel remainder is
\begin{equation}
        \widehat{\mathcal E}_{\rm Duh}
        (\boldsymbol\xi,L-\epsilon,t)
        :=
        -
        \int_0^\epsilon
        e^{-(\epsilon-u)\mathcal C(\boldsymbol\xi)}
        \widehat{\mathcal E}_{\rm bc}(\boldsymbol\xi,L-u,t)\,du .
        \label{eq:SM_Duhamel_remainder}
\end{equation}
This is the same remainder term denoted by \(\widehat{\mathcal R}_{\rm Duh}\) in the main text. Under the regularity assumption quantified in Sec.~\ref{sec:S7_Duhamel},
this remainder is \(O(\epsilon^2)\) modewise as \(\epsilon\to0\). Hence it has no
zeroth- or first-order contribution at the detecting plane. The leading inward normal response
is therefore determined by the homogeneous boundary symbol.

\subsection{Tangential symbol and spin--momentum branches}
\label{subsec:SM_symbol_branches}

The tangential Fourier transform is used only to expose the local symbol of the
boundary operator. In the finite-difference calculation, the analogous object is the
discrete tangential symbol used in Sec.~\ref{sec:S7_boundary_diagnostics}. For a tangential
Fourier mode
\begin{equation}
        \boldsymbol\xi=(\xi_x,\xi_y),
        \qquad
        \xi_\pm=\xi_x\pm i\xi_y,
\end{equation}
Eq.~\eqref{eq:SM_C_operator} gives
\begin{equation}
        \mathcal C(\boldsymbol\xi)
        =
        \begin{pmatrix}
        i\kappa & -i\xi_-\\
        i\xi_+ & i\kappa
        \end{pmatrix}.
        \label{eq:SM_C_symbol}
\end{equation}
Separate the scalar impedance from the tangential spin coupling:
\begin{equation}
        \mathcal J(\boldsymbol\xi)
        :=
        \mathcal C(\boldsymbol\xi)-i\kappa I .
        \label{eq:SM_J_def}
\end{equation}
Then
\begin{equation}
        \mathcal J(\boldsymbol\xi)
        =
        (\hat{\boldsymbol z}\times\boldsymbol\xi)\cdot\boldsymbol\sigma
        =
        -\xi_y\sigma_x+\xi_x\sigma_y,
        \qquad
        \mathcal J(\boldsymbol\xi)^2=R^2I,
        \qquad
        R=|\boldsymbol\xi|.
        \label{eq:SM_J_symbol}
\end{equation}
For \(R>0\), define
\begin{equation}
        \Gamma(\boldsymbol\xi)
        =
        \frac{\mathcal J(\boldsymbol\xi)}{R}
        =
        (\hat{\boldsymbol z}\times\hat{\boldsymbol\xi})\cdot\boldsymbol\sigma,
        \qquad
        \Gamma(\boldsymbol\xi)^2=I .
        \label{eq:SM_Gamma_def}
\end{equation}

The zero mode \(R=0\) is degenerate and carries no \(R\Gamma\) contribution; it is
understood by continuity. Therefore
\begin{equation}
        \mathcal C(\boldsymbol\xi)
        =
        i\kappa I+R\Gamma(\boldsymbol\xi).
        \label{eq:SM_C_branch_form}
\end{equation}
The two boundary eigenbranches are
\begin{equation}
        \lambda_\pm(\boldsymbol\xi)=i\kappa\pm R,
        \label{eq:SM_branch_eigenvalues}
\end{equation}
with projectors
\begin{equation}
        \Pi_\pm(\boldsymbol\xi)
        =
        \frac12\bigl(I\pm\Gamma(\boldsymbol\xi)\bigr).
        \label{eq:SM_projectors}
\end{equation}
The projectors select two tangential spin--momentum branches, with in-plane spin
direction perpendicular to the tangential momentum. This is the key local difference
between the spinor ABC and a scalar absorbing boundary.

\subsection{Auxiliary normal response}
\label{subsec:SM_auxiliary_normal_response}
Since \(\Gamma^2=I\), the homogeneous exponential in
Eq.~\eqref{eq:SM_Duhamel_identity} is explicit:\footnote{For \(R>0\), \(\mathcal C(\boldsymbol{\xi})\) is
diagonalizable with eigenvalues \(\lambda_\pm(\boldsymbol{\xi})=i\kappa\pm R\)
and spectral projectors \(\Pi_\pm(\boldsymbol{\xi})\). The Lagrange--Sylvester
functional calculus gives
\(f(\mathcal C(\boldsymbol{\xi})) =
f(\lambda_+)\Pi_+(\boldsymbol{\xi})+
f(\lambda_-)\Pi_-(\boldsymbol{\xi})\)
for analytic \(f\). Taking \(f(\lambda)=e^{(z-L)\lambda}\), and then
\(z=L-\epsilon\), gives Eq.~\eqref{eq:SM_expC}. The degenerate \(R=0\) mode is obtained by
continuity.}
\begin{equation}
\begin{aligned}
        e^{-\epsilon\mathcal C(\boldsymbol\xi)}
        &=
        e^{-i\kappa\epsilon}
        e^{-\epsilon R\Gamma(\boldsymbol\xi)}\\
        &=
        e^{-i\kappa\epsilon}
        \left[
        e^{-R\epsilon}\Pi_+(\boldsymbol\xi)
        +
        e^{R\epsilon}\Pi_-(\boldsymbol\xi)
        \right].
\end{aligned}
        \label{eq:SM_expC}
\end{equation}
Putting \(z=L-\epsilon\) in Eq.~\eqref{eq:SM_Duhamel_identity} gives
\begin{equation}
        \widehat\Psi_\omega(\boldsymbol\xi,L-\epsilon,t)
        =
        e^{-i\kappa\epsilon}
        B_{\rm br}(\epsilon,\boldsymbol\xi)W_\omega(\boldsymbol\xi,t)
        +
        \widehat{\mathcal R}_{\rm Duh}
        (\boldsymbol\xi,L-\epsilon,t),
        \label{eq:SM_near_roof_response}
\end{equation}
where the two-branch normal-response map is
\begin{equation}
        B_{\rm br}(\epsilon,\boldsymbol\xi)
        =
        e^{-R\epsilon}\Pi_+(\boldsymbol\xi)
        +
        e^{R\epsilon}\Pi_-(\boldsymbol\xi).
        \label{eq:SM_BR_def}
\end{equation}
The map resolves the auxiliary inward continuation into
exponentially decaying and growing factors
$e^{-R\epsilon}\Pi_+$ and $e^{R\epsilon}\Pi_-$,
with characteristic length $R^{-1}$ for $R>0$.
These factors describe the homogeneous boundary
response, not separate detector models or physical
amplification. They are distinct from the stationary
evanescent channels of Sec.~\ref{sec:evan}.

Using Eq.~\eqref{eq:SM_projectors},
\begin{equation}
B_{\rm br}(\epsilon,\boldsymbol\xi)
=
\cosh(R\epsilon)I-\sinh(R\epsilon)\Gamma(\boldsymbol\xi)
=
e^{-R\epsilon\Gamma(\boldsymbol\xi)} .
\end{equation}
Thus the branch structure is not visible as a multiplicative suppression of the
detector density itself, since \(B_{\rm br}(0,\boldsymbol \xi)=I\). It enters through the first inward
normal response:
\begin{equation}
\partial_\epsilon B_{\rm br}(\epsilon,\boldsymbol\xi)|_{\epsilon=0}
=
-R\Gamma(\boldsymbol\xi)
=
-\mathcal J(\boldsymbol\xi).
\label{eqn:1stVariation}
\end{equation}
Equation~\eqref{eqn:1stVariation} shows that \(R\) sets the scale of the first inward normal response for each tangential Fourier component. For the incident transverse ground-state family,
\(\ell_\perp=\omega^{-1/2}\), giving the characteristic preparation
scale
\begin{equation}
R=|\boldsymbol\xi|\sim \ell_\perp^{-1}\sim \sqrt{\omega}.
\end{equation}
The same confinement scale appears in the coupling between transverse oscillator states, as shown in Sec.~S5.5.

The evolved boundary trace \(W_\omega(\boldsymbol\xi,t)\) is determined
by the detector-present dynamics and need not retain this incident
momentum distribution.

\textit{Connection with local scattering.}
The same projectors diagonalize the local plane-wave matching
problem in Sec.~4(iii) of the main text. On a fixed tangential
branch, the boundary relation at $z=L$ becomes
\begin{equation}
    \partial_z\psi_\pm=(i\kappa\pm R)\psi_\pm.
\end{equation}
The corresponding local reflection operator is
\begin{equation}
    \mathcal{R}_{\rm loc}(k,\boldsymbol{\xi})
    =
    r_+(k,R)\Pi_+(\boldsymbol{\xi})
    +
    r_-(k,R)\Pi_-(\boldsymbol{\xi}),
\end{equation}
where $k>0$ is the incident normal wave number and the
amplitudes $r_\pm$ are given in Eq.~(27) of the main text.
For $R>0$, writing
$r_\pm=\sqrt{1-A}\,e^{i\phi_\pm}$ gives
\begin{equation}
    \mathcal{R}_{\rm loc}^{\dagger}\mathcal{R}_{\rm loc}
    =(1-A)I,
    \qquad
    \mathcal{R}_{\rm loc}
    =
    \sqrt{1-A}\,
    e^{i\phi_0}
    \exp\left(
        \frac{i\Delta\phi}{2}\Gamma(\boldsymbol{\xi})
    \right),
\end{equation}
where $A=A_\pm(k,R)$,
$\phi_\pm=\arg r_\pm(k,R)$,
$\phi_0=(\phi_++\phi_-)/2$, and
$\Delta\phi=\phi_+-\phi_-$.

Thus, at fixed $(k,\boldsymbol{\xi})$, the reflected
amplitude undergoes equal attenuation in both branches,
while their relative phase produces a spin rotation
about $\boldsymbol{n}_{\boldsymbol{\xi}}$.
The reflection operator maps incident to reflected
amplitudes, whereas $B_{\rm br}$ continues a prescribed
boundary trace inward. Both are governed by the same
boundary symbol, but neither replaces the full
confined-wavepacket evolution.

\subsection{Covariance form of the local boundary response}
\label{subsec:SM_covariance_source}

The previous subsection identifies the local boundary scale. We now express its
first-order effect on a normalized auxiliary time moment in covariance form. This
auxiliary moment is only a diagnostic of the local boundary response; the physical
detector statistic remains the detector flux at \(z=L\).

Since the scalar phase \(e^{-i\kappa\epsilon}\) drops out of densities, the relevant
first-order density response is obtained after removing this phase. By
Eq.~\eqref{eq:SM_near_roof_response} and the Duhamel estimate of
Sec.~\ref{sec:S7_Duhamel},
\begin{equation}
e^{i\kappa\epsilon}\widehat\Psi_\omega(\boldsymbol\xi,L-\epsilon,t)
=
B_{\rm br}(\epsilon,\boldsymbol\xi)W_\omega(\boldsymbol\xi,t)
+
O_{\rm Duh}(\epsilon^2).
\end{equation}
Thus the exact phase-removed trace and the homogeneous branch probe have the same
zeroth and first inward variations at the detecting plane. We therefore define
\begin{equation}
\widetilde W_{\omega,\epsilon}(\boldsymbol\xi,t)
:=
B_{\rm br}(\epsilon,\boldsymbol\xi)W_\omega(\boldsymbol\xi,t),
\end{equation}
and compute the first normal variation from this homogeneous branch probe. We keep
the branch exponential exact, rather than replacing it by a finite-depth Taylor
approximation. Since \(B_{\rm br}(\epsilon,\boldsymbol\xi)=e^{-\epsilon R\Gamma(\boldsymbol\xi)}\)
and \(\Gamma(\boldsymbol\xi)\) is Hermitian with \(\Gamma^2=I\), we have
\begin{equation}
B_{\rm br}(\epsilon,\boldsymbol\xi)^\dagger
B_{\rm br}(\epsilon,\boldsymbol\xi)
=
e^{-2\epsilon R\Gamma(\boldsymbol\xi)}
=
\cosh(2R\epsilon)I-\sinh(2R\epsilon)\Gamma(\boldsymbol\xi).
\end{equation}
Therefore the auxiliary homogeneous boundary-layer spectral density is exactly
\begin{equation}
\rho^{\rm bl}_{\omega,\epsilon}(\boldsymbol\xi,t)
:=
\left|\widetilde W_{\omega,\epsilon}(\boldsymbol\xi,t)\right|^2
=
W_\omega^\dagger e^{-2\epsilon R\Gamma(\boldsymbol\xi)}W_\omega .
\end{equation}
Where \(W^\dagger W=0\), \(a_\omega\) may be defined arbitrarily, e.g. \(a_\omega=0\). Writing
\begin{equation}
\rho^{\rm bl}_{\omega,0}(\boldsymbol\xi,t)=W_\omega^\dagger W_\omega, \quad\text{and}  \qquad  a_\omega(\boldsymbol\xi,t)
:=
-2\frac{W_\omega^\dagger\Gamma(\boldsymbol\xi)W_\omega}
        {W_\omega^\dagger W_\omega},
\end{equation}
we obtain the exact reweighting form
\begin{equation}
\rho^{\rm bl}_{\omega,\epsilon}(\boldsymbol\xi,t)
=
\rho^{\rm bl}_{\omega,0}(\boldsymbol\xi,t)\,
b_{\omega,\epsilon}(\boldsymbol\xi,t),
\end{equation}
where
\begin{equation}
b_{\omega,\epsilon}(\boldsymbol\xi,t)
=
\cosh(2R\epsilon)
+
\frac{a_\omega(\boldsymbol\xi,t)}{2}\sinh(2R\epsilon).
\label{eq:SM_Filter2}
\end{equation}
The quantity \(a_\omega\) is real because \(\Gamma\) is Hermitian. Since the eigenvalues of \(\Gamma\) are \(\pm1\),
\begin{equation}
|a_\omega(\boldsymbol\xi,t)|\le 2.
\end{equation}
In particular,
\begin{equation}
b_{\omega,0}=1,
\qquad
\left.\partial_\epsilon b_{\omega,\epsilon}\right|_{\epsilon=0}
=
R a_\omega(\boldsymbol\xi,t).
\label{eq:SM_filterAt0}
\end{equation}
No finite-depth smallness assumption \(R\epsilon\ll1\) is used here: the branch map has been kept exact, and only its derivative at the physical detecting plane is taken. We now pass from this pointwise density reweighting to a normalized auxiliary time moment. Let
\begin{equation}
d\mu^{(0)}_\omega(\boldsymbol\xi,t)
:=
\rho^{\rm bl}_{\omega,0}(\boldsymbol\xi,t)\,d^2\boldsymbol\xi\,dt,
\qquad
Z^{(0)}_\omega
:=
\int d\mu^{(0)}_\omega ,
\end{equation}
where the time integral is over the chosen finite observation window. Define the
normalized boundary-trace measure
\begin{equation}
d\nu^{(0)}_\omega
:=
\frac{d\mu^{(0)}_\omega}{Z^{(0)}_\omega}.
\end{equation}
For \(D_T>0\), constant \(\kappa\), and the same observation window, the temporal marginal of \(\nu_\omega^{(0)}\) is \(g(t;\omega)/D_T(\omega)\). The branch-reweighted auxiliary measure at depth \(\epsilon\) is
\begin{equation}
d\mu^{\rm bl}_{\omega,\epsilon}
=
\rho^{\rm bl}_{\omega,\epsilon}(\boldsymbol\xi,t)\,d^2\xi\,dt
=
b_{\omega,\epsilon}(\boldsymbol\xi,t)\,d\mu^{(0)}_\omega .
\end{equation}
After normalization,
\begin{equation}
d\nu^{\rm bl}_{\omega,\epsilon}
=
\frac{d\mu^{\rm bl}_{\omega,\epsilon}}
     {\int d\mu^{\rm bl}_{\omega,\epsilon}}
=
\frac{b_{\omega,\epsilon}(\boldsymbol\xi,t)}
     {\mathbb E_{\nu^{(0)}_\omega}[b_{\omega,\epsilon}]}
\,d\nu^{(0)}_\omega .
\end{equation}
This identity is exact for the homogeneous branch probe. Let \(F(\boldsymbol\xi,t)\) be real and independent of \(\epsilon\). Assume \(0<Z_\omega^{(0)}<\infty\), that the following expectations are finite, and that the auxiliary measures are normalizable near \(\epsilon=0^+\), with differentiation permitted under the integrals. Then
\begin{equation}
\mathbb E_{\nu^{\rm bl}_{\omega,\epsilon}}[F]
=
\frac{\mathbb E_{\nu^{(0)}_\omega}[F\,b_{\omega,\epsilon}]}
     {\mathbb E_{\nu^{(0)}_\omega}[b_{\omega,\epsilon}]}.
\label{eq:SM_TimeMoment}     
\end{equation}
Differentiating Eq.~\eqref{eq:SM_TimeMoment} at \(\epsilon=0\), and using Eq.~\eqref{eq:SM_filterAt0}
gives
\begin{equation}
\left.
\partial_\epsilon
\mathbb E_{\nu^{\rm bl}_{\omega,\epsilon}}[F]
\right|_{\epsilon=0}
=
\mathbb E_{\nu^{(0)}_\omega}[F\,R a_\omega]
-
\mathbb E_{\nu^{(0)}_\omega}[F]\,
\mathbb E_{\nu^{(0)}_\omega}[R a_\omega].
\end{equation}
Thus
\begin{equation}
\left.
\partial_\epsilon
\mathbb E_{\nu^{\rm bl}_{\omega,\epsilon}}[F]
\right|_{\epsilon=0}
=
\operatorname{Cov}_{\nu^{(0)}_\omega}
\!\left(F,R a_\omega\right).
\end{equation}
Taking \(F=t\) to calculate the normalized first time moment, the local first-order boundary-response coefficient is
\begin{equation}
\Lambda_\omega
:=
\left.
\partial_\epsilon
\bar t_{\rm bl}(\omega,\epsilon)
\right|_{\epsilon=0},
\qquad
\bar t_{\rm bl}(\omega,\epsilon)
:=
\mathbb E_{\nu^{\rm bl}_{\omega,\epsilon}}[t],
\end{equation}
and therefore
\begin{equation}
\Lambda_\omega
=
\operatorname{Cov}_{\nu^{(0)}_\omega}
\!\left(t,R a_\omega\right).
\end{equation}
This is the desired covariance form of the local boundary response. The derivation used
the exact branch exponential through \(b_{\omega,\epsilon}\), and the only derivative taken
was the derivative at the physical detecting plane, \(\epsilon=0\).  For the harmonic transverse family we write
\begin{equation}
R=\sqrt{\omega}\,s,
\end{equation}
where \(s\) is the dimensionless tangential radius in oscillator units. Hence
\begin{equation}
\Lambda_\omega
=
\sqrt{\omega}\,
\operatorname{Cov}_{\nu^{(0)}_\omega}
\!\left(t,s a_\omega\right)
=
\sqrt{\omega}\,\beta^{\rm bl}_\omega,
\label{eq:SM_source2}
\end{equation}
with
\begin{equation}
\beta^{\rm bl}_\omega
:=
\operatorname{Cov}_{\nu^{(0)}_\omega}
\!\left(t,s a_\omega\right).
\end{equation}
Since \(|a_\omega|\le 2\) and \(0\le t\le T\) on the finite observation window, Cauchy--Schwarz gives, for example,
\begin{equation}
|\beta_\omega^{\rm bl}|
\le
2T\,
\sqrt{\mathbb E_{\nu_\omega^{(0)}}[s^2]} .
\label{eqn:covBound}
\end{equation}
At fixed \(T\), Eq.~\eqref{eqn:covBound} bounds \(\beta_\omega^{\mathrm{bl}}\) uniformly on any family for which the scaled second moment is uniformly bounded. Table~\ref{tab:SM_one_grid_rechecked} reports the moments and covariances at the sampled confinements; a bound on the magnitude does not fix the sign or variation of the coefficient.
For \(D_T>0\), constant \(\kappa\), and the same observation window, the temporal marginal of \(\nu_\omega^{(0)}\) is \(g(t;\omega)/D_T(\omega)\), so \(\bar t_{\mathrm{bl}}(\omega,0)=\bar\tau_T^{\mathrm{cond}}(\omega)\). The derivative \(\Lambda_\omega\) measures the initial change of this normalized moment under the prescribed inward branch reweighting. Its sign is set by the correlation of time with the spin-resolved tangential factor \(Ra_\omega\). Equation~\eqref{eq:SM_source2} separates the oscillator wave-number scale from this trace-dependent correlation. Confinement-dependent changes of the physical detection law are obtained by comparing the corresponding detector-present evolutions.

\subsection{Stationary multichannel scattering in the transverse oscillator basis}
\label{sec:evan}
We give the oscillator-basis formulation and numerical
validation of the stationary scattering problem introduced
in Sec.~4(iv) of the main text. The semi-infinite harmonic
guide $z\leq L$ has transverse domain $\mathbb{R}^2$ and
is terminated by the spinor ABC.

Let $\chi_{n_x,n_y}^{(\omega)}$ denote the normalized Cartesian
eigenfunctions of the transverse harmonic oscillator centered at
$(x_c,y_c)$. With $H_\perp=H_{xy}(\omega)$, they satisfy
\begin{equation}
H_\perp\chi_{n_x,n_y}^{(\omega)}
=
E_{n_x,n_y}^{\perp}\chi_{n_x,n_y}^{(\omega)},
\qquad
E_{n_x,n_y}^{\perp}=\omega(n_x+n_y+1).
\end{equation}
The ground state $\chi_{00}^{(\omega)}$ is the untruncated
counterpart of the transverse preparation in Eq.~(S4).

We consider an incident particle in the transverse ground state
with spin up and longitudinal wave number $k>0$.
Its incident wave is proportional to
$\chi_{00}^{(\omega)}|\uparrow\rangle e^{ik(z-L)}$.
Although the bulk Hamiltonian is spin independent, the tangential
derivatives in the spinor ABC couple different oscillator states
and spin components at the detecting plane.

Using $p_j=-i\partial_j$ and the oscillator ladder operators
$a_j$ and $a_j^\dagger$, $j\in\{x,y\}$, the tangential boundary
operator $\mathcal{J}=C-i\kappa I$ takes the form
\begin{equation}
\mathcal{J}
=
\sigma_y p_x-\sigma_x p_y
=
i\sqrt{\frac{\omega}{2}}
\left[
\sigma_y(a_x^\dagger-a_x)
-
\sigma_x(a_y^\dagger-a_y)
\right].
\end{equation}

For the incident transverse ground state, the derivatives
$D_\pm=\partial_x\pm i\partial_y$ in Eq.~(S44) satisfy
\begin{equation}
D_\pm\chi_{00}^{(\omega)}
=
-\sqrt{\omega}\,\chi_\pm^{(\omega)},
\qquad
\chi_\pm^{(\omega)}
=
\frac{\chi_{10}^{(\omega)}
\pm i\chi_{01}^{(\omega)}}{\sqrt{2}}.
\end{equation}
Consequently,
\begin{equation}
\mathcal{J}
\left(\chi_{00}^{(\omega)}|\uparrow\rangle\right)
=
-\sqrt{\omega}\,
\chi_+^{(\omega)}|\downarrow\rangle.
\end{equation}
The boundary thus couples neighboring oscillator shells
with accompanying spin flips and matrix elements
proportional to $\sqrt{\omega}$ for fixed oscillator
indices.

The total incident energy is $E=\omega+k^2/2$.
For a transverse channel $\boldsymbol{n}=(n_x,n_y)$,
the longitudinal wave number is
\begin{equation}
q_{\boldsymbol{n}}
=
\sqrt{k^2-2\omega(n_x+n_y)}.
\end{equation}
The square-root branch is chosen such that
$\operatorname{Re}q_{\boldsymbol{n}}\geq0$ and
$\operatorname{Im}q_{\boldsymbol{n}}\geq0$.
Open channels have positive real wave numbers, whereas
closed channels have $q_{\boldsymbol{n}}=i\gamma_{\boldsymbol{n}}$,
with
$\gamma_{\boldsymbol{n}}
=\sqrt{2\omega(n_x+n_y)-k^2}>0$.
In particular, all excited transverse channels are closed
when $\omega>k^2/2$.
None of the sampled parameter combinations lies exactly
at a channel threshold.

For the numerical calculation, we retain complete oscillator
shells with $n_x+n_y\leq M$.
Suppressing the common time factor $e^{-iEt}$, we expand
the stationary spinor as
\begin{equation}
\Psi_E^{(M)}(x,y,z)
=
\sum_{\substack{n_x,n_y\geq0\\n_x+n_y\leq M}}
\sum_{\alpha=\uparrow,\downarrow}
u_{\boldsymbol{n},\alpha}(z)
\chi_{n_x,n_y}^{(\omega)}(x,y)|\alpha\rangle.
\end{equation}

Since the bulk Hamiltonian is separable, the longitudinal
amplitude in each channel can be written as
\begin{equation}
u_{\boldsymbol{n},\alpha}(z)
=
a_{\boldsymbol{n},\alpha}e^{iq_{\boldsymbol{n}}(z-L)}
+
b_{\boldsymbol{n},\alpha}e^{-iq_{\boldsymbol{n}}(z-L)}.
\end{equation}
The incident coefficients are
$a_{\boldsymbol{n},\alpha}
=\delta_{n_x,0}\delta_{n_y,0}\delta_{\alpha,\uparrow}$.
The second term describes reflected waves in open channels
and exponentially decaying contributions in closed channels.

Let $K_M$ be the diagonal matrix of longitudinal
wave numbers, including spin degeneracy, and let
$\mathcal{C}_M=i\kappa I+\mathcal{J}_M$ be the boundary operator
projected onto the retained oscillator states.
Writing $\mathbf{a}$ and $\mathbf{b}$ for the incident and
reflected coefficient vectors, the spinor ABC gives
\begin{equation}
iK_M(\mathbf{a}-\mathbf{b})
=
\mathcal{C}_M(\mathbf{a}+\mathbf{b}).
\end{equation}
The boundary-trace coefficient vector
$\mathbf{W}=\mathbf{a}+\mathbf{b}$ therefore satisfies
\begin{equation}
(iK_M+\mathcal{C}_M)\mathbf{W}
=
2iK_M\mathbf{a}.
\end{equation}

The incident longitudinal flux is $k$.
Using the detector flux law in Eq.~(S7), the absorbed and
reflected flux fractions are
\begin{equation}
A_{\mathrm{osc}}
=
\frac{\kappa}{k}\|\mathbf{W}\|^2,
\qquad
R_{\mathrm{refl}}
=
\frac{1}{k}
\sum_{\boldsymbol{n},\alpha:\,\mathrm{open}}
q_{\boldsymbol{n}}|b_{\boldsymbol{n},\alpha}|^2.
\end{equation}
Closed channels carry no asymptotic reflected longitudinal
flux, but they contribute to the boundary density and hence
to absorption.
Hermiticity of $\mathcal{J}_M$ and the boundary matching
relation imply $A_{\mathrm{osc}}+R_{\mathrm{refl}}=1$.

The ideal harmonic guide and the spinor ABC are invariant
under simultaneous spatial and spin rotations about the
guide axis. The calculation can therefore be restricted
to the sector $L_z+\sigma_z/2=1/2$, where
$L_z=(x-x_c)p_y-(y-y_c)p_x$.
In the circular oscillator basis, defined by
$a_\pm^\dagger=(a_x^\dagger\pm ia_y^\dagger)/\sqrt{2}$,
this sector contains one spin-orbital state per oscillator shell:
\begin{equation}
|2j\rangle
=
|j,j\rangle_{\mathrm{circ}}|\uparrow\rangle,
\qquad
|2j+1\rangle
=
|j+1,j\rangle_{\mathrm{circ}}|\downarrow\rangle,
\end{equation}
where $j=0,1,\ldots$.
After rephasing the channel states, the matrix of
$\mathcal{J}_M$ is tridiagonal, with adjacent-shell
coupling magnitudes
\begin{equation}
\left|\langle n+1|\mathcal{J}|n\rangle\right|
=
\sqrt{\omega\left(\left\lfloor n/2\right\rfloor+1\right)},
\end{equation}
where $n$ is the total oscillator shell index.

For $k=\kappa=\pi$ and incident spin up, the resulting
stationary absorption probabilities are
\begin{equation}
\begin{array}{c|cccccccc}
\omega & 1 & 25 & 50 & 75 & 100 & 150 & 200 & 300 \\
\hline
A_{\mathrm{osc}}
& 0.976 & 0.851 & 0.743 & 0.667
& 0.611 & 0.532 & 0.478 & 0.406
\end{array}.
\end{equation}

Increasing the maximum shell index from $M=127$ to 255
changes the calculated absorption probabilities by less
than $1.6\times10^{-13}$, with no further change at the
displayed precision for $M=511$.
Independent complete-shell Cartesian-basis calculations
agree with the reduced representation to within
$6\times10^{-16}$ at matched cutoffs through $M=24$.

These results confirm confinement-dependent absorption
suppression in the absence of the reflecting lower wall.
The stationary absorption probability $A_{\mathrm{osc}}$
characterizes a monochromatic incident state and is not
identified with the finite-window detected fraction
$D_T(\omega)$.

\section{Finite-grid boundary-symbol diagnostic and Duhamel estimate}
\label{sec:S7_boundary_diagnostics}

We use the eight reference evolutions of Secs.~\ref{sec:SM_numerics}--\ref{sec:SM_finite_window} over
$0\leq t\leq40$ to evaluate the boundary-weighted covariance, check the
recorded detector flux, and compare the auxiliary inward continuation with
the neighbouring numerical plane. Section~\ref{sec:S7_Duhamel} gives the continuum
remainder estimate; Sec.~\ref{sec:S7_spin_branches} examines the spin and branch composition of the detector signal. Section~\ref{sec:refinement-T40} reports a separate \(T=40\) refinement series at \(\omega=300\); Sec.~\ref{sec:SM_initial_boundary_compatibility} examines sensitivity to initial boundary compatibility.

\noindent\textit{Boundary trace and discrete symbol.}
The auxiliary probe depth is one recorded longitudinal spacing,
\begin{equation}
 \epsilon_g=h_z,
 \label{eq:S7_probe_depth}
\end{equation}
with $h_z\simeq0.013$. The boundary trace is evaluated at the numerical
detecting plane $z_D$ of Table~\ref{tab:SM_reference_numerics}, and the inward trace at the preceding
stored plane $z_{N_z-2}$. The continuum estimate in Sec.~\ref{sec:S7_Duhamel} retains the
boundary coordinate $L$. We use the notation of Sec.~\ref{sec:SM_boundary_symbol} for the numerical
boundary trace,
\begin{equation}
 W_\omega(\boldsymbol\xi,t)
 =\widehat\Psi_\omega(\boldsymbol\xi,z_D,t),
 \label{eq:S7_boundary_trace}
\end{equation}
where $\boldsymbol\xi=(\xi_x,\xi_y)$ labels the orthonormal tangential
Fourier transform. The central-difference symbol is
\begin{equation}
 \boldsymbol\xi_h(\boldsymbol\xi)
 =\left(\frac{\sin(\xi_xh_x)}{h_x},
         \frac{\sin(\xi_yh_y)}{h_y}\right),
 \qquad R_h=|\boldsymbol\xi_h|,\qquad s=\frac{R_h}{\sqrt\omega}.
 \label{eq:S7_discrete_symbol}
\end{equation}
Set $J_h=(\widehat{\mathbf z}\times\boldsymbol\xi_h)\cdot\boldsymbol\sigma$.
Then $J_h^2=R_h^2I$ and
\begin{equation}
 \widehat J_h=\frac{J_h}{R_h}\qquad (R_h>0).
 \label{eq:S7_normalized_symbol}
\end{equation}
This is the local stencil symbol, not an exact diagonalization of the
finite transverse Dirichlet problem. Numerically, we treat
$R_h\leq r_{0,h}$ as neutral, with
$r_{0,h}=64\epsilon_{\rm mach}\max(1,\max R_h)$ in double precision.
The even $100\times100$ Fourier grid has four zero/Nyquist combinations.
On nonneutral modes with nonzero density, the spin factor is
\begin{equation}
 a_\omega(\boldsymbol\xi,t)
 =-2\frac{W_\omega^\dagger\widehat J_hW_\omega}
           {W_\omega^\dagger W_\omega},
 \qquad |a_\omega|\leq2.
 \label{eq:S7_spin_factor}
\end{equation}
We set $a_\omega=0$ on neutral or zero-density modes. Neutral modes remain
in the total boundary weight but contribute neither to the $R_ha_\omega$
nor to the $sa_\omega$ numerator.

\noindent\textit{Boundary-weighted moments.}
For a horizon $T$ with nonzero integrated boundary density, define
\begin{equation}
 \mathbb E_{\nu_\omega^{(0)}}[F]
 =\frac{\displaystyle\int_0^T
       \sum_{\boldsymbol\xi}\|W_\omega(\boldsymbol\xi,t)\|^2
       F(\boldsymbol\xi,t)\,dt}
       {\displaystyle\int_0^T
       \sum_{\boldsymbol\xi}\|W_\omega(\boldsymbol\xi,t)\|^2\,dt}.
 \label{eq:S7_weighted_mean}
\end{equation}
The transverse quadrature factor $h_xh_y$ cancels in this ratio.
The recorded density and moment numerators are interpolated linearly on
their actual diagnostic times. Both their integrals and first time moments
are evaluated from these interpolants. The finite-grid counterparts of
the coefficients in Sec.~\ref{sec:SM_boundary_symbol} are
\begin{equation}
 \beta_\omega^{\rm bl}
 =\operatorname{Cov}_{\nu_\omega^{(0)}}(t,sa_\omega),
 \qquad
 \Lambda_\omega
 =\operatorname{Cov}_{\nu_\omega^{(0)}}(t,R_ha_\omega)
 =\sqrt\omega\,\beta_\omega^{\rm bl}.
 \label{eq:S7_covariance}
\end{equation}
Their dependence on $T$ is implicit. The dimensionless probe scale
\begin{equation}
 q_\omega
 =\mathbb E_{\nu_\omega^{(0)}}[(h_zR_h)^2]
 =(h_z\sqrt\omega)^2\mathbb E_{\nu_\omega^{(0)}}[s^2]
 \label{eq:S7_probe_scale}
\end{equation}
quantifies the sampled tangential scale at this depth; it is not an error
estimate or a smallness assumption in the covariance derivation.

\noindent\textit{Flux consistency.} Let $P(t)$ be the stored squared norm of Sec.~\ref{sec:SM_numerics}, $P_0=P(0)$, and
$N_D(t)=h_xh_y\sum_{\boldsymbol\xi}\|W_\omega(\boldsymbol\xi,t)\|^2$
the transverse squared-norm integral at the detecting plane.
The endpoint-density budget check is
\begin{equation}
 \Delta_{\rm bud}^{\rm rel}
 =\frac{\left|\displaystyle\int_0^T\kappa N_D(t)\,dt
                 -[P_0-P(T)]\right|}{P_0-P(T)}.
 \label{eq:S7_endpoint_budget}
\end{equation}
The boundary-density integral is evaluated by trapezoidal quadrature on
the native plane-record times, separately from the accepted-step
Crank--Nicolson (CN) event sum of Sec.~\ref{sec:SM_finite_window}.

We also compare the normalized boundary-density rate
$J_{\rho,h}=\kappa N_D/P_0$ with the accepted-step detector rate
$g_{\rm CN}$. On the native plane-record intervals $I_j$, of duration
$\Delta_j$, partitioning $[0,40]$, define
\begin{equation}
 e(f,g)=\frac{1}{D_{40}}\sum_j\Delta_j
           |\overline f_j-\overline g_j|.
\end{equation}
Here $D_{40}$ is from the same run. Bars denote interval-averaged rates:
CN event probabilities are accumulated with interval overlap and divided
by $\Delta_j$; endpoint rates are trapezoid-averaged. The absolute values
prevent cancellation between intervals.  For the $\omega=300$ reference evolution, the interval-averaged discrepancy is $e(J_{\rho,h},g_{\rm CN})\simeq 9.43\times10^{-4}$.

For the unrounded semidiscrete last-row stencil, the normalized current
entering the final numerical cell satisfies
\begin{equation}
 J_{{\rm link},h}=J_{\rho,h}+\frac{dP_{\rm cell}}{dt},
 \qquad P_{\rm cell}=\frac{h_zN_D}{2P_0}.
\end{equation}
Thus current entering that cell includes the change of its stored
probability as well as permanent detector loss.  
At $\omega=300$, the CN midpoint link current, integrated over the transverse grid and both spin components, has discrepancy $5.29\times10^{-3}$ from $g_{\rm CN}$ on the native plane-record intervals. Subtracting the normalized terminal-cell probability change divided by the interval duration reduces this discrepancy to $5.64\times10^{-7}$. This is an interval-averaged consistency check of the discrete probability balance within the same evolution.

\noindent\textit{One-grid inward continuation.}
We compare the homogeneous continuation generated by the boundary symbol
with the recorded field on the preceding longitudinal plane:
\begin{equation}
 \begin{aligned}
 W_{\rm hom}(\boldsymbol\xi,t;h_z)
   &=e^{-i\kappa h_z}e^{-h_zJ_h}W_\omega(\boldsymbol\xi,t),\\
 W_{\rm in}(\boldsymbol\xi,t)
   &=\widehat\Psi_\omega(\boldsymbol\xi,z_{N_z-2},t).
 \end{aligned}
 \label{eq:S7_continuation}
\end{equation}
The reported discrepancy is
\begin{equation}
 \mathcal E_{2\rm br}(h_z)
 =\left\langle
 \frac{\|W_{\rm in}(\cdot,t)-W_{\rm hom}(\cdot,t;h_z)\|_{\ell^2_\xi}}
      {\|W_{\rm in}(\cdot,t)\|_{\ell^2_\xi}}
 \right\rangle_{\rm valid}.
 \label{eq:S7_continuation_defect}
\end{equation}
Here
$\langle f\rangle_{\rm valid}
 =\int_V N_D(t)f(t)\,dt/\int_V N_D(t)\,dt$,
where $V$ consists of diagnostic intervals with finite, valid discrepancy
values at both endpoints. We integrate the product of the separate linear
density and discrepancy interpolants. Excluded intervals enter neither
integral, and no additional signal-amplitude threshold is applied.

\begin{table}[H]
\centering
\caption{Finite-grid diagnostics at $T=40$. The endpoint budget uses the
native plane-record times; the moments and continuation discrepancy use
the native spin/branch times. The valid-interval average is defined in
Eq.~\eqref{eq:S7_continuation_defect}. Displayed digits do not specify
discretization accuracy.}
\label{tab:SM_one_grid_rechecked}

\begingroup
\small

\setlength{\tabcolsep}{5pt}

\renewcommand{\arraystretch}{0.80}

\setlength{\heavyrulewidth}{0.12em}
\setlength{\lightrulewidth}{0.05em}


\begin{tabular*}{0.89\linewidth}
{@{\extracolsep{\fill}}rrrrrrrr@{}}

\toprule

$\omega$
& $h_z\sqrt{\omega}$
& $q_\omega$
& $\Delta_{\rm bud}^{\rm rel}$
& $\mathbb E[s^2]$
& $\beta_\omega^{\rm bl}$
& $\Lambda_\omega$
& ${\mathcal E}_{2\rm br}$ \\

\midrule

1
& 0.013
& $1.75\times10^{-4}$
& $1.81\times10^{-5}$
& 0.982
& -0.16
& -0.16
& $8.90\times10^{-4}$ \\

25
& 0.067
& $8.74\times10^{-3}$
& $1.61\times10^{-4}$
& 1.966
& -0.255
& -1.274
& $3.47\times10^{-3}$ \\

50
& 0.094
& $2.44\times10^{-2}$
& $5.15\times10^{-4}$
& 2.745
& -0.060
& -0.424
& $6.55\times10^{-3}$ \\

75
& 0.115
& $4.44\times10^{-2}$
& $2.67\times10^{-4}$
& 3.333
& -0.015
& -0.134
& $9.78\times10^{-3}$ \\

100
& 0.133
& $6.77\times10^{-2}$
& $7.13\times10^{-4}$
& 3.808
& 0.001
& 0.012
& $1.32\times10^{-2}$ \\

150
& 0.163
& $1.21\times10^{-1}$
& $3.84\times10^{-4}$
& 4.522
& 0.008
& 0.099
& $2.07\times10^{-2}$ \\

200
& 0.189
& $1.78\times10^{-1}$
& $6.69\times10^{-4}$
& 5.009
& 0.004
& 0.053
& $2.87\times10^{-2}$ \\

300
& 0.231
& $2.95\times10^{-1}$
& $9.43\times10^{-4}$
& 5.524
& 0.001
& 0.025
& $4.56\times10^{-2}$ \\

\bottomrule

\end{tabular*}

\endgroup
\end{table}

Table~\ref{tab:SM_one_grid_rechecked} gives $0.982\leq\mathbb E[s^2]\leq5.524$. On the reference grid, the covariance $\beta_\omega^{\rm bl}$ is negative at $\omega=1,25,50,75$ and positive at $\omega=100,150,200,300$. The identity $\Lambda_\omega=\sqrt{\omega}\,\beta_\omega^{\rm bl}$ separates the oscillator scale from a trace-dependent correlation; it does not determine the restricted-mean fit coefficient. The largest tabulated endpoint-budget mismatch is $9.43\times10^{-4}$, and the largest averaged continuation discrepancy is $4.56\%$.

\subsection{Duhamel estimate}
\label{sec:S7_Duhamel}

We now give the continuum estimate underlying the auxiliary first
variation. For fixed $(\boldsymbol\xi,t)$, assume that the boundary-condition
defect $\widehat E_{\rm bc}$ of Sec.~\ref{subsec:SM_exact_near_roof_identity} is $C^1$ in a normal neighbourhood
of $L$, and define
\begin{equation}
 \mathcal{G}_\omega(\boldsymbol\xi,t;\epsilon)
 =\sup_{0\leq u\leq\epsilon}
   \|\partial_z\widehat E_{\rm bc}(\boldsymbol\xi,L-u,t)\|.
 \label{eq:S7_defect_derivative}
\end{equation}
Since $\widehat E_{\rm bc}(\boldsymbol\xi,L,t)=0$,
\begin{equation}
 \|\widehat E_{\rm bc}(\boldsymbol\xi,L-u,t)\|
 \leq u\mathcal{G}_\omega(\boldsymbol\xi,t;\epsilon).
 \label{eq:S7_bc_bound}
\end{equation}
The Hermitian part of $C(\boldsymbol\xi)$ has eigenvalues $\pm R$, so
\begin{equation}
 \|e^{-(\epsilon-u)C(\boldsymbol\xi)}\|
 \leq e^{(\epsilon-u)R}\leq e^{\epsilon R}.
 \label{eq:S7_exponential_bound}
\end{equation}
Inserting these bounds into the Duhamel remainder of Sec.~\ref{subsec:SM_exact_near_roof_identity} gives
\begin{equation}
 \begin{aligned}
 \|\widehat E_{\rm Duh}(\boldsymbol\xi,L-\epsilon,t)\|
 &\leq e^{\epsilon R}\mathcal{G}_\omega(\boldsymbol\xi,t;\epsilon)
           \int_0^\epsilon u\,du\\
 &=\frac{\epsilon^2}{2}e^{\epsilon R}
             \mathcal{G}_\omega(\boldsymbol\xi,t;\epsilon).
 \end{aligned}
 \label{eq:S7_Duhamel_bound}
\end{equation}
Local boundedness of $G_\omega$ therefore yields, modewise,
\begin{equation}
 \widehat E_{\rm Duh}(\boldsymbol\xi,L-\epsilon,t)
 =O(\epsilon^2)\qquad(\epsilon\to0^+),
 \label{eq:S7_Duhamel_order}
\end{equation}
and
\begin{equation}
 \widehat E_{\rm Duh}(\boldsymbol\xi,L,t)=0,
 \qquad
 \left.\partial_\epsilon
 \widehat E_{\rm Duh}(\boldsymbol\xi,L-\epsilon,t)
 \right|_{\epsilon=0^+}=0.
 \label{eq:S7_Duhamel_first_variation}
\end{equation}
The remainder thus leaves the first inward normal variation unchanged.
This is a modewise estimate; no bound uniform in $\boldsymbol\xi$ or
$\omega$ is asserted. The one-depth comparison in
Table~\ref{tab:SM_one_grid_rechecked} does not test the asymptotic
$\epsilon^2$ order.

\subsection{Boundary branches and particle-spin polarization}
\label{sec:S7_spin_branches}

\begin{figure}[H]
 \centering
 \includegraphics[width=0.92\linewidth]{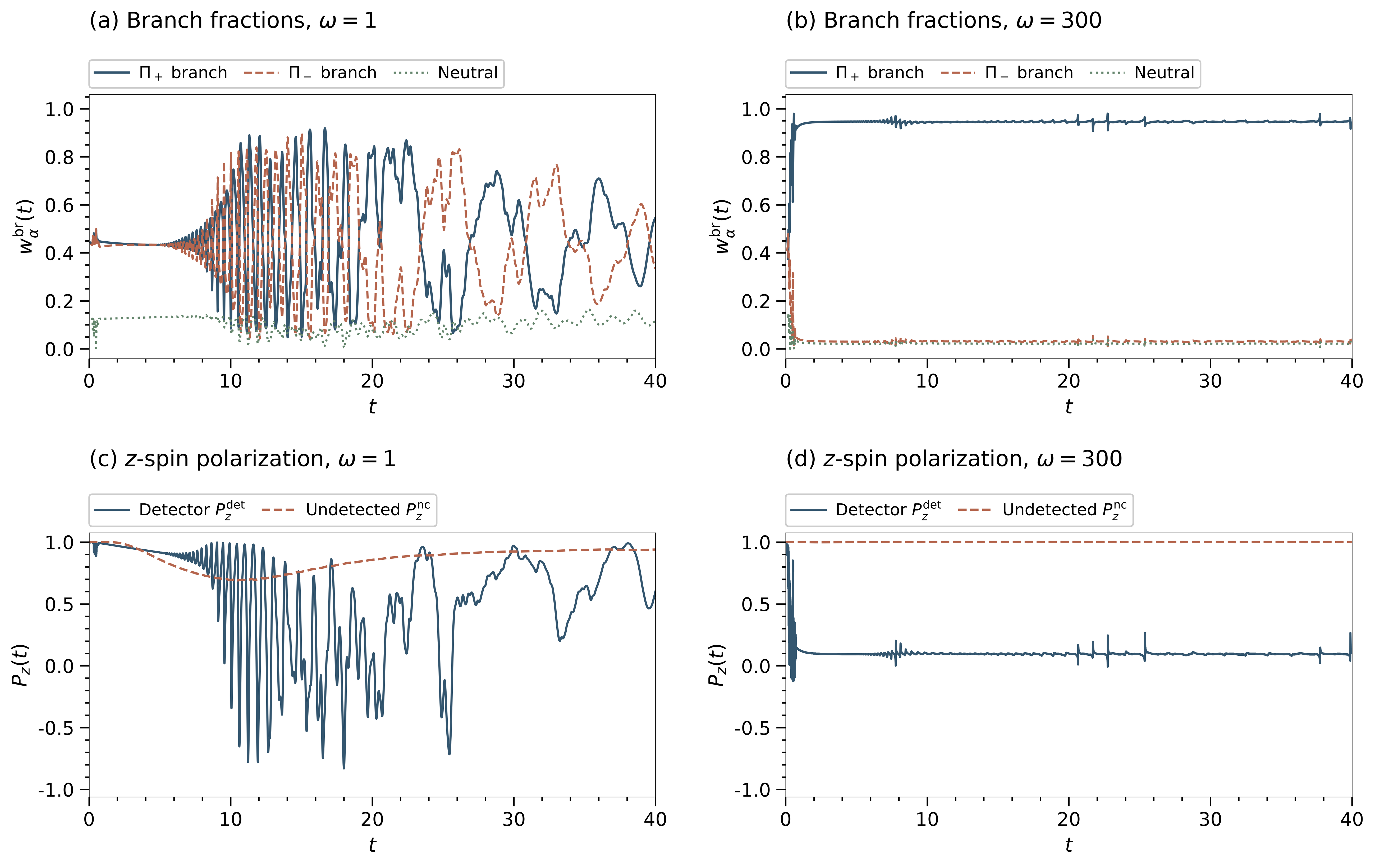}
 \caption{Time-dependent branch fractions and particle-spin polarizations
 for the reference preparation of Table~\ref{tab:SM_reference_numerics},
 $\eta=(1,0)^{\mathsf T}$, $k_z=\kappa=\pi$, $L=20$, and
 $\sigma_\parallel=0.5$. Left: $\omega=1$; right: $\omega=300$.
 (a),(b) Boundary fractions in the $+$, $-$ and neutral sectors.
 (c),(d) Detector-associated and whole-survivor laboratory $z$-spin
 polarizations. Curves connect defined recorded ratios without smoothing;
 the diagnostics use the reference discretization of Table~\ref{tab:SM_reference_numerics}. The absolute detector rates are shown in main-text Fig.~1, which uses the time-step-refined history at $\omega=300$.}
 \label{fig:S7_spin_branches}
\end{figure}

For the same eight evolutions, initialized with $\eta=(1,0)^{\mathsf T}$,
we compare two distinct quantities: the tangential branch composition at
the detecting plane and the particle-spin polarization associated with
detection or survival. Table~\ref{tab:SM_branch_spin_rechecked} summarizes all eight
confinements; Fig.~\ref{fig:S7_spin_branches} shows the time dependence for
$\omega=1$ and $300$.

\noindent\textit{Tangential branch composition.}
For $R_h>r_{0,h}$, the finite-difference counterparts of the projectors in
Eq.~\eqref{eq:SM_projectors} (main-text Eq.~(23)) are
\begin{equation}
 \Pi_{\pm,h}=\frac12\left(I\pm\frac{J_h}{R_h}\right).
\end{equation}
They act separately on each Fourier component of the evolved boundary
trace. With the orthonormal Fourier convention, define
\begin{equation}
 N_\pm(t)=h_xh_y\sum_{R_h>r_{0,h}}
     \|\Pi_{\pm,h}W_\omega(\boldsymbol\xi,t)\|^2,
 \qquad
 N_0(t)=h_xh_y\sum_{R_h\leq r_{0,h}}
     \|W_\omega(\boldsymbol\xi,t)\|^2.
\end{equation}
The neutral sector $N_0$ contains numerically zero-symbol modes, not a
third spin orientation. On the same stored field,
$N(t)=N_+(t)+N_-(t)+N_0(t)=N_D(t)$. The absolute component rates and their
sum are
\begin{equation}
 g_{\alpha,h}^{\rm br}(t)=\frac{\kappa N_\alpha(t)}{P_0},
 \qquad \sum_{\alpha\in\{+,-,0\}}g_{\alpha,h}^{\rm br}(t)=J_{\rho,h}(t).
\end{equation}
For positive denominators, the instantaneous and window-integrated
fractions are
\begin{equation}
 w_\alpha^{\rm br}(t)=\frac{N_\alpha(t)}{N(t)},
 \qquad
 \overline w_\alpha^{\rm br}(T)
   =\frac{\int_0^T N_\alpha(t)\,dt}{\int_0^T N(t)\,dt},
 \qquad\alpha\in\{+,-,0\}.
\end{equation}
Each triplet sums to one; the integrated fractions weight times by boundary density and include the neutral sector. The same projectors resolve the spin factor in
Eq.~\eqref{eq:S7_spin_factor}: on nonneutral modes with nonzero density,
\begin{equation}
 a_\omega(\boldsymbol\xi,t)
 =-2\frac{\|\Pi_{+,h}W_\omega\|^2-\|\Pi_{-,h}W_\omega\|^2}
          {\|W_\omega\|^2}.
\end{equation}
The covariance retains the momentum- and time-resolved information in
this expression, which is lost in the integrated branch fractions.

\noindent\textit{Particle-spin polarization.}
For normalized initial data, we define the detector-associated spin-matrix rate \(G_\omega(t)\) and the subnormalized reduced spin matrix \(Q_\omega(t)\) of the state surviving no click up to time \(t\),  in continuum notation
\begin{equation}
 G_\omega(t)=\kappa\int_{\Sigma_L}\Psi_\omega\Psi_\omega^\dagger\,dA,
 \qquad
 Q_\omega(t)=\int_{\Omega_\omega}\Psi_\omega\Psi_\omega^\dagger\,dV.
\end{equation}
Their traces are $g(t;\omega)$ and $S(t;\omega)$, respectively. The diagonal
entries of $G_\omega$ resolve the detector rate in the chosen laboratory
spin basis. The corresponding Pauli polarizations are
\begin{equation}
 P_i^{\rm det}(t)=\frac{\operatorname{tr}[\sigma_iG_\omega(t)]}{g(t;\omega)},
 \qquad
 P_i^{\rm nc}(t)=\frac{\operatorname{tr}[\sigma_iQ_\omega(t)]}{S(t;\omega)},
 \qquad i\in\{x,y,z\}.
\end{equation}
Here ``nc'' denotes conditioning on no click; $P_z=+1$ is the prepared
$z$-spin orientation. These quantities characterize particle-spin content
within the model without assuming an additional apparatus spin measurement.
The detector polarization accumulated over the window is
\begin{equation}
 \overline P_i^{\rm det}(T)
 =\frac{\int_0^T\operatorname{tr}[\sigma_iG_\omega(t)]\,dt}{D_T(\omega)}.
\end{equation}

\noindent\textit{Numerical evaluation.}
Detector polarizations use accepted-step CN event matrices, with interval
overlap for window integrals and step-midpoint plotting times. Survivor
polarizations use the recorded diagnostic-time matrices. Each ratio uses
the trace of the same matrix record; integrated polarizations use the
trace of the integrated matrix. Branch integrals use piecewise-linear
histories on their actual recording times. Undefined ratios are omitted;
no smoothing or additional signal threshold is applied.

\begin{table}[H]
\centering
\caption{Integrated branch fractions and particle-spin polarizations at \(T=40\). Bars denote ratios of integrals over \([0,40]\). Branch fractions include the neutral sector and use native boundary-density records; detector polarizations use accepted-step event matrices. \(P_z^{\rm nc}(40)\) describes the entire final surviving state. Displayed digits do not specify discretization accuracy; rounded fractions need not sum exactly to one.}
\label{tab:SM_branch_spin_rechecked}

\begingroup
\small


\setlength{\tabcolsep}{6pt}

\renewcommand{\arraystretch}{0.80}

\setlength{\tabcolsep}{4pt}

\setlength{\heavyrulewidth}{0.12em}
\setlength{\lightrulewidth}{0.05em}


\begin{tabular*}{0.58\linewidth}
{@{\extracolsep{\fill}}rrrrrr@{}}

\toprule

$\omega$
& $\overline w_+^{\rm br}(40)$
& $\overline w_-^{\rm br}(40)$
& $\overline w_0^{\rm br}(40)$
& $\overline P_z^{\rm det}(40)$
& $P_z^{\rm nc}(40)$ \\

\midrule

1
& 0.442
& 0.430
& 0.128
& 0.907
& 0.940 \\

25
& 0.819
& 0.114
& 0.067
& 0.307
& 0.998 \\

50
& 0.879
& 0.073
& 0.049
& 0.232
& 1.000 \\

75
& 0.902
& 0.058
& 0.040
& 0.193
& 1.000 \\

100
& 0.916
& 0.049
& 0.035
& 0.169
& 1.000 \\

150
& 0.931
& 0.040
& 0.029
& 0.143
& 1.000 \\

200
& 0.938
& 0.036
& 0.026
& 0.128
& 1.000 \\

300
& 0.946
& 0.032
& 0.022
& 0.097
& 1.000 \\

\bottomrule

\end{tabular*}

\endgroup
\end{table}
\noindent\textit{Results and interpretation.} At $\omega=1$, the integrated nonneutral branch fractions are nearly balanced despite substantial instantaneous variation. At $\omega=300$, the $+$ branch accounts for approximately $95\%$ of the integrated boundary density in both diagnostic windows, $[0,3.6]$ and $(3.6,40]$. The detected and surviving ensembles have different laboratory-$z$ spin content: at $\omega=300$, $\overline P_z^{\mathrm{det}}(40)\simeq0.097$, while $P_z^{\mathrm{nc}}(40)$ is close to unity. The former averages detections over the window; the latter describes the entire state surviving at its endpoint. The integrated detector $z$-polarization decreases across the sampled confinements. Branch fractions resolve momentum-dependent transverse spin axes, whereas $P_z$ uses a fixed laboratory axis. Unequal weights in the evolved boundary trace are compatible with equal local absorption probabilities for equally normalized incident branches at fixed normal and tangential wave numbers in the matching problem of main-text Sec.~4(iii).

\FloatBarrier

\subsection{Refinement checks at $\omega=300$ and $T=40$}
\label{sec:refinement-T40}

This five-run series tests separate transverse, longitudinal, and time-step
refinements with the reference preparation (Table~\ref{tab:SM_convergence}).
The baseline and spatial refinements use $\Delta t=2\times10^{-4}$.
The numerical detector and lateral-wall positions are held fixed.
The $N_z=2999$ grid halves the baseline $h_z$, while the non-nested
$N_z=4000$ grid reduces it further to $h_z\simeq0.004998$.
All statistics below use $T=40$.
For baseline $b$, refinement $r$, and a comparison window
$W\subseteq[0,T]$, define the following percentage discrepancies:
\begin{equation}
\begin{aligned}
\epsilon_{\rm rel}
&=\frac{\bigl|\mu^{*,(r)}(T)-\mu^{*,(b)}(T)\bigr|}
 {\bigl|\mu^{*,(r)}(T)\bigr|}\times100\%,\\[3pt]
\epsilon_{\rm CN}(W)
&=\frac{\displaystyle\int_W
 \bigl|g_{\rm CN}^{(r)}(t)-g_{\rm CN}^{(b)}(t)\bigr|\,dt}
 {\displaystyle\int_W g_{\rm CN}^{(b)}(t)\,dt}\times100\%,\\[3pt]
\epsilon_{\rm br}(W)
&=\frac{\displaystyle\sum_{\alpha\in\{+,-,0\}}\int_W
 \bigl|g_{\alpha,h}^{\mathrm{br},(r)}(t)
       -g_{\alpha,h}^{\mathrm{br},(b)}(t)\bigr|\,dt}
 {\displaystyle\int_W J_{\rho,h}^{(b)}(t)\,dt}\times100\%.
\end{aligned}
\label{eq:SM_refinement_discrepancies}
\end{equation}
Here $\epsilon_{\rm rel}$ uses the \emph{refined} mean in the denominator.
The accepted-step rate $g_{\rm CN}$ is the recorded detector probability
divided by the step duration and launch norm. The sampled branch rates
$g_{\alpha,h}^{\rm br}$, including the neutral sector, sum to $J_{\rho,h}$
as defined above; all branch differences use this same total baseline
denominator. CN rates are piecewise constant on accepted steps, while
sampled branch densities are interpolated linearly on their native times.
Absolute differences are integrated on the union of the time grids,
splitting zero crossings of the linear differences; no smoothing or time
shift is applied.

\begin{table}[H]
\centering
\begingroup
\small
\setlength{\tabcolsep}{5pt}
\renewcommand{\arraystretch}{0.75}
\caption{Spatial and time-step refinement checks at \(\omega=300\) and \(T=40\), with the reference preparation of Table~\ref{tab:SM_reference_numerics} and fixed numerical boundary positions. (a) Detection statistics, integrated $+$-branch fraction, and full-window rate discrepancies. (b) Integrated detector and final whole-survivor $z$-polarizations, boundary-weighted momentum moment and covariance, and late-window rate discrepancies. Discrepancies are defined in Eq.~\eqref{eq:SM_refinement_discrepancies}; unmarked rate discrepancies use $[0,40]$, and late discrepancies use $[8,40]$. All $\epsilon$ entries compare with the baseline and are expressed as
percentages. Displayed digits do not specify discretization accuracy.}
\label{tab:SM_convergence}

\begin{tabular}{@{}lrrrrrrrrr@{}}
\toprule
(a) Refinement & $N_x=N_y$ & $N_z$ & $\Delta t$ & $D_T$
& $\mu^*(T)$ & $\epsilon_{\rm rel}$ & $\bar w_+^{\rm br}(T)$
& $\epsilon_{\rm CN}$ & $\epsilon_{\rm br}$\\
\midrule
Baseline & 100 & 1500 & $2\times10^{-4}$ & 0.740 & 18.631 & 0 & 0.946 & 0 & 0 \\
Transverse & 120 & 1500 & $2\times10^{-4}$ & 0.743 & 18.510 & 0.654 & 0.946 & 0.742 & 0.752 \\
Longitudinal I & 100 & 2999 & $2\times10^{-4}$ & 0.733 & 18.977 & 1.824 & 0.952 & 6.188 & 6.320 \\
Longitudinal II & 100 & 4000 & $2\times10^{-4}$ & 0.729 & 19.148 & 2.697 & 0.953 & 6.841 & 6.999 \\
Time step & 100 & 1500 & $1\times10^{-4}$ & 0.740 & 18.626 & 0.031 & 0.946 & 2.647 & 2.647 \\

\bottomrule
\noalign{\vskip 10pt}
\toprule

(b) Refinement & $N_x=N_y$ & $N_z$ & $\Delta t$
& $\overline P_z^{\rm det}(T)$ & $P_z^{\rm nc}(T)$
& $\mathbb E_{\nu_\omega^{(0)}}[s^2]$ & $10^3\beta_\omega^{\rm bl}$
& $\epsilon_{\rm CN}^{\rm late}$ & $\epsilon_{\rm br}^{\rm late}$\\
\midrule
Baseline & 100 & 1500 & $2\times10^{-4}$ & 0.097 & 1.0 & 5.524 & 1.47 & 0 & 0 \\
Transverse & 120 & 1500 & $2\times10^{-4}$ & 0.100 & 1.0 & 5.607 & 0.99 & 0.588 & 0.610 \\
Longitudinal I & 100 & 2999 & $2\times10^{-4}$ & 0.104 & 1.0 & 6.167 & 14.856 & 9.580 & 9.857 \\
Longitudinal II & 100 & 4000 & $2\times10^{-4}$ & 0.099 & 1.0 & 6.311 & 16.745 & 9.665 & 10.001 \\
Time step & 100 & 1500 & $1\times10^{-4}$ & 0.097 & 1.0 & 5.524 & 1.661 & 5.332 & 5.314 \\
\bottomrule
\end{tabular}

\endgroup
\end{table}

All four refinements in Table~\ref{tab:SM_convergence} preserve the
integrated $+$-branch predominance and the contrast between
detector-associated and final in-guide $z$-polarizations.
Comparing $N_z=2999$ and $4000$ as $b$ and $r$, respectively, in
Eq.~\eqref{eq:SM_refinement_discrepancies}, gives a restricted-mean
discrepancy of $0.890\%$. The detector- and branch-rate discrepancies
are $0.884\%$ and $0.913\%$ over $[0,40]$, and $0.503\%$ and
$0.565\%$ over $[8,40]$, respectively.
These smaller differences between the two finer grids support the
stability of the finite-window rate histories under further longitudinal
refinement. The covariance nevertheless changes from
$10^3\beta_\omega^{\rm bl}=14.856$ to $16.745$, and its magnitude
remains resolution-sensitive. These comparisons quantify changes between
the tested discretizations; they do not bound the continuum error.

To assess whether the changing spacing of prominent late-time maxima is sensitive to longitudinal resolution, we analyzed the native accepted-step detector-rate records at \(\omega=300\) for \(N_z=2999\) and \(N_z=4000\), with identical physical boundary positions and \(\Delta t=2\times10^{-4}\). Using an absolute peak-prominence threshold of 0.003 and a minimum separation of 0.1 time units, we identified 41 corresponding prominent maxima across the intervals \(8\le t<16\), \(16\le t<27\), and \(27\le t<40\). The median adjacent-peak spacings were respectively 0.5234, 0.7516, and 1.2127 for \(N_z=2999\), compared with 0.5236, 0.7528, and 1.2140 for \(N_z=4000\). The maximum shift of a matched peak time was 0.0074. The normalized integrated absolute detector-rate discrepancy over \(8\le t\le40\) was approximately 0.50\%. Linearly detrended, Hann-windowed spectral analysis yields dominant nonzero-frequency Fourier bins at approximately \(1.375\), \(1.000\), and \(0.615\) cycles per time unit over the respective intervals \(8\le t<16\), \(16\le t<27\), and \(27\le t<40\), with consistent results on both refined longitudinal grids. These are finite-window spectral peaks, not independently identified eigenmode frequencies. The changing spectral composition is consistent with multimode interference, but a unique modal explanation is not established.

The increase in median adjacent-peak spacing persists under transverse refinement and time-step halving, as well as longitudinal refinement. At \(N_z=1500\), halving the time step from \(2\times10^{-4}\) to \(10^{-4}\) changes the median spacings in the three observation windows by less than \(0.45\%\). Across all five numerical configurations, the increase persists for the tested absolute prominence thresholds \(\{0.001,0.002,0.003,0.004,0.005\}\) and minimum peak separations \(\{0.05,0.10,0.15,0.20\}\). The same conclusion holds for window-relative prominence thresholds of \(2.5\%\), \(5\%\), \(10\%\), \(15\%\), and \(20\%\) of the corresponding detector-rate range. The precise spacing ratio nevertheless depends on the peak-selection criterion. The same dominant nonzero-frequency Fourier bins are found across all five configurations. These results support the evolution toward longer characteristic time scales without establishing a unique recurrence period or identifying the responsible eigenmodes.

\subsection{Sensitivity to initial boundary compatibility}
\label{sec:SM_initial_boundary_compatibility}

We repeat the baseline calculation of Table S5 at \(\omega=300\) and \(T=40\), modifying only the initial preparation. The sampled spinor of Sec. S2 is multiplied by the smooth cutoff
\begin{equation}
\begin{aligned}
\Psi_0^{\rm cut}(\mathbf r)
&=\mathcal N_{\rm cut}F_{\rm cut}(\mathbf r)
\Psi_0(\mathbf r),\\
F_{\rm cut}(\mathbf r)
&=\prod_{j\in\{x,y,z\}}
c\left(\frac{d_j-0.25}{0.75}\right),
\end{aligned}
\end{equation}

where \(d_j\) is the distance to the nearer of the first and last stored coordinate planes in direction \(j\), measured in oscillator units for \(x,y\) and in simulation length units for \(z\). Here \(c(s)=0\) for \(s\leq0\), \(c(s)=1\) for \(s\geq1\), and
\begin{equation}
c(s)=\frac{e^{-1/s}}{e^{-1/s}+e^{-1/(1-s)}},
\qquad 0<s<1.
\end{equation}

The modified state is stored in the same numerical precision and renormalized using Eq. (S12). The corresponding smooth profile vanishes in a neighborhood of every boundary, satisfying both the Dirichlet and spinor ABCs at \(t=0\).

Let \(u_h\) and \(v_h\) denote the reference and modified launch states, normalized in the inner product associated with Eq. (S12), with \(u_h\) reconstructed using the reference initialization procedure. Their pure-state trace distance is
\begin{equation}
\delta_h=\sqrt{1-|\langle u_h,v_h\rangle_h|^2}
\simeq4.09\times10^{-5}.
\end{equation}

For the ideal unrounded CN evolution with the same prescribed sequence of time steps for both preparations and with exact linear solves, the successive detector-loss effects and the final survival effect form a POVM by the discrete norm-loss identity of Sec. S2. The trace-distance bound on measurement statistics therefore gives
\begin{equation}
\begin{gathered}
|\Delta D_T|\leq\delta_h,\qquad
|\Delta\mu^*(T)|\leq T\delta_h,\\
\int_0^T|\Delta g_{\rm CN}(t)|\,dt
\leq2\delta_h.
\end{gathered}
\end{equation}

These bounds do not include finite-precision propagation errors.

At \(T=40\), the numerical differences in \(D_{40}\), \(\mu^*(40)\), and \(\bar{\tau}_{40}^{\rm cond}\) are \(1.03\times10^{-6}\), \(1.59\times10^{-5}\), and \(1.43\times10^{-5}\) in absolute value, respectively. Using Eq. (S130), the detector-rate discrepancies \(\epsilon_{\rm CN}\) are \(2.24\times10^{-4}\%\) over \([0,40]\) and \(3.51\times10^{-4}\%\) over \([8,40]\). The corresponding branch-rate discrepancies \(\epsilon_{\rm br}\) are \(5.42\times10^{-4}\%\) and \(8.17\times10^{-4}\%\). These changes are substantially smaller than those observed under spatial and temporal refinement in Sec. S6.3, demonstrating the insensitivity of the recorded detector and branch-rate profiles to this preparation modification at the tested resolution.

\section{Reduced return closure}
\label{sec:S8_return_closure}

Section~\ref{sec:S7_boundary_diagnostics} characterizes the signal at the detecting plane. Here we
separate exact probability balance near that plane from an assumed
scalar description of later returns of undetected amplitude. The physical
detection density remains $g(t;\omega)=J_L(t)$.

\noindent\textit{Exact probability balance.} For $0<\ell<L$, introduce a matching plane and the adjoining control
region within the original guide:
\begin{equation}
 \Sigma_* = \{z=z_*\},\qquad z_*=L-\ell,\qquad
 \Omega_{\rm bl}^{(\ell)}
 =\{\mathbf r\in\Omega_\omega:z_*<z<L\}.
 \label{eq:S8_control_region}
\end{equation}
Its probability content is
\begin{equation}
 P_{\rm bl}^{(\ell)}(t)
 =\int_{\Omega_{\rm bl}^{(\ell)}}\rho(\mathbf r,t)\,dV.
 \label{eq:S8_layer_probability}
\end{equation}
Define the matching-plane and detector fluxes by
\begin{equation}
 J_*(t)=\int_{\Sigma_*}j_z(x,y,z_*,t)\,dx\,dy,
 \qquad
 J_L(t)=\int_{\Sigma_L}j_z(x,y,L,t)\,dx\,dy,
 \label{eq:S8_plane_fluxes}
\end{equation}
where $j_z$ is the normal component of the Pauli current and $J_*>0$
means net flow into the control region from below. The spinor ABC gives
\begin{equation}
 J_L(t)=\kappa\int_{\Sigma_L}\rho(x,y,L,t)\,dx\,dy\geq0.
 \label{eq:S8_detector_law}
\end{equation}
With no lateral flux, continuity gives the exact balance
\begin{equation}
 \frac{d}{dt}P_{\rm bl}^{(\ell)}(t)=J_*(t)-J_L(t).
 \label{eq:S8_exact_balance}
\end{equation}
Probability entering the region can be detected, remain temporarily
stored, or leave again through $\Sigma_*$. The signed net flux $J_*$
does not separate these transfers into individual encounters.

\noindent\textit{Assumed return closure.}
Let $F_{\rm out}^{\rm nc}(t;\omega,\ell)$ denote the unresolved outgoing
flux of undetected amplitude through $\Sigma_*$. A causal kernel
$k_{\omega,\ell}^{\rm ret}(u)$ represents its later contribution to the
detector signal, where $u\geq0$ is the lag and $t-u$ the earlier departure
time. The branch fractions of Sec.~\ref{sec:S7_spin_branches} describe spin--momentum
composition rather than separate encounters; neither they nor the
early/late split of Sec.~\ref{sec:SM_finite_window} determines $F_{\rm out}^{\rm nc}$ or this
kernel. We write
\begin{equation}
 g_{\rm box}(t;\omega)=g_{\rm first}(t;\omega)
 +\int_0^t k_{\omega,\ell}^{\rm ret}(u)
 F_{\rm out}^{\rm nc}(t-u;\omega,\ell)\,du.
 \label{eq:S8_return_ansatz}
\end{equation}
Here $g_{\rm box}$ is the finite-guide detector signal within the reduced
description, and $g_{\rm first}$ its first-encounter contribution,
including the boundary response. Equation~\eqref{eq:S8_return_ansatz}
compresses unresolved phase, branch, and mode information into a scalar
kernel; it is not an exact TDSE decomposition.

For a one-channel moment calculation, we make the additional assumption
of a scalar convolution closure with a causal lag-only kernel
$k_{\omega,\ell}^{\rm eff}$, distinct from $k_{\omega,\ell}^{\rm ret}$:
\begin{equation}
 \begin{aligned}
 g_{\rm box}&=g_{\rm first}+K_{\omega,\ell}^{\rm eff}g_{\rm box},\\
 (K_{\omega,\ell}^{\rm eff}f)(t)
 &=\int_0^t k_{\omega,\ell}^{\rm eff}(u)f(t-u)\,du.
 \end{aligned}
 \label{eq:S8_volterra}
\end{equation}
The kernel summarizes the unresolved no-click channel; the occurrence
of $g_{\rm box}$ on the right does not mean that detected probability
returns. If $g_{\rm first}\in L^1(0,\infty)$ and
$\|k_{\omega,\ell}^{\rm eff}\|_{L^1(0,\infty)}<1$, the Neumann series
\begin{equation}
 g_{\rm box}=(I-K_{\omega,\ell}^{\rm eff})^{-1}g_{\rm first}
 =\sum_{n=0}^{\infty}(K_{\omega,\ell}^{\rm eff})^n g_{\rm first}
 \label{eq:S8_neumann}
\end{equation}
converges in $L^1(0,\infty)$.

\noindent\textit{Time moments of the closure.}
Assume finite absolute zeroth and first moments of both fluxes and of
the effective kernel, and define
\begin{equation}
 M_0[f]=\int_0^\infty f(t)\,dt,\qquad
 M_1[f]=\int_0^\infty t f(t)\,dt.
 \label{eq:S8_flux_moments}
\end{equation}
The kernel moments are
\begin{equation}
 m_{\rm ret}^{(0)}=\int_0^\infty k_{\omega,\ell}^{\rm eff}(u)\,du,
 \qquad
 m_{\rm ret}^{(1)}=\int_0^\infty u k_{\omega,\ell}^{\rm eff}(u)\,du.
 \label{eq:S8_kernel_moments}
\end{equation}
For $m_{\rm ret}^{(0)}\neq0$, write
\begin{equation}
 \overline U_{\rm ret}(\omega;\ell)
 =\frac{m_{\rm ret}^{(1)}}{m_{\rm ret}^{(0)}}.
 \label{eq:S8_mean_lag}
\end{equation}
For a nonnegative kernel, $m_{\rm ret}^{(0)}$ is an effective return
weight and $\overline U_{\rm ret}$ its mean lag. A detection-density
interpretation additionally requires nonnegative fluxes with total
weights at most one. In general, these are the zeroth moment and the
normalized first moment of the assumed kernel.

For positive zeroth flux moments, set
\begin{equation}
 \overline\tau_{\rm first}=\frac{M_1[g_{\rm first}]}{M_0[g_{\rm first}]},
 \qquad
 \overline\tau_{\rm box}=\frac{M_1[g_{\rm box}]}{M_0[g_{\rm box}]}.
\end{equation}
The convolution identities are
\begin{align}
 M_0[K_{\omega,\ell}^{\rm eff}f]
 &=m_{\rm ret}^{(0)}M_0[f],
 \label{eq:S8_convolution_mass}\\
 M_1[K_{\omega,\ell}^{\rm eff}f]
 &=m_{\rm ret}^{(0)}M_1[f]+m_{\rm ret}^{(1)}M_0[f].
 \label{eq:S8_convolution_first_moment}
\end{align}
Applying them to Eq.~\eqref{eq:S8_volterra} gives
\begin{equation}
 \overline\tau_{\rm box}=\overline\tau_{\rm first}
 +\frac{m_{\rm ret}^{(1)}}{1-m_{\rm ret}^{(0)}}.
 \label{eq:S8_mean_identity}
\end{equation}
For $m_{\rm ret}^{(0)}\neq0$, this is equivalently
\begin{equation}
 \overline\tau_{\rm box}=\overline\tau_{\rm first}
 +\alpha_{\rm ret}(\omega;\ell)\overline U_{\rm ret}(\omega;\ell),
 \qquad
 \alpha_{\rm ret}(\omega;\ell)
 =\frac{m_{\rm ret}^{(0)}(\omega;\ell)}{1-m_{\rm ret}^{(0)}(\omega;\ell)}.
 \label{eq:S8_weight_delay}
\end{equation}
This separates the effective returned weight from its lag in the nonnegative-kernel interpretation. It is a moment identity of the assumed closure, not a fit to the simulations. The local scale \(R=\sqrt{\omega}s\) and the computed \(\beta_\omega^{\mathrm{bl}}\) do not determine the kernel moments.

\textit{Connection to the finite-window statistic.}
For $D_T>0$, Eq.~\eqref{eqn:balance} combines
detected-event timing with the probability still undetected at $T$.
The recorded histories do not determine the untruncated means
entering the closure. The fixed-source lower-wall comparison in
Sec.~\ref{subsec:SM_length_scan} provides separate numerical evidence
that the late detector response depends on the reflecting boundary.
The closure therefore provides a conditional interpretation
of possible return contributions, whereas the reported
finite-window statistics come from the full spinor-ABC evolution.

\section{Control Checks}
\label{sec:SM_controls}

This section tests the sensitivity of the confinement contrast to selected bulk-spin couplings and transverse preparations, and compares the spinor ABC with finite absorbing layers. The comparisons concern the stated configurations and observation windows; they do not establish a universal confinement law or uniqueness of the detector model.

\subsection{Bulk spin and transverse-state controls}
\label{sec:SM_S91_controls}
\label{sec:SM_S91_global_controls}
We compare \(\omega=1\) and \(100\) using \(H=H_{\mathrm{waveguide}}(\omega)+H_a\), retaining the longitudinal preparation, spatial grid, spinor ABC, and initial spinor \(\eta=(1,0)^T\) of Sec.~\ref{sec:SM_numerics}. The Hermitian additions in Table~\ref{tab:SM_S91_controls} act throughout the guide from \(t=0\), and the transverse state is initially \(\chi_{00}^{(\omega)}\).

To test whether the confinement contrast requires transverse
ground-state preparation, we instead set $H_a=0$ and use
the excited chiral profile introduced in Sec.~S5.5,
\begin{equation}
    \chi_+^{(\omega)}
    =
    \sqrt{\omega}(X+iY)\chi_{00}^{(\omega)},
\end{equation}
restricted and normalized on the same grid.
Here $X=x-L_x/2$, $Y=y-L_y/2$, and $p_j=-i\partial_j$.

Each tested control gives a smaller \(D_{20}\) and larger \(\bar{\tau}_{20}^{\mathrm{cond}}\) and \(\mu^*(20)\) at \(\omega=100\) than at \(\omega=1\), with late detections at both confinements. Relative to the same-\(\omega\) reference, the uniform and radial terms change \(\mu^*(20)\) by at most approximately \(1.2\%\) in magnitude. The Rashba and Rashba–Dresselhaus terms increase it by approximately \(3.0\)–\(6.0\%\) across the four comparisons. The chiral preparation increases it by approximately \(5.5\%\) at \(\omega=1\) and \(24.7\%\) at \(\omega=100\). Thus the confinement contrast persists for these couplings and this excited preparation, although its magnitude remains dependent on the dynamics and initial state. These two-confinement tests do not establish a scaling law or independence from arbitrary transverse preparations.

\begin{table}[H]
\centering
\begingroup
\small

\setlength{\tabcolsep}{6pt}
\renewcommand{\arraystretch}{0.75}

\setlength{\heavyrulewidth}{0.10em}
\setlength{\lightrulewidth}{0.045em}

\caption{Controls at \(T=20\), with two confinement rows per control. From normalized, piecewise-linear survival histories we evaluate \(D_{20}=1-S(20)\), \(\mu^*(20)\), and the absolute late probability \(D_L=S(3.6)-S(20)\). The conditional mean follows from Eq.~\eqref{eq:s34draft:conditional_from_survival}. Entries are rounded independently.}
\label{tab:SM_S91_controls}
\label{tab:SM_S91_global_controls}

\begin{tabular}{@{}lrrrrr@{}}
\toprule
Control and added Hamiltonian
& $\omega$
& $D_{20}$
& $\bar\tau_{20}^{\mathrm{cond}}$
& $\mu^*(20)$
& $D_L$ \\
\midrule

Reference: $H_a=0$
& 1
& 0.925
& 3.947
& 5.151
& 0.348 \\ 

& 100
& 0.724
& 5.548
& 9.534
& 0.307 \\ 

\midrule

Uniform: $H_a=\pi\sigma_x+\tfrac12\sigma_z$
& 1
& 0.928
& 3.938
& 5.091
& 0.350 \\ 

& 100
& 0.724
& 5.549
& 9.536
& 0.307 \\ 

\midrule

Radial: $H_a=0.2\sqrt{\omega}(X\sigma_x+Y\sigma_y)$
& 1
& 0.923
& 3.935
& 5.165
& 0.346 \\ 

& 100
& 0.723
& 5.556
& 9.554
& 0.307 \\ 

\midrule

Rashba: $H_a=0.5(\sigma_xp_y-\sigma_yp_x)$
& 1
& 0.920
& 4.111
& 5.375
& 0.356 \\ 

& 100
& 0.707
& 5.690
& 9.883
& 0.306 \\ 

\midrule

\begin{tabular}[c]{@{}ll@{}}
Rashba--Dresselhaus:
& $H_a=0.4(\sigma_xp_y-\sigma_yp_x)$ \\
&$\phantom{H_a={}}+0.3(\sigma_xp_x-\sigma_yp_y)$
\end{tabular}
& 1
& 0.917
& 4.147
& 5.460
& 0.355 \\ 

& 100
& 0.710
& 5.663
& 9.818
& 0.307 \\ 

\midrule

Chiral: $H_a=0$, profile $\chi_+^{(\omega)}$
& 1
& 0.914
& 4.060
& 5.435
& 0.348 \\ 

& 100
& 0.598
& 6.434
& 11.89
& 0.287 \\ 

\bottomrule
\end{tabular}

\endgroup
\end{table}

\subsection{Hamiltonian-layer mimics}
\label{subsec:SM_layer_mimics}

We compare the spinor ABC with finite absorbing layers, first using selected spin-dependent additions and then calibrating the complex entrance response. For the layer calculations, the ABC is removed and the guide extends to a nominal Dirichlet backing wall at \(z=22\).\footnote{Related scalar ABC limits involving complex potentials and Dirichlet-backed layers are recently discussed by Kaimal and Tumulka (see the main-text bibliography).} The source and initial spin are retained: \(z_c=10\), \(k_z=\kappa=\pi\), \(\sigma_{\parallel}=0.5\), and \(\eta=(1,0)^T\). The transverse grid is unchanged, and \(h_z=20/1500\) gives \(N_z=1650\) in the extended guide. For \(H=H_{\mathrm{Herm}}-iW_{\mathrm{loss}}\), with \(W_{\mathrm{loss}}\geq0\), the detection density is
\begin{equation}
 g_{\mathrm{layer}}(t)
 =2\langle\Psi_t,W_{\mathrm{loss}}\Psi_t\rangle
 =-\partial_t\|\Psi_t\|^2.
 \label{eq:s82-revised-loss}
\end{equation}
All calculations use \(T=20\). The spinor-ABC reference values in Tables~\ref{tab:S82_generic} and~\ref{tab:S82_inverse_T20} are taken from Table~\ref{tab:SM_S91_controls}. Layer statistics are evaluated from normalized, piecewise-linear survival histories, with \(D_L=S(3.6)-S(20)\). When the launch sample is absent, normalization uses the first recorded norm, with \(S=1\) before that sample.

\textit{Generic layers.} The controls in Table~\ref{tab:S82_generic} use the absorption profile
\begin{equation}
\chi(z)=\tfrac12[1+\tanh((z-20)/0.165)],
\end{equation}
whose entrance tail extends below \(z=20\). Write \(K=\sigma_xp_y-\sigma_yp_x\), \(R=\sqrt{K^2}\), and \(Q_s=(I+sK/R)/2\) for \(R>0\) and \(s=\pm1\). The projector \(Q_s\) selects \(K=sR\); its tangential Fourier symbol is \(\Pi_{-s}\).  At the stated strengths, the spin-dependent additions change the scalar-layer statistics only weakly. Their restricted means lie between approximately \(6.37\) and \(6.51\), compared with \(9.534\) for the spinor ABC at \(\omega=100\).

\begin{table}[H]
\centering\small
\caption{Generic absorbing-layer controls at \(\omega=100\) and \(T=20\). The spinor-ABC reference is taken from Table~\ref{tab:SM_S91_controls}. The layer calculations use the listed Hamiltonian additions in the extended guide without the spinor ABC, with \(\Delta t=4\times10^{-4}\). Statistics use the survival-history convention specified in the text. Entries are rounded independently and do not specify discretization accuracy.}
\label{tab:S82_generic}
\label{tab:S82_generic_completed}
\setlength{\tabcolsep}{5pt}\renewcommand{\arraystretch}{0.70}
\begin{tabular}{@{}lrrrr@{}}
\toprule
Layer addition / reference & $D_{20}$ & $\bar\tau_{20}^{\rm cond}$ & $\mu^*(20)$ & $D_L$\\
\midrule
Spinor ABC reference & 0.724 & 5.548 & 9.534 & 0.307 \\
$-i40\chi I$ & 0.857 & 4.240 & 6.487 & 0.304 \\
$\sigma_x-i40\chi I$ & 0.857 & 4.240 & 6.487 & 0.304 \\
$\chi\sigma_x-i40\chi I$ & 0.857 & 4.241 & 6.487 & 0.304 \\
$-i\chi(40I+20\sigma_y)$ & 0.862 & 4.198 & 6.374 & 0.306 \\
$0.5\chi K-i40\chi I$ & 0.857 & 4.241 & 6.487 & 0.304 \\
$-i\chi(40I+0.5RQ_+)$ & 0.856 & 4.251 & 6.514 & 0.304 \\
\bottomrule
\end{tabular}
\end{table}

\textit{Impedance-calibrated layers.} We next use a uniform layer with a separately chosen complex potential in each tangential branch. The added Hamiltonian term is
\begin{equation}
 H_{\mathrm{layer}}=\mathbf{1}_{[20,22)}(z)V_{\mathrm{imp}},
 \qquad
 V_{\mathrm{imp}}=\sum_{s=\pm1}V_s(R)Q_s,
 \qquad -\operatorname{Im}V_s\geq0.
 \label{eq:s82-revised-inverse}
\end{equation}
For fixed branch \(s\) and longitudinal energy \(E\) in the auxiliary stationary problem, the Dirichlet-backed field is proportional to \(\sin[q_s(22-z)]\). Its derivative-to-value ratio at the entrance is
\begin{equation}
 \begin{aligned}
 \Lambda_s(E,R)&=-q_s\cot(q_s\ell),
 &q_s^2&=2[E-V_s(R)],\\
 Z_s(R)&=i\kappa-sR,
 &\ell&=2.
 \end{aligned}
 \label{eq:s82-revised-entrance}
\end{equation}
Here \(Z_s\) is the target ABC impedance. Central-exact calibration imposes \(\Lambda_s(E_{\mathrm{cal}},R)=Z_s(R)\) at \(E_{\mathrm{cal}}=k_z^2/2\). Weighted-broadband calibration starts from this solution and reduces the normalized squared complex-impedance mismatch over the positive-\(k\) Gaussian band, using flux weights proportional to \(k|\widetilde{\phi}_0(k)|^2\) and enforcing passivity. The objective also penalizes the mismatch at \(E_{\mathrm{cal}}\), with dimensionless weight \(w_c\) relative to the normalized band contribution. We use \(w_c=2\) at \(\omega=1\) and \(100\), and additionally test \(w_c=0.5\) at \(\omega=1\). The potentials are calibrated up to \(R_{\mathrm{cal}}=4\sqrt{\omega}\) and held at their cutoff values for larger \(R\). Functions of \(K\) are evaluated in the spectral basis of the centered transverse Dirichlet finite-difference operators. This calibration concerns the auxiliary stationary problem; the full transverse harmonic evolution also couples the branches.

\begin{table}[H]
\centering
\begingroup
\small
\setlength{\tabcolsep}{5pt}
\renewcommand{\arraystretch}{0.70}
\setlength{\heavyrulewidth}{0.10em}
\setlength{\lightrulewidth}{0.045em}
\caption{Finite-window statistics at \(T=20\) for the spinor ABC, scalar complex absorbing potential layer (CAP, $-i40\chi(z)I$), and impedance-calibrated layers.
The absorbing layers have nominal thickness $\ell=2$.
The weighted-broadband cases are labelled by the central-energy
penalty weight $w_{\mathrm c}$. At $\omega=1$, the broadband
calculations use $\Delta t=8\times10^{-4}$ and the remaining
cases use $10^{-3}$; all $\omega=100$ calculations use
$4\times10^{-4}$. Entries are rounded independently and do
not specify discretization accuracy.}
\label{tab:S82_inverse_T20}
\label{tab:S82_broadband_common}
\begin{tabular}{@{}lrrrrr@{}}
\toprule
Model & $\omega$ & $D_{20}$ & $\bar\tau_{20}^{\mathrm{cond}}$ & $\mu^*(20)$ & $D_L$ \\
\midrule
Spinor ABC reference & 1 & 0.925 & 3.947 & 5.151 & 0.348 \\ 
Inverse: central-exact & 1 & 0.805 & 4.817 & 7.773 & 0.379 \\ 
Inverse: weighted-broadband ($w_{\mathrm c}=2$) & 1 & 0.911 & 4.064 & 5.480 & 0.382 \\ 
Inverse: weighted-broadband ($w_{\mathrm c}=0.5$) & 1 & 0.905 & 4.003 & 5.522 & 0.363 \\ 
\midrule
Spinor ABC reference & 100 &  0.724 & 5.548 & 9.534 & 0.307 \\ 
Scalar CAP & 100 & 0.857 & 4.240 & 6.487 & 0.304 \\ 
Inverse: central-exact & 100 & 0.530 & 6.645 & 12.927 & 0.254 \\ 
Inverse: weighted-broadband ($w_{\mathrm c}=2$) & 100 & 0.636 & 5.904 & 11.031 & 0.272 \\ 
\bottomrule
\end{tabular}
\endgroup
\end{table}

\textit{Comparison and interpretation.} At \(\omega=100\), the central-exact and weighted-broadband layers give \(\mu^*(20)=12.927\) and \(11.031\), respectively, compared with \(9.534\) for the spinor ABC in Table~\ref{tab:SM_S91_controls}. Their relative restricted-mean discrepancies are therefore approximately \(35.6\%\) and \(15.7\%\). Both broadband variants at \(\omega=1\) also improve the restricted-mean agreement relative to central-exact calibration. Broadband matching thus improves this integrated comparison, but does not establish equivalence of the full detection-time laws.

\begin{figure}[H]
  \centering
  \includegraphics[width=0.70\linewidth]
  {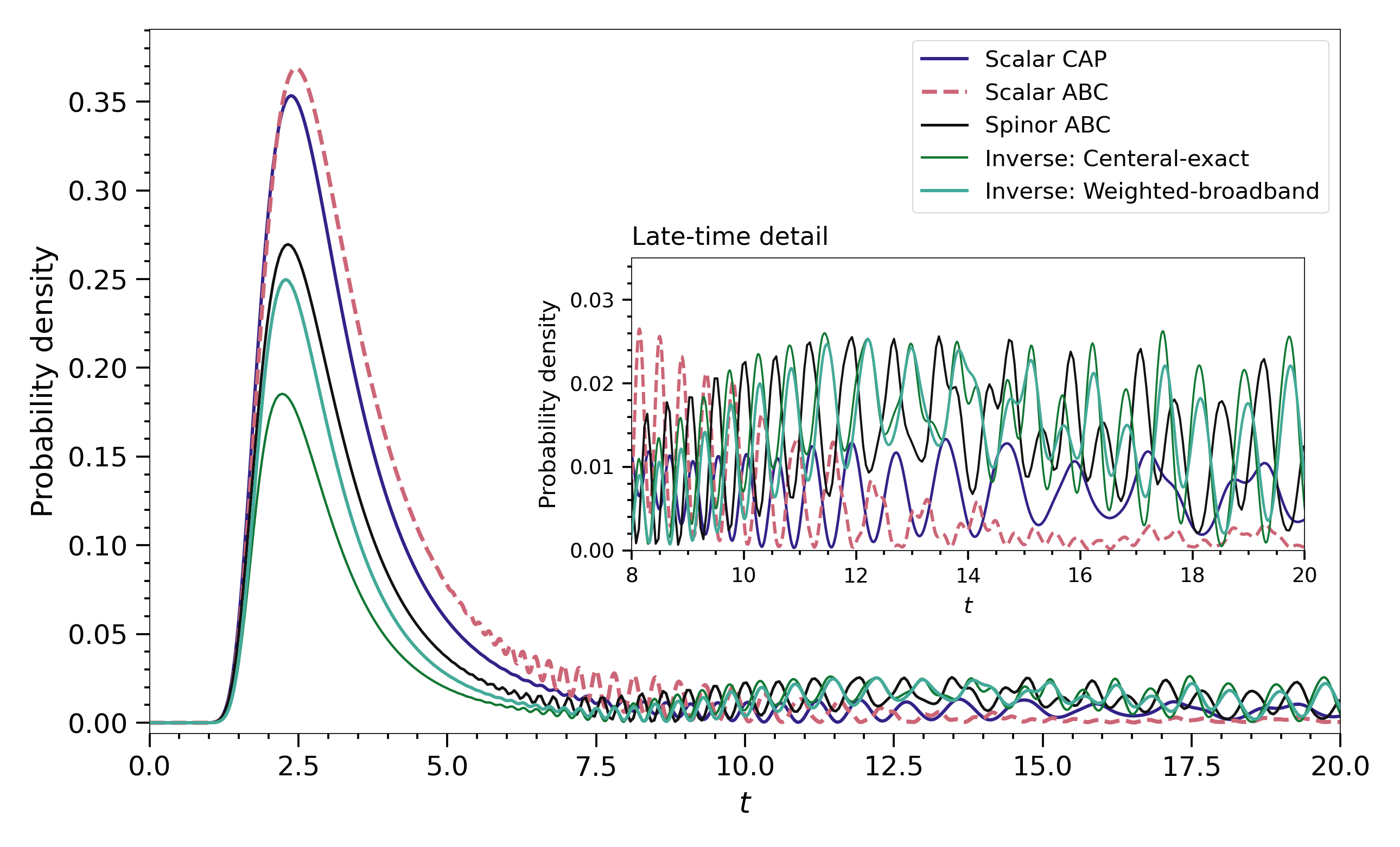}
\caption{Detection-time densities at $\omega=100$ over $0\leq t\leq20$
for the spinor ABC, CAP, scalar ABC, central-exact
inverse layer, and weighted-broadband inverse layer ($w_c=2$),
using common time bins of width $0.04$. The inset resolves
$8\leq t\leq20$.  No time shift, amplitude
rescaling, or conditioning on detection is applied.}
\label{fig:S82_layer_early}
\end{figure}

At \(\omega=100\), the broadband layer has a smaller detected fraction than the centered ABC reference (\(0.636\) versus \(0.724\)) and a larger conditional mean (\(5.904\) versus \(5.548\)). Both differences increase its restricted mean through Eq.~\eqref{eqn:balance}. Figure~\ref{fig:S82_layer_early} shows the time profiles for these model classes using common bins of width \(0.04\). The full profiles distinguish temporal differences that are not specified by the restricted mean alone.

For the uniform inverse layer,
\begin{equation}
J_{\mathrm{entrance}}
=g_{\mathrm{layer}}+\partial_tP_{\mathrm{layer}},
\end{equation}
where \(P_{\mathrm{layer}}\) is the probability stored in \(20\leq z<22\). The entrance current and detection rate therefore differ while the stored probability changes. The complex entrance impedance is a useful calibration target, but matching the auxiliary stationary response does not reproduce the full confined-wavepacket signal by itself.

\section{Source, detector, and guide-length dependence}
\label{sec:SM_parameter_scans}

We examine the dependence of the detector response on $\kappa$, $k_z$, $\sigma_\parallel$, and the guide geometry. Fixed-confinement scans characterize changes in the time profiles; separate confinement sweeps test the dependence beyond the reference preparation. We use the detection density and finite-window statistics defined in Secs.~\ref{sec:SM_numerics}–\ref{sec:SM_finite_window}. The parameter comparisons use $T=20$, while the guide-length comparisons and fixed-source lower-wall controls use $T=40$; these calculations are distinct from the main-text $T=40$ reference sweep.

Where indicated, the confinement dependence is summarized by
\begin{equation}
\mu^*(T;\omega)\simeq A_T+B_T\sqrt{\omega},
\label{eqn:mu_T}
\end{equation}
with root-mean-square residual
\begin{equation}
r_{\mathrm{rms}}=
\left[\frac1n\sum_{i=1}^{n}
\left(\mu^*(T;\omega_i)-A_T-B_T\sqrt{\omega_i}\right)^2\right]^{1/2}.
\label{eq:s9draft:rms}
\end{equation}
The coefficients depend on the preparation, geometry, horizon, and fitted confinement set. These fits summarize a finite range of the bounded statistic $\mu^*(T)\le T$; $r_{\mathrm{rms}}$ measures fit agreement separately from numerical discretization accuracy.

The fixed-confinement profiles at $\omega=1$ in
Figs.~\ref{fig:SM_kappa_kz_confinement},~\ref{fig:SM_kz_densities}, and ~\ref{fig:SM_sigmaz_density} retain the earlier one-sided normal-boundary closure. At the checked parameter values, comparisons with the centered
implementation showed no appreciable change in the qualitative
trends. The confinement sweeps in Figs.~\ref{fig:SM_kappa_confinement},~\ref{fig:SM_kz_fixedpi}, and~\ref{fig:width-confinement} use the centered implementation described in Sec.~\ref{sec:SM_numerics}.

\subsection{Detector parameter $\kappa$}

\begin{figure}[H]
  \centering
  \begin{minipage}[t]{0.48\linewidth}
    \centering
    \includegraphics[width=\linewidth]{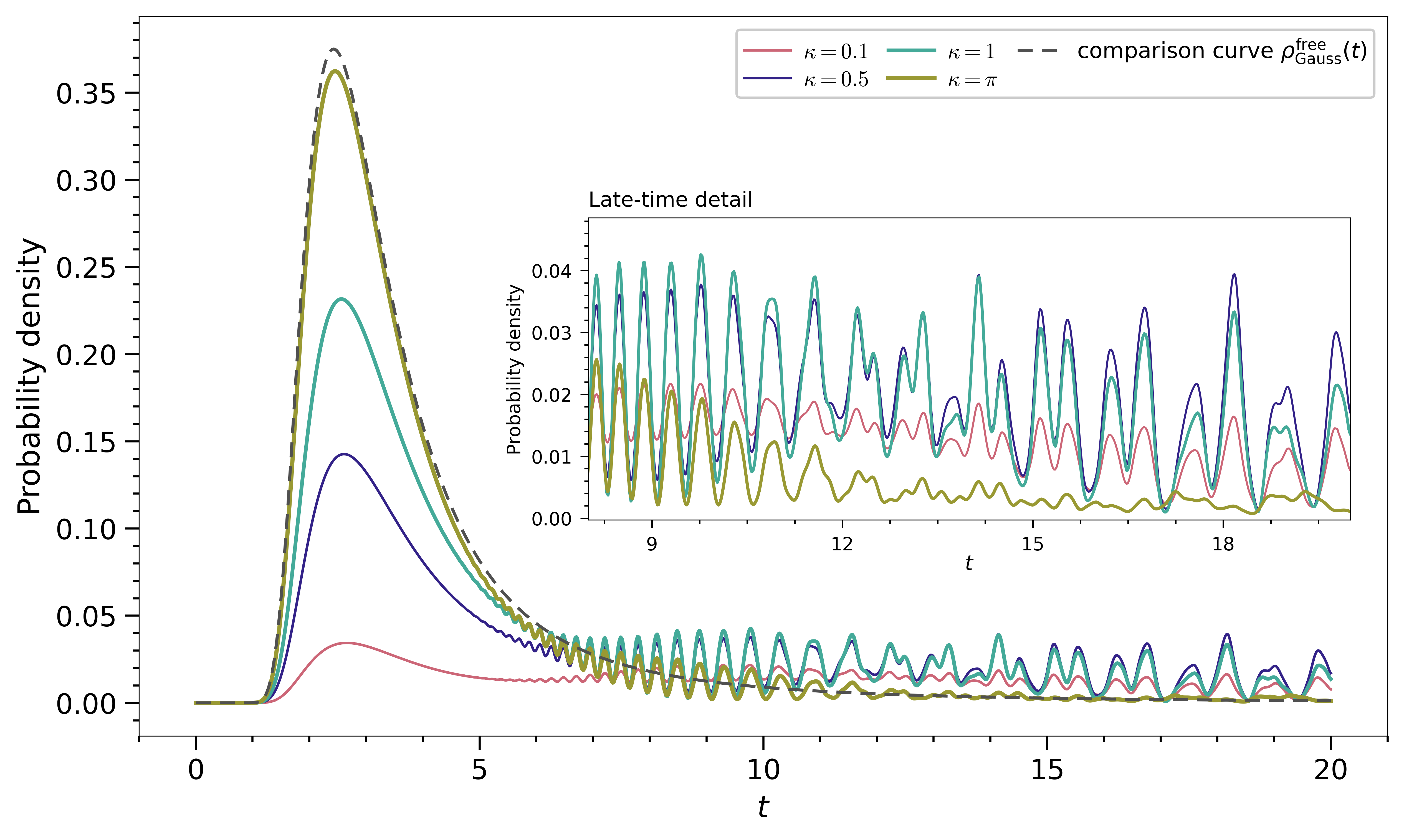}
    \par\vspace{0.3em}
    {\small (a)}
  \end{minipage}
  \hfill
  \begin{minipage}[t]{0.48\linewidth}
    \centering
    \includegraphics[width=\linewidth]{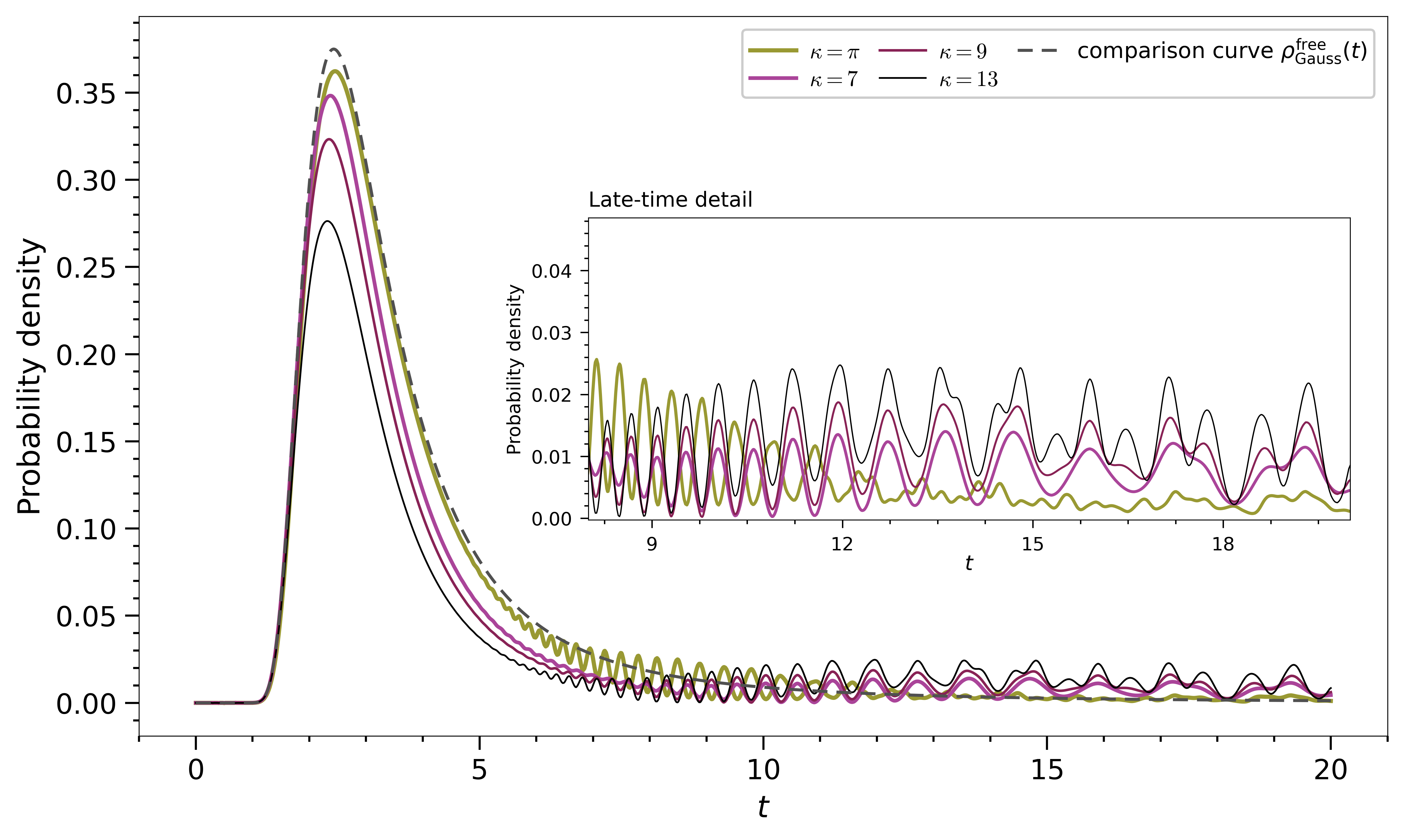}
    \par\vspace{0.3em}
    {\small (b)}
  \end{minipage}
  \caption{Detector-time profiles over $0\le t\le20$ for the detector parameters $\kappa$ indicated in the legends, at $\omega=1$, $k_z=\pi$, and $\sigma_\parallel=0.5$. The gray dashed curves show the detector-free Gaussian comparison current for the fixed longitudinal preparation. The insets magnify the late-time signal.}
  \label{fig:SM_kappa_kz_confinement}
\end{figure}

At $\omega=1$, we vary $\kappa$ with $k_z=\pi$ and $\sigma_\parallel=0.5$. Figure~\ref{fig:SM_kappa_kz_confinement} shows the time profiles, and Table~\ref{tab:SM_kappa_scan} reports the restricted means at $T=20$. Among the tabulated values, $\mu^*(20)$ is smallest at $\kappa=\pi$, the scalar matching point for the central longitudinal component [main-text Sec.~4(iii)].

\begin{table}[H]
\centering
\begingroup
\small

\setlength{\tabcolsep}{8pt}
\renewcommand{\arraystretch}{0.75}

\setlength{\heavyrulewidth}{0.10em}
\setlength{\lightrulewidth}{0.045em}

\caption{Finite-window statistic \(\mu^*(T;\kappa)\) for the \(\kappa\)-scan at
fixed \(\omega=1\), \(k_z=\pi\), and \(\sigma_\parallel=0.5\) over $0\le t\le20$.}
\label{tab:SM_kappa_scan}

\begin{tabular}{@{}cccccccccc@{}}
\toprule
\(\kappa\)
& 0.1
& 0.5
& 1
& 2
& \(\pi\)
& 5
& 7
& 9
& 13 \\
\midrule
\(\mu^*(20;\kappa)\)
& 17.016
& 11.709
& 8.247
& 5.653
& 5.151
& 5.729
& 6.717
& 7.687
& 9.343 \\
\bottomrule
\end{tabular}

\endgroup
\end{table}

\begin{figure}[H]
  \centering
  \includegraphics[width=0.65\linewidth]{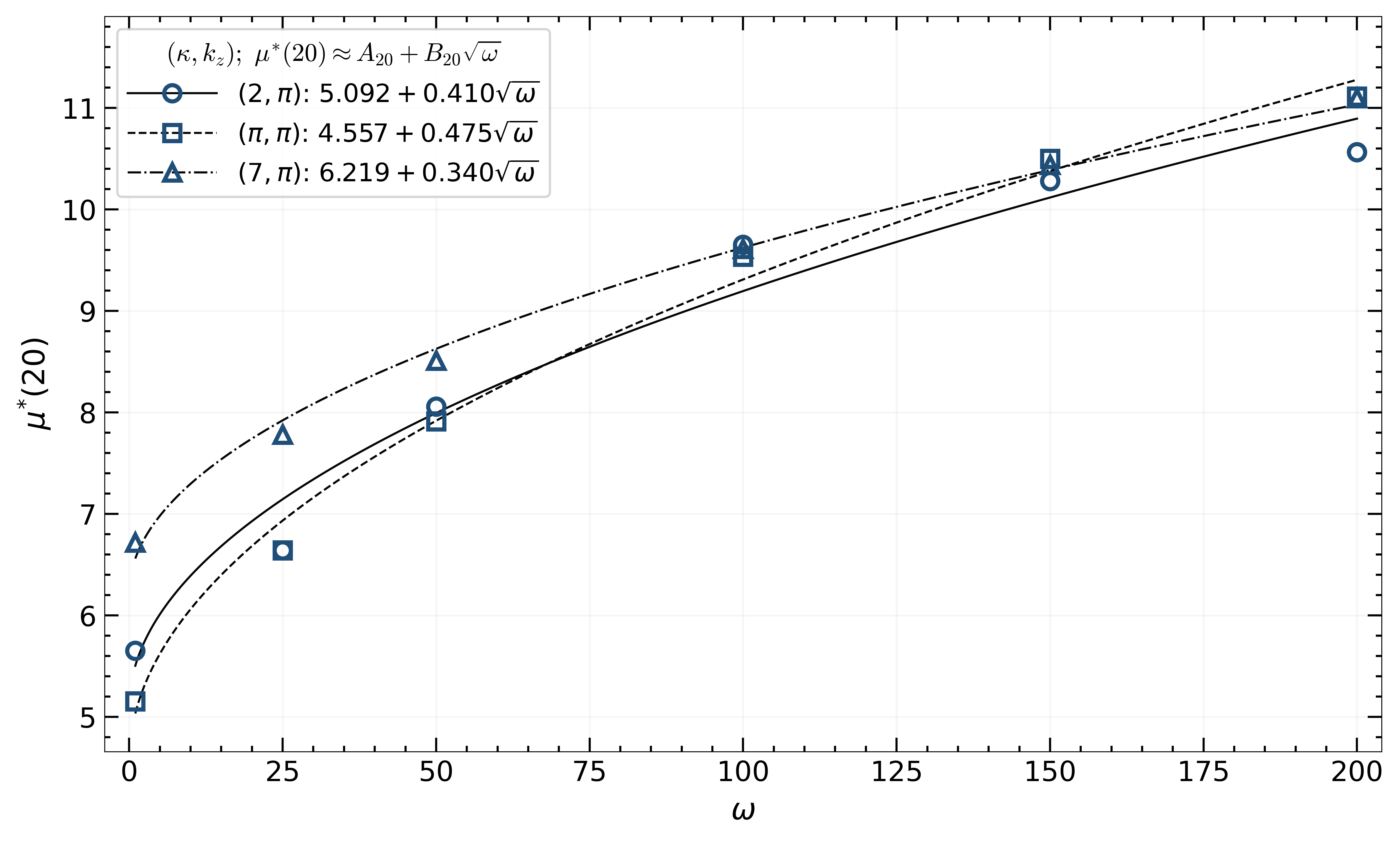}
 \caption{Restricted means at $T=20$ for $\kappa=2,\pi,7$, with $k_z=\pi$ and $\sigma_\parallel=0.5$. Markers show the calculated values at $\omega=1,25,50,100,150,200$; lines show the empirical fits written in the legend, whose parameter pairs are ordered as $(\kappa,k_z)$.}
\label{fig:SM_kappa_confinement}
\end{figure}

The confinement sweeps at $\kappa=2,\pi,7$ all show increasing restricted means, while their ordering changes with confinement (Fig.~\ref{fig:SM_kappa_confinement}). Thus the detector setting giving the smallest restricted mean at $\omega=1$ need not retain that ordering throughout the sweep.

\subsection{Longitudinal wave number $k_z$}
\label{subsec:SM_kz_scan}

\begin{figure}[H]
  \centering
  \includegraphics[width=0.70\linewidth]{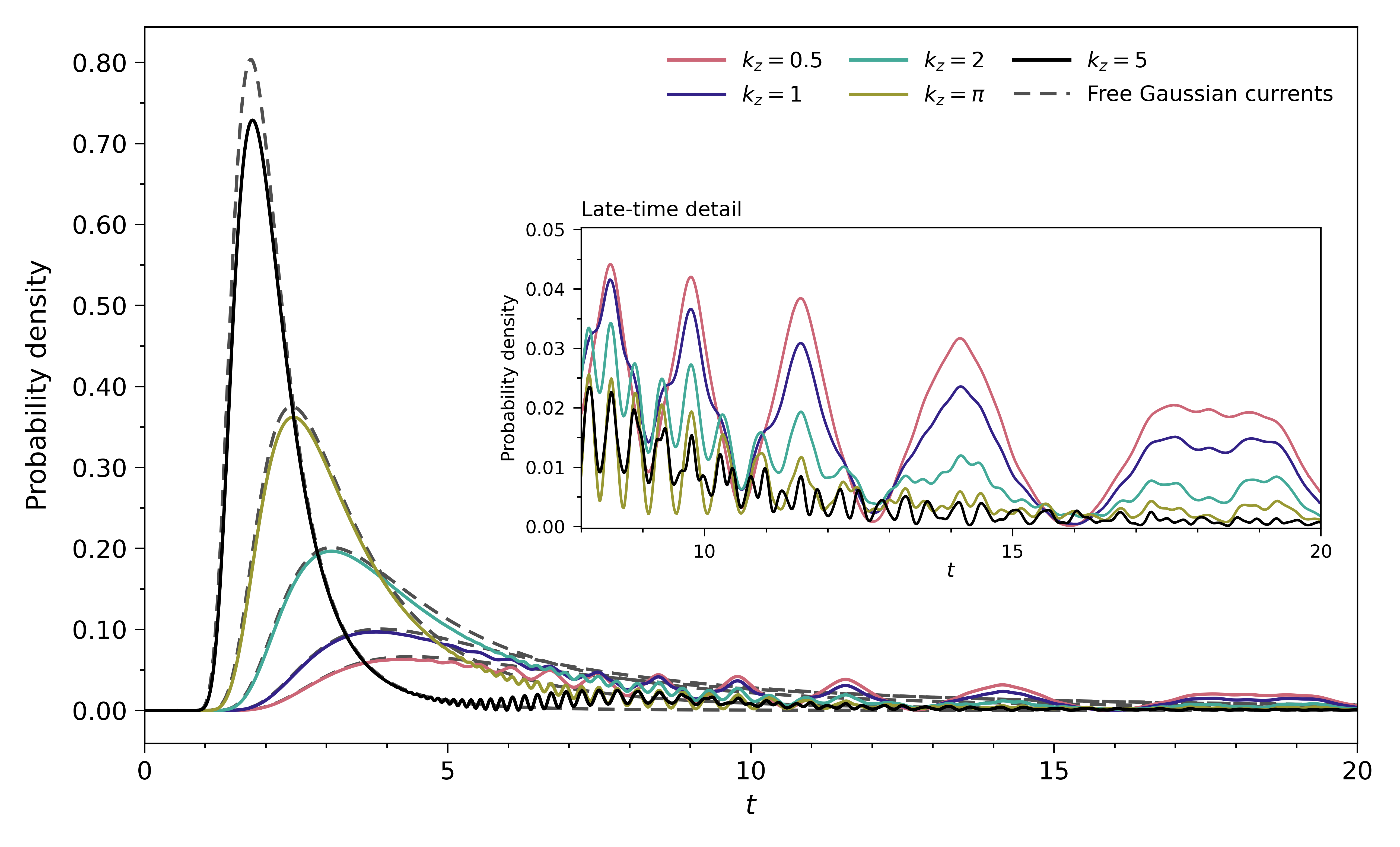}
  \caption{Detector-time profiles over $0\le t\le20$ for $k_z=0.5,1,2,\pi,5$ at $\omega=1$, $\kappa=\pi$, and $\sigma_\parallel=0.5$. The gray dashed curves are the detector-free Gaussian comparison currents for the corresponding preparations. The inset magnifies the late-time signal.}
\label{fig:SM_kz_densities}
\label{fig:SM_kz_confinement}
\end{figure}

\begin{figure}[H]
  \centering
 \includegraphics[width=0.65\linewidth]
{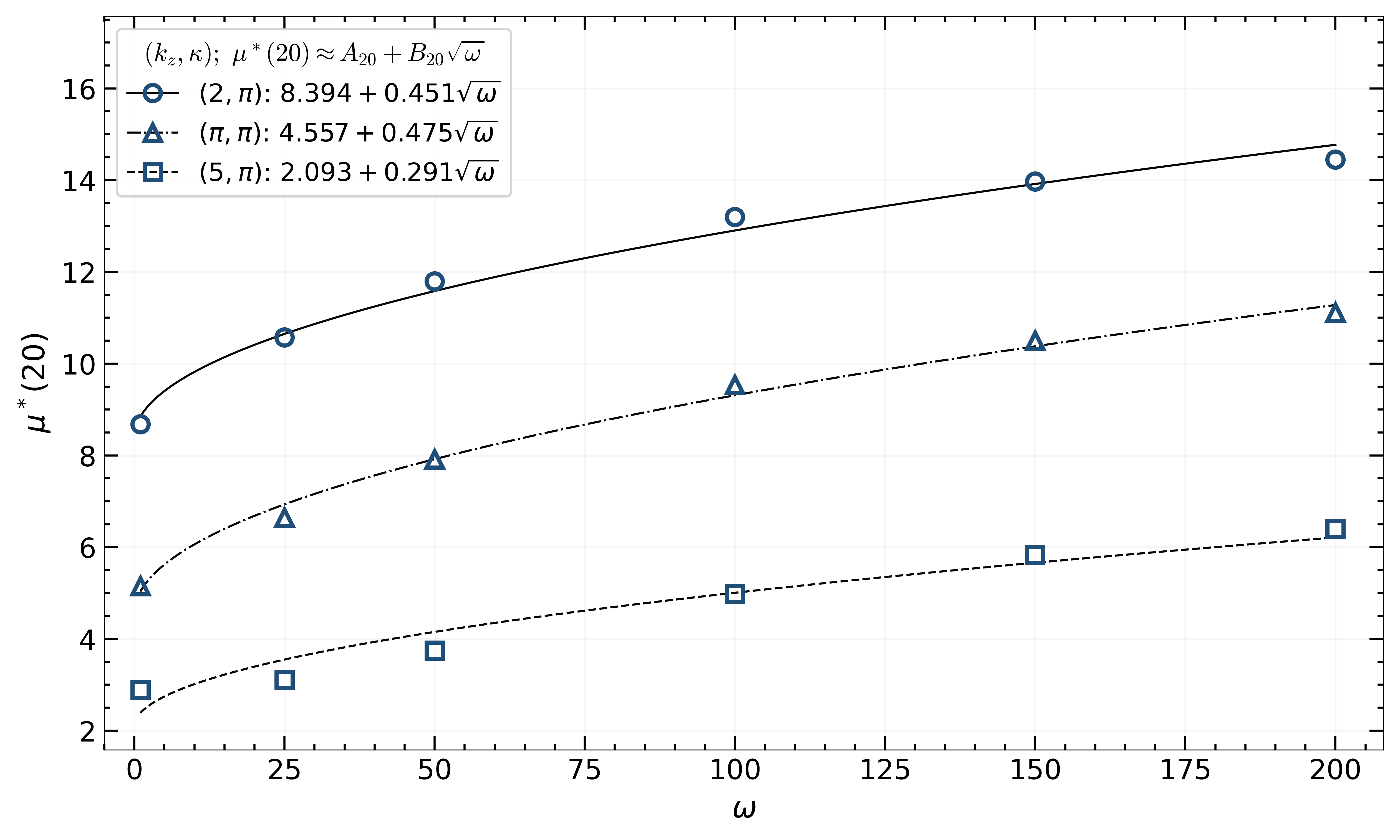}
\caption{
Restricted means at $T=20$ for $k_z=2,\pi,5$, with $\kappa=\pi$ and $\sigma_\parallel=0.5$. Markers show the calculated values at $\omega=1,25,50,100,150,200$; lines show the empirical fits written in the legend, whose parameter pairs are ordered as $(k_z,\kappa)$ in Table~\ref{tab:SM_kz_fits}.}
\label{fig:SM_kz_fixedpi}
\end{figure}

At $\omega=1$, $\kappa=\pi$, and $\sigma_\parallel=0.5$, Fig.~\ref{fig:SM_kz_confinement} compares detector-time profiles for $k_z=0.5,1,2,\pi,5$ over $0\le t\le20$.

\begin{table}[H]
\centering
\begingroup
\small

\setlength{\tabcolsep}{5pt}
\renewcommand{\arraystretch}{0.65}

\setlength{\heavyrulewidth}{0.10em}
\setlength{\lightrulewidth}{0.045em}

\setlength{\belowcaptionskip}{5pt}

\caption{
Unweighted fits of \(\mu^*(20;\omega)\) to
\(A_{20}+B_{20}\sqrt{\omega}\).
The three rows correspond to the centered spinor ABC
at fixed \(\kappa=\pi\) in Fig.~\ref{fig:SM_kz_fixedpi},
using \(\omega=1,25,50,100,150,200\).
The residual \(r_{\rm rms}\) is the root mean square
of the six fit residuals.}
\label{tab:SM_kz_fits}

\begin{tabular}{@{}c@{\hspace{1.2cm}}rrr@{}}
\toprule
$(k_z,\kappa)$
& $A_{20}$
& $B_{20}$
& $r_{\rm rms}$ \\
\midrule

$(2,\pi)$
& 8.394
& 0.451
& 0.212 \\

$(\pi,\pi)$
& 4.557
& 0.475
& 0.179 \\

$(5,\pi)$
& 2.093
& 0.291
& 0.335 \\

\bottomrule
\end{tabular}

\par\smallskip
\endgroup
\end{table}

\subsection{Longitudinal packet width $\sigma_\parallel$}
\label{subsec:SM_sigmaz_scan}

At $\omega=1$ and $k_z=\kappa=\pi$, Fig.~\ref{fig:SM_sigmaz_density} compares $\sigma_\parallel=0.1,0.25,0.5,0.75,1$ over $0\le t\le20$. Changing this Gaussian amplitude width changes both spatial localization and the longitudinal momentum spread, $\Delta k_z=1/(\sqrt{2}\,\sigma_\parallel)$. The narrower preparations show more pronounced oscillatory structure in these profiles. The restricted mean increases with confinement for all three widths
tested in Fig.~\ref{fig:width-confinement}, with width-dependent
empirical fit coefficients.

\begingroup
\setlength{\intextsep}{6pt plus 2pt minus 2pt}
\begin{figure}[H]
  \centering
  \includegraphics[width=0.70\linewidth]{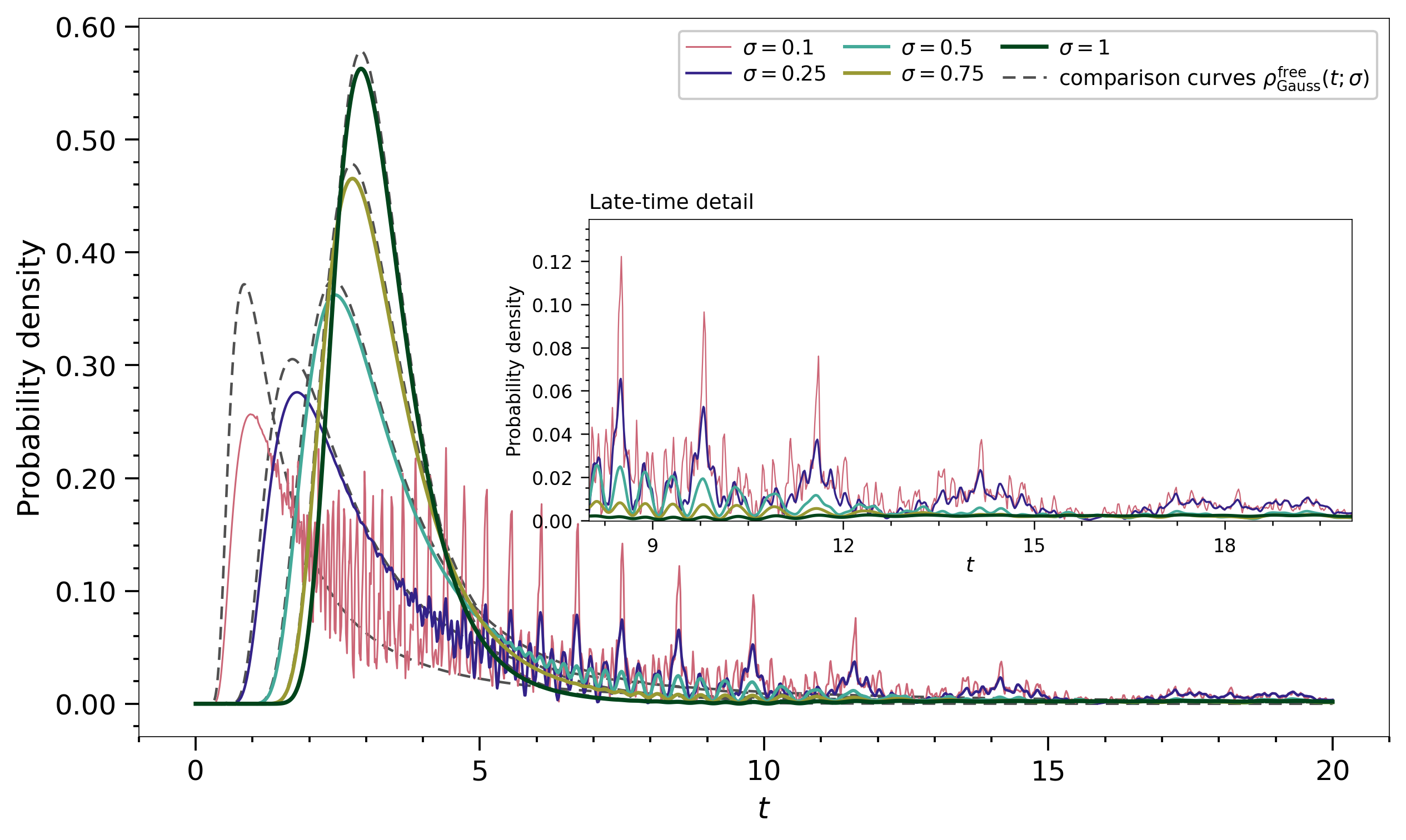}
  \caption{Detector-time profiles over $0\le t\le20$ for $\sigma_\parallel=0.1,0.25,0.5,0.75,1$ at $\omega=1$ and $k_z=\kappa=\pi$. The widths refer to the Gaussian amplitude. Gray dashed curves show the detector-free Gaussian comparison currents for the respective preparations; the inset magnifies the late-time signal.}
\label{fig:SM_sigmaz_density}
\end{figure}
\endgroup

\begin{figure}[H]
  \centering
  \includegraphics[width=0.65\linewidth]
  {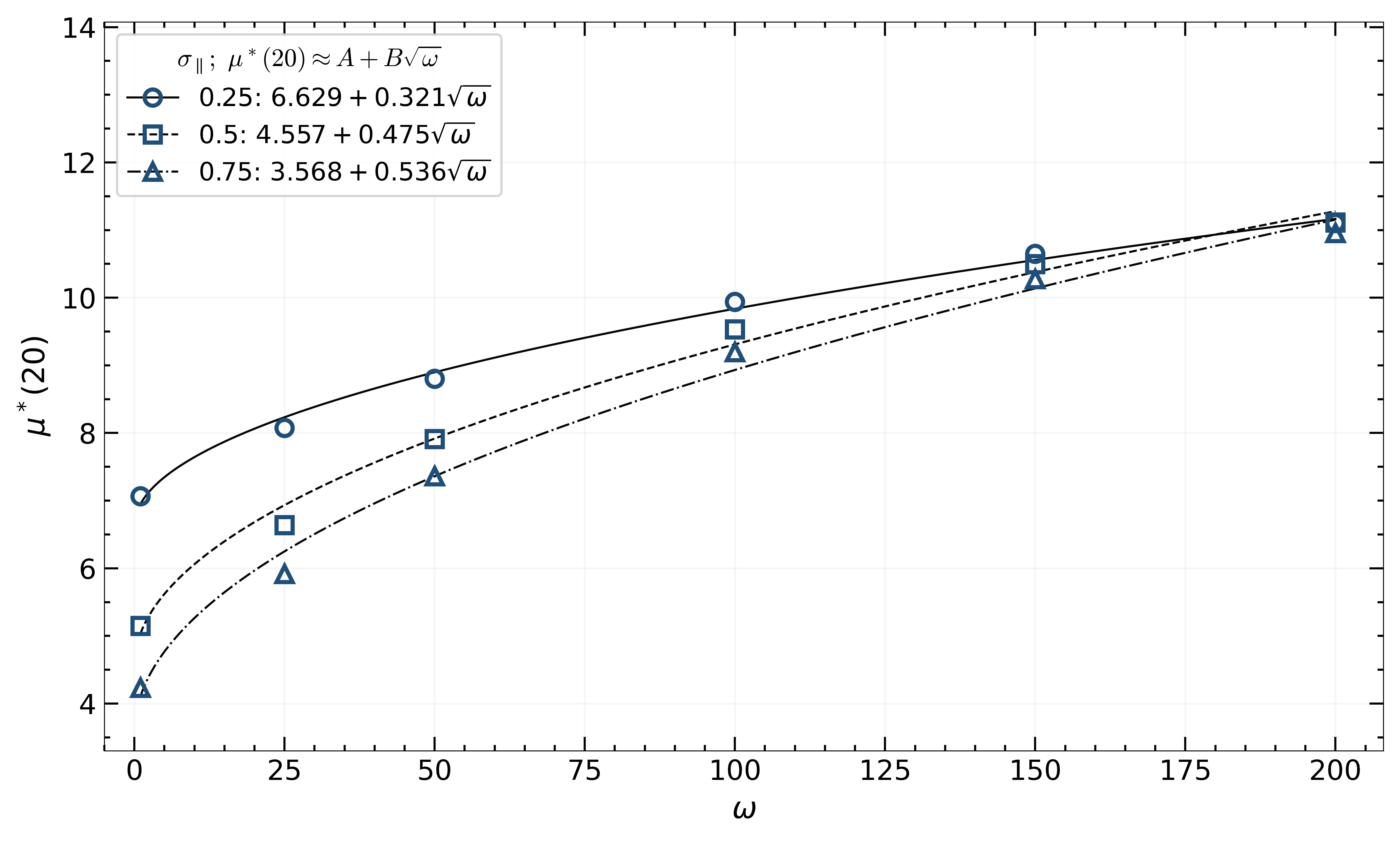}
  \caption{Restricted mean detection times $\mu^*(20;\omega)$ for
$\sigma_\parallel=0.25,0.5,0.75$, with $L=20$ and
$k_z=\kappa=\pi$. Markers show the calculated values at
$\omega=1,25,50,100,150,200$ for all three widths; lines show the empirical fits
$A_{20}+B_{20}\sqrt{\omega}$ given in the legend.}
\label{fig:width-confinement}
\end{figure}

\subsection{Guide geometry and lower-wall dependence}
\label{subsec:SM_length_scan}

\begin{figure}[H]
  \centering
  \includegraphics[width=0.70\linewidth]{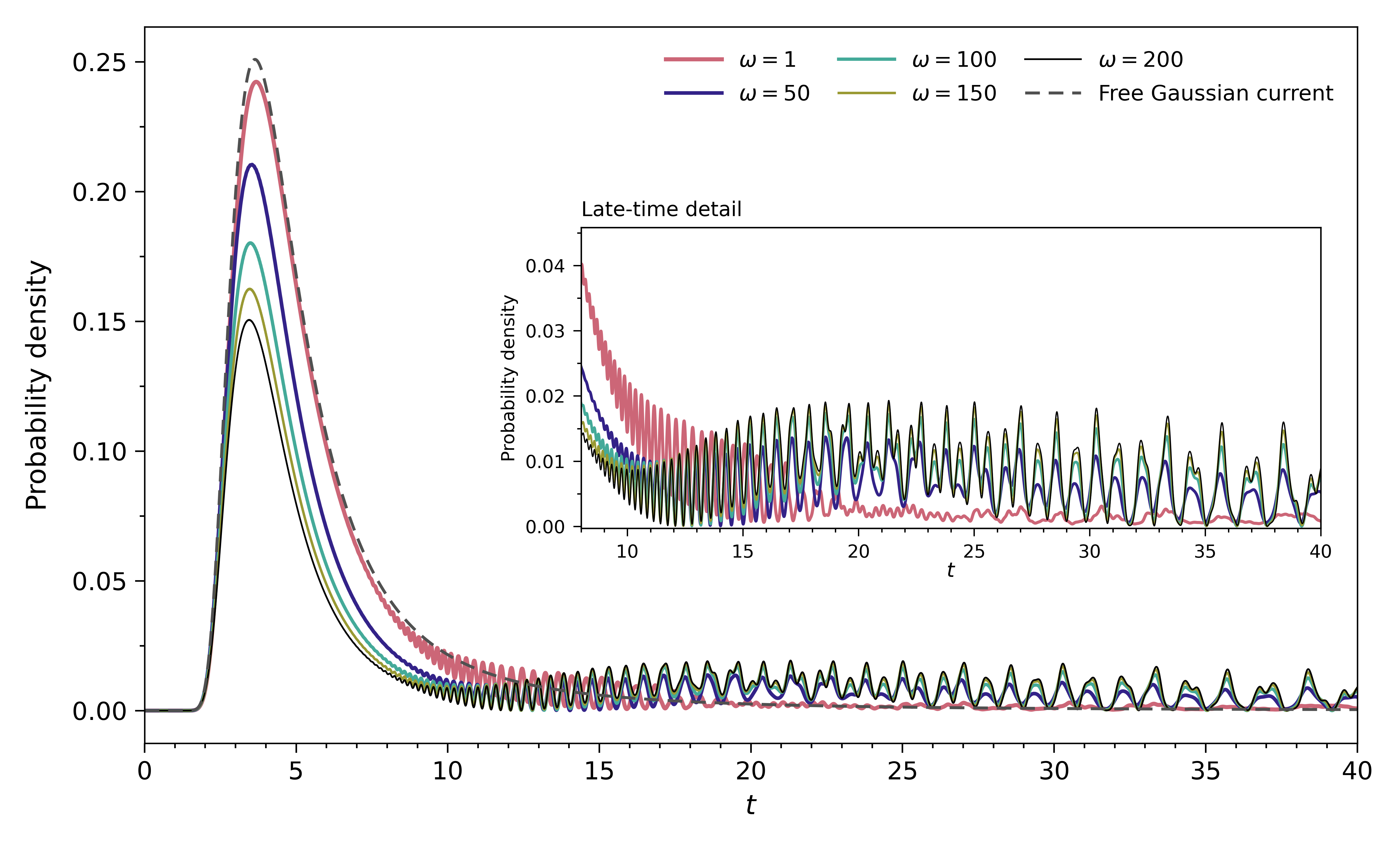}
  \caption{
Detector-time profiles for the nominal $L=30$ guide,
with $z_c=15$, $k_z=\kappa=\pi$, and
$\sigma_\parallel=0.5$, over $0\leq t\leq40$.
The displayed confinements are $\omega=1,50,100,150,200$.
Rates are reconstructed by numerical differentiation of the
raw saved squared norms.
The gray dashed curve is the detector-free Gaussian comparison
current; the inset magnifies the late-time signal.
}
\label{fig:SM_length_density_comparison}
\end{figure}

\paragraph{Midpoint-source comparison.} 
Figure~\ref{fig:SM_length_density_comparison} shows representative
detector-time profiles for the nominal $L=30$ guide, with
$z_c=L/2=15$, over $0\le t\le40$.

\begin{figure}[H]
  \centering
  \includegraphics[width=0.65\linewidth]
  {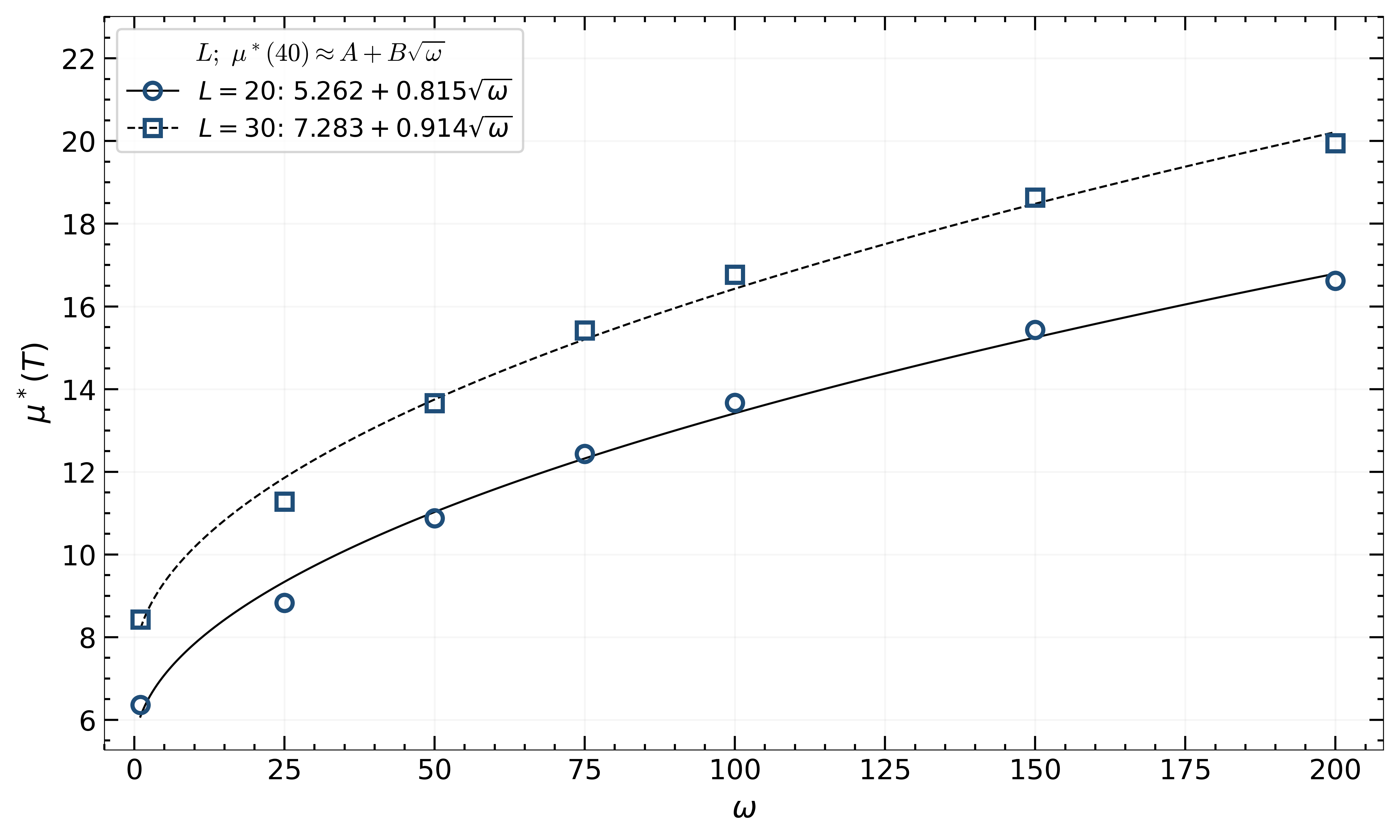}
  \caption{Restricted means for nominal guide lengths \(L=20\) and \(30\) at the common horizon \(T=40\), with displayed markers at \(\omega=1,25,50, 75, 100,150,200\) for both series. Markers are calculated values and lines are unweighted fits to \(A_{40}+B_{40}\sqrt{\omega}\). The source center is \(z_c=L/2\).}
  \label{fig:SM_length_common_T20}
\end{figure}

Figure~\ref{fig:SM_length_common_T20} compares the restricted
means for nominal guide lengths $L=20$ and $30$ at the common
horizon $T=40$, with the source at $z_c=L/2$ in each guide.

Therefore the reported fits for $L=20$ and $30$ over
$\omega\in\{1,25,50,75, 100,150,200\}$ are
\begin{equation}
\mu^*(40;\omega)\simeq
\begin{cases}
5.262+0.815\sqrt{\omega}, & L=20,\quad r_{\mathrm{rms}}=0.27,\\
7.283+0.914\sqrt{\omega}, & L=30,\quad r_{\mathrm{rms}}=0.30.
\end{cases}
\end{equation}

\paragraph{Fixed-source lower-wall control.} We vary the reflecting lower-wall position at $\omega=200$, keeping the source at $z_c=10$ and the spinor-ABC detector at $z_D\simeq19.986$, and evolve to $T=40$. The preparation is $k_z=\kappa=\pi$, $\sigma_\parallel=0.5$, and $\eta=(1,0)^{\mathsf T}$. Moving the wall to $z_B=0,-5,-10,-15$ increases $N_z$ while preserving $h_z\simeq0.013$, the transverse grid, and the original nonnegative-$z$ nodes (Table~\ref{tab:SM_R4_fixed_source_wall}). All four displaced-wall controls use $\Delta t=3\times10^{-4}$.

\begin{table}[H]
\centering
\caption{Fixed-source lower-wall controls at \(\omega=200\) and \(T=40\). $D_{40}$ and conditional means use accepted-step upper-boundary events; $\mu^*(40)$ uses piecewise-linear normalized survival. The finite-window identity in Eq.~\eqref{eqn:balance} is checked independently. Here \(D_{L,40}=S(3.6)-S(40)\) and \(t_{L,40}\) is the mean conditioned on detection in \((3.6,40]\). Values are rounded independently.}
\label{tab:SM_R4_fixed_source_wall}

\begingroup
\small


\setlength{\tabcolsep}{5pt}

\renewcommand{\arraystretch}{0.70}

\setlength{\heavyrulewidth}{0.12em}
\setlength{\lightrulewidth}{0.05em}


\begin{tabular*}{0.69\linewidth}
{@{\extracolsep{\fill}}rrrrrrr@{}}

\toprule

$z_B$
& $N_z$
& $D_{40}$
& $\bar\tau^{\rm cond}_{40}$
& $\mu^*(40)$
& $D_{L,40}$
& $t_{L,40}$ \\

\midrule

0
& 1500
& 0.784
& 10.179
& 16.620
& 0.438
& 16.234 \\

$-5$
& 1875
& 0.750
& 10.431
& 17.833
& 0.404
& 17.216 \\

$-10$
& 2250
& 0.716
& 10.382
& 18.795
& 0.370
& 17.740 \\

$-15$
& 2625
& 0.687
& 10.263
& 19.571
& 0.341
& 18.125 \\

\bottomrule

\end{tabular*}

\endgroup
\end{table}

All four runs give $D_E\simeq0.346$ at displayed precision. Moving the wall from $z_B=0$ to $-15$ reduces $D_{40}$ from $0.784$ to $0.687$ and $D_{L,40}$ from $0.438$ to $0.341$, while increasing the late conditional mean $t_{L,40}$ from $16.234$ to $18.125$. The overall detected conditional mean is nonmonotonic across the four positions. In the finite-window identity, the larger survival contribution $40S(40)$ more than compensates for the smaller detected-event contribution between these endpoints, raising the restricted mean from $16.620$ to $19.571$.

\begin{figure}[H]
  \centering
  \includegraphics[width=0.9\linewidth]{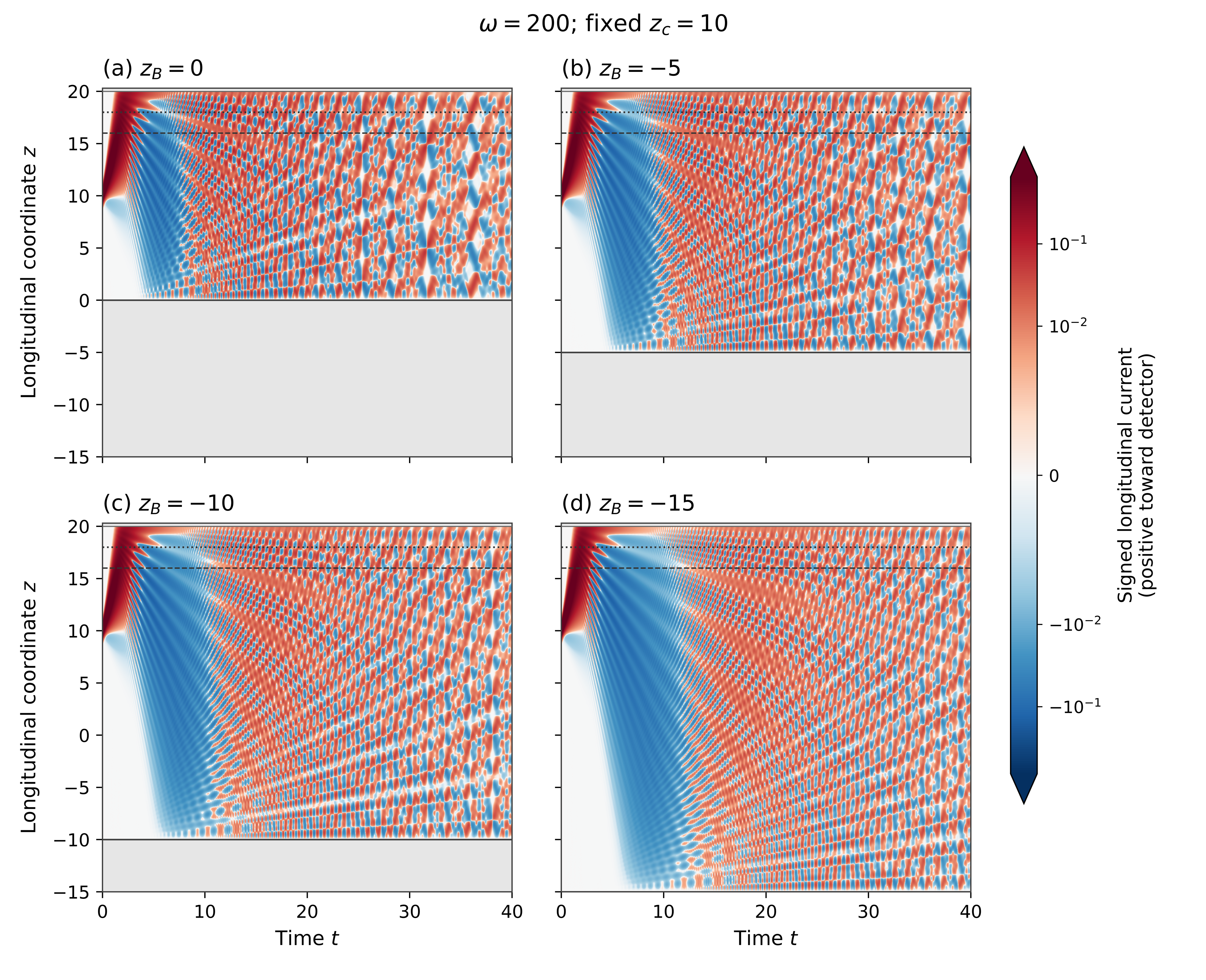}
 \caption{Signed longitudinal-current maps at $\omega=200$, over $0\le t\le40$, with fixed $z_c=10$ and $z_D\simeq19.9867$. Panels (a)--(d): $z_B=0,-5,-10,-15$. Red/blue indicate net current toward/away from the detector; gray lies outside the guide. Dashed/dotted lines mark the monitors at $z\simeq16.007,18.007$.}
\label{fig:sign}
\end{figure}

Figure~\ref{fig:sign} shows the corresponding signed longitudinal currents. These are finite-difference link currents, summed over the transverse grid and spin components and normalized by the initial norm, with positive direction toward the detector. The maps show net transport of undetected probability away from the upper region after the prompt signal. Their later evolution depends on the wall position.

The wall-displacement comparisons show lower-wall sensitivity of the late finite-window signal at the stated numerical settings, without identifying individual oscillations with particular round trips.

%
\subsection{Absorbing lower-boundary control}
\label{sec:SM_lower_absorber_control}

We replace the lower Dirichlet wall by the componentwise scalar ABC
\begin{equation}
 \partial_z\Psi(x,y,0,t)
 =-i\kappa_{\mathrm{lower}}\Psi(x,y,0,t),
 \qquad \kappa_{\mathrm{lower}}=\pi.
 \label{eq:SM_lower_scalar_abc}
\end{equation}
The upper spin-coupled detector, bulk Hamiltonian, preparation,
boundary positions, lateral walls, and spatial grid are retained.
The controls use $\omega=1,200$, $T=40$, and
$\Delta t=10^{-3},\,2.5\times10^{-4}$, respectively.
Both absorbing endpoints use centered ghost-point elimination
and half longitudinal quadrature weight.

For incident normal wave number $k>0$, the lower reflection
probability is
\begin{equation}
 \mathcal R_{\mathrm{lower}}(k)
 =\left(
   \frac{k-\kappa_{\mathrm{lower}}}
        {k+\kappa_{\mathrm{lower}}}
  \right)^2.
 \label{eq:SM_lower_reflectivity}
\end{equation}
For the ideal centered longitudinal stencil, $k$ is replaced by
$\sin(kh_z)/h_z$. The termination is therefore partially absorbing,
not transparent to the complete wavepacket.

Upper-detection and lower-escape probabilities are accumulated
separately from accepted-step events and satisfy
\begin{equation}
 S_{\mathrm{in}}(t)+D_{\mathrm{top}}(t)
 +E_{\mathrm{lower}}(t)=1.
 \label{eq:SM_lower_two_channel_balance}
\end{equation}
Here $S_{\mathrm{in}}(t)$ is the probability that neither upper
detection nor lower escape has occurred by time $t$, whereas the
probability of no upper click is
$1-D_{\mathrm{top}}(t)
=S_{\mathrm{in}}(t)+E_{\mathrm{lower}}(t)$.  The maximum recorded balance defect is below $4\times10^{-7}$ in
both runs; this checks accounting, not convergence.
Total norm loss includes escape and is not the upper-click probability.

\begin{figure}[H]
  \centering
  \includegraphics[width=0.70\linewidth]{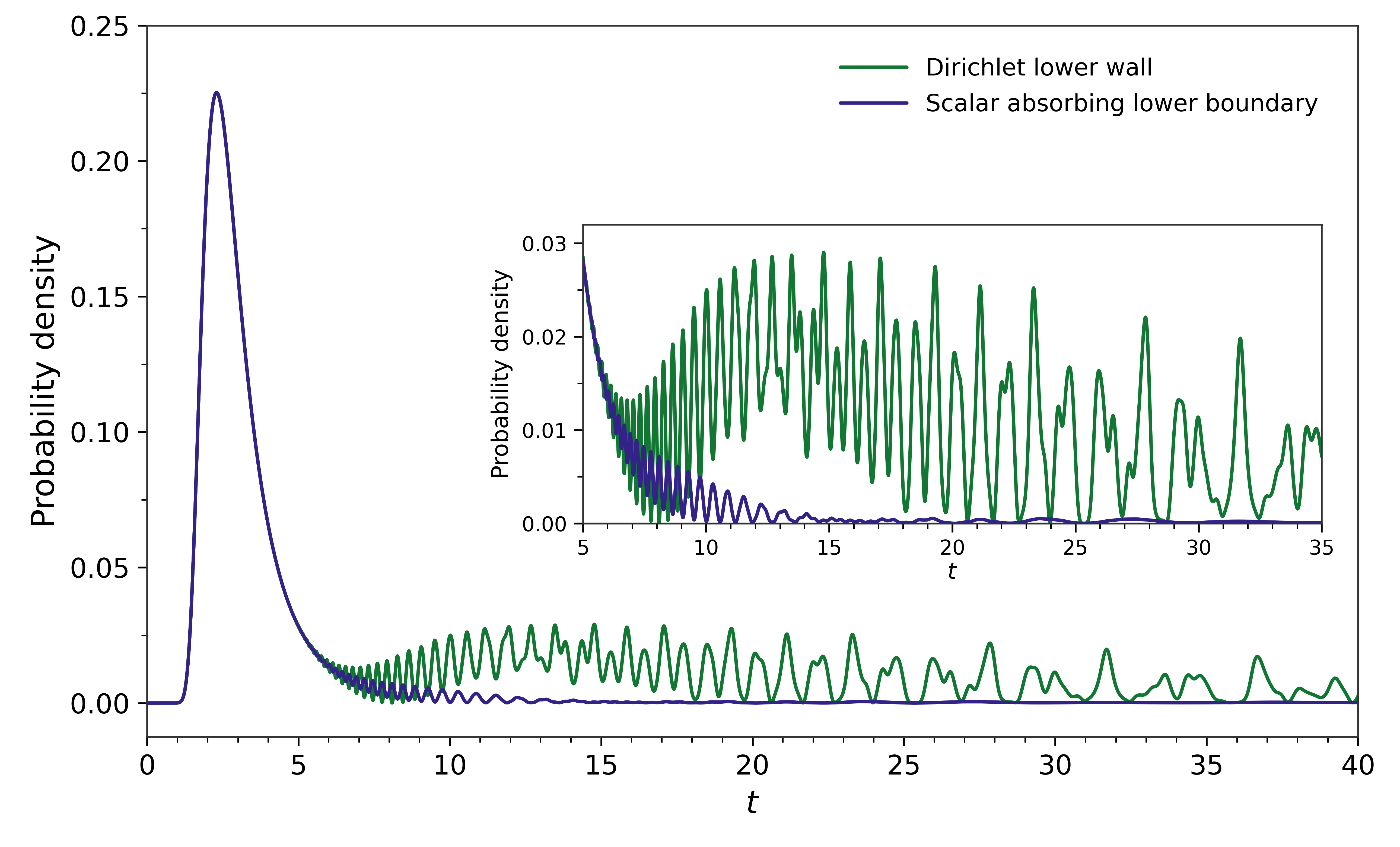}
\caption{Upper-detector click-time densities at $\omega=200$
for a Dirichlet lower wall and a scalar absorbing lower boundary
with $\kappa_{\mathrm{lower}}=\pi$. Both retain the spin-coupled
upper detector and the reference preparation of Table~S1.
Curves use accepted-step upper-boundary events without conditioning
on detection; their areas equal $D_{\mathrm{top}}(40)$.
The inset resolves $5\le t\le35$.}
\label{fig:SM_lower_absorber_control}
\label{fig:lowerWall}
\end{figure}

Figure~\ref{fig:lowerWall} and Table~\ref{tab:SM_lower_absorber_control} show that the partially absorbing lower
boundary reduces late upper detections and their conditional mean
time, while the early detected probability and mean time agree with
the Dirichlet references at the reported precision. At $\omega=200$,
$D_L$ decreases from $0.438$ to $0.130$ and $t_L$ from $16.234$
to $6.231$. In this lower absorbing-boundary run, the window-integrated
$+$-branch fraction and detector-associated $z$-polarization remain
approximately $0.938$ and $0.128$, respectively, close to the
Dirichlet values in Table~\ref{tab:SM_branch_spin_rechecked}; the final in-guide $z$-polarization
remains close to unity. Thus the late signal changes substantially
without removing these branch and spin contrasts. With the absorbing lower boundary, $D_{\mathrm{top}}(40)$ decreases
from $0.905$ at $\omega=1$ to $0.476$ at $\omega=200$.
Confinement-dependent suppression of upper detection therefore
persists at these two frequencies without a perfectly reflecting
lower wall. Residual oscillations remain, and the termination is
only partially absorbing; these controls establish lower-boundary
sensitivity but do not identify individual returns.

\begin{table}[H]
\centering
\caption{Lower-boundary controls at $T=40$.
Dirichlet rows reproduce Table~S2, with $D_L=D_{40}w_L$
and $S_{\mathrm{in}}=1-D_{40}$; scalar-ABC rows use
accepted-step boundary events.
Here $D_L=D_{\mathrm{top}}(40)-D_{\mathrm{top}}(3.6)$
and $t_L$ are the probability and conditional mean time
of upper detection in $(3.6,40]$.}
\label{tab:SM_lower_absorber_control}

\begingroup
\small

\setlength{\tabcolsep}{5pt}
\renewcommand{\arraystretch}{0.70}
\setlength{\heavyrulewidth}{0.12em}
\setlength{\lightrulewidth}{0.05em}

\begin{tabular}{@{}rlrrrrrr@{}}

\toprule

$\omega$
& Lower boundary
& $D_{\mathrm{top}}$
& $E_{\mathrm{lower}}$
& $S_{\mathrm{in}}$
& $\bar\tau_{\mathrm{top}}^{\mathrm{cond}}$
& $D_L$
& $t_L$ \\

\midrule

1
& Dirichlet
& 0.952
& 0
& 0.048
& 4.663
& 0.375
& 7.839 \\

1
& Scalar ABC
& 0.905
& 0.062
& 0.033
& 3.865
& 0.329
& 6.092 \\

200
& Dirichlet
& 0.784
& 0
& 0.216
& 10.178
& 0.438
& 16.234 \\

200
& Scalar ABC
& 0.476
& 0.472
& 0.052
& 3.525
& 0.130
& 6.231 \\

\bottomrule

\end{tabular}

\endgroup
\end{table}

\subsection{Long-time evolution at $\omega=200$}
\label{sec:SM_long_time}

\begin{figure}[H]
\centering
\includegraphics[width=0.8\linewidth]
{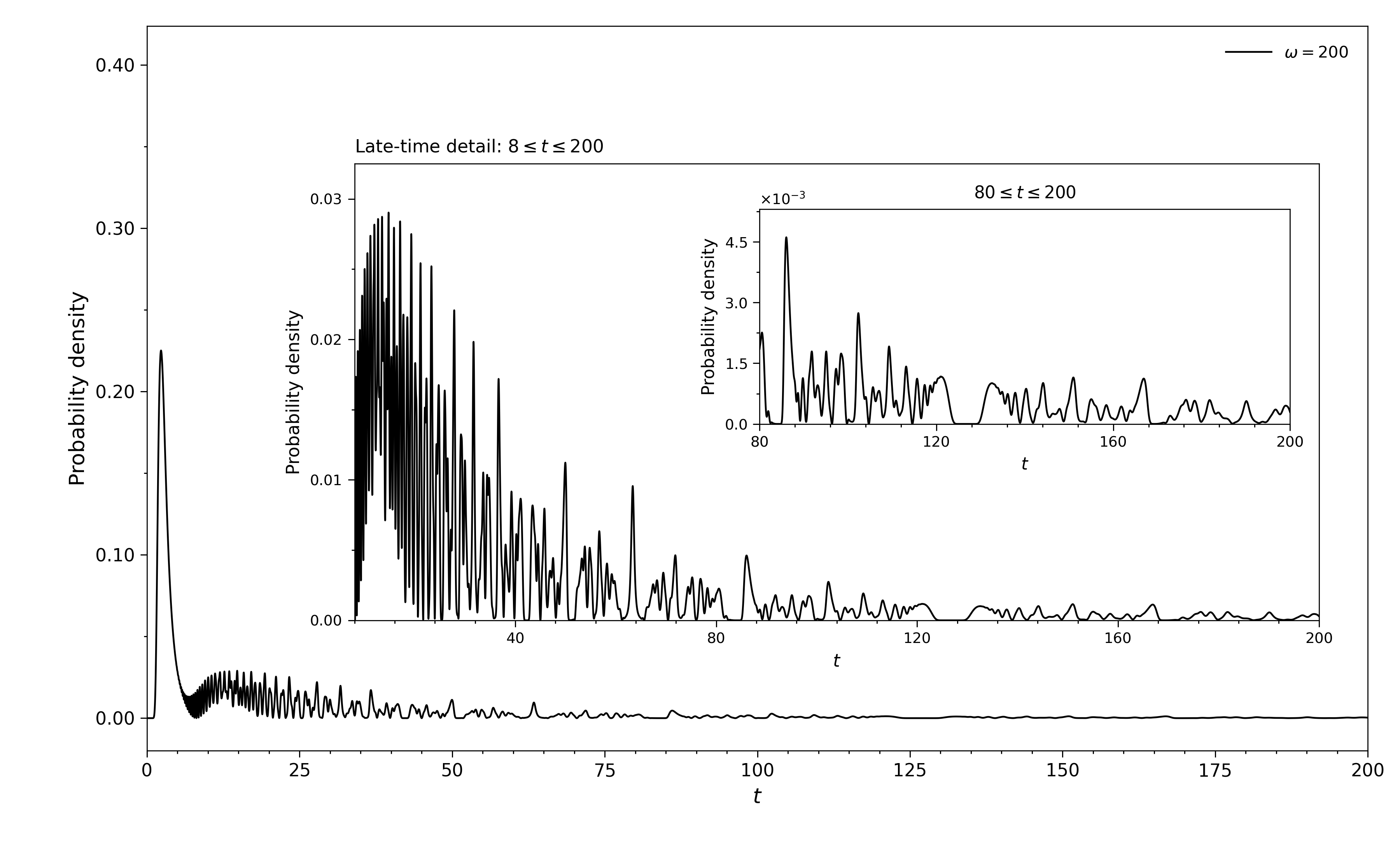}
\caption{
Long-time detection density over $0\leq t\leq200$
at $\omega=200$, using the centered spinor ABC and
a reflecting lower wall. The insets resolve
$8\leq t\leq200$ and $80\leq t\leq200$.}
\label{fig:SM_long_time}
\end{figure}

To examine detection beyond the main observation window, we perform
an additional centered spinor-ABC calculation at $\omega=200$ up to
$T=200$. The physical preparation and spatial grid are those of
Table~\ref{tab:SM_reference_numerics}, including the reflecting lower wall, with time step
$\Delta t=2.5\times10^{-4}$.
Figure~\ref{fig:SM_long_time} shows an oscillatory detector signal
extending well beyond $T=40$, with late structure still visible
in the interval $80\le t\le200$.

The accepted-step detector records give \(D_{40}=0.784\) and \(D_{200}=0.943\), with \(S(200)=0.057\). Thus, approximately \(0.1590\) of the initial probability is detected during \(40<t\le200\), corresponding to \(73.6\%\) of the probability surviving at \(T=40\). The window-averaged detector rate decreases from \(7.13\times10^{-3}\) over \(20\le t\le40\) to \(2.71\times10^{-4}\) over \(160\le t\le200\), although individual peaks remain nonmonotonic. The restricted mean increases from \(\mu^*(40)=16.618\) to \(\mu^*(200)=32.466\), consistent with \(\partial_T\mu^*(T)=S(T)\). These results demonstrate appreciable subsequent
detection and a decreasing finite-window average rate,
but establish neither an asymptotic decay law nor the
unrestricted mean. The time profile does not resolve
individual encounters or establish progressive filtering
over successive reflections, and this single-confinement
extension does not determine the confinement dependence
at longer horizons.

\section{Physical scales}
\label{sec:SM_physical_scales}

The analysis above is dimensionless. Let \(a_0\) be the physical length represented by one dimensionless unit. Then
\begin{equation}
\mathbf{r}_{\mathrm{phys}}=a_0\mathbf{r},\qquad t_{\rm phys}=T_*t,\qquad
T_*=\frac{m_{\rm phys}a_0^2}{\hbar}.
\end{equation}
The transverse trap frequency and longitudinal wave numbers scale as
\begin{equation}
\Omega_\perp=\frac{\omega}{T_*},\qquad
k_{z,{\rm phys}}=\frac{k_z}{a_0},\qquad
\kappa_{\rm phys}=\frac{\kappa}{a_0}.
\end{equation}
The physical transverse oscillator length is
\begin{equation}
\ell_{\perp,{\rm phys}}=\frac{a_0}{\sqrt{\omega}}.
\end{equation}
As an illustrative scale, for single \(^{87}{\rm Rb}\) atoms and \(a_0=2\,\mu{\rm m}\), one obtains
\begin{equation}
T_*\simeq 5.47\,{\rm ms}.
\end{equation}
Thus the observation windows \(T=20\) and \(T=40\) correspond to approximately \(109\,{\rm ms}\) and \(219\,{\rm ms}\), respectively. The values \(\omega=1,100,300\) correspond to transverse oscillator lengths approximately \(2.0\,\mu{\rm m}\), \(0.20\,\mu{\rm m}\), and \(0.115\,\mu{\rm m}\), respectively.

A literal implementation of the spinor ABC would require an effective absorbing boundary whose entrance response contains the tangential spin--momentum branches
\begin{equation}
i\kappa\pm|\boldsymbol\xi|.
\end{equation}
Generic spin-dependent loss or bulk spin--orbit terms should not be expected to reproduce this response without engineering the corresponding boundary impedance. The layer comparisons in Sec.~\ref{subsec:SM_layer_mimics} use these branch impedances as calibration targets and quantify the resulting
finite-window agreement. A microscopic physical realization
remains outside the present study.